\documentclass[twocolumn]{aastex61}

\newcommand\aastex{AAS\TeX}

\shorttitle{\aastex\ Comparing treatments of weak reactions with nuclei in simulations of core-collapse supernovae}
\shortauthors{Nagakura et al.}

\begin{document}
\title{Comparing treatments of weak reactions with nuclei in simulations of core-collapse supernovae}
\correspondingauthor{Hiroki Nagakura}
\email{hirokin@astro.princeton.edu}

\author{Hiroki Nagakura}
\affiliation{Department of Astrophysical Sciences, Princeton University, Princeton, NJ 08544}
\affiliation{TAPIR, Walter Burke Institute for Theoretical Physics, Mailcode 350-17, California Institute of Technology, Pasadena, CA 91125, USA}

\author{Shun Furusawa}
\affiliation{Interdisciplinary Theoretical and Mathematical Sciences Program (iTHEMS), RIKEN 2-1 Hirosawa, Wako, Saitama 351-0198, Japan}

\author{Hajime Togashi}
\affiliation{Nishina Center for Accelerator-based Science, RIKEN 2-1 Hirosawa, Wako, Saitama 351-0198, Japan}

\author{Sherwood Richers}
\affiliation{North Carolina State University, Raleigh, NC 27607}

\author{Kohsuke Sumiyoshi}
\affiliation{Numazu College of Technology, Ooka 3600, Numazu, Shizuoka 410-8501, Japan}

\author{Shoichi Yamada}
\affiliation{Advanced Research Institute for Science \&
Engineering, Waseda University, 3-4-1 Okubo,
Shinjuku, Tokyo 169-8555, Japan}
\affiliation{Department of Science and Engineering, Waseda University, 3-4-1 Okubo, Shinjuku, Tokyo 169-8555, Japan}

\begin{abstract}
We perform an extensive study of the influence of nuclear weak interactions on core-collapse supernovae (CCSNe), paying particular attention to consistency between nuclear abundances in the equation of state (EOS) and nuclear weak interactions. We compute properties of uniform matter based on the variational method. For inhomogeneous nuclear matter, we take a full ensemble of nuclei into account with various finite-density and thermal effects and directly use the nuclear abundances to compute nuclear weak interaction rates. To quantify the impact of a consistent treatment of nuclear abundances on CCSN dynamics, we carry out spherically symmetric CCSN simulations with full Boltzmann neutrino transport, systematically changing the treatment of weak interactions, EOSs, and progenitor models. We find that the inconsistent treatment of nuclear abundances between the EOS and weak interaction rates weakens the EOS dependence of both the dynamics and neutrino signals. We also test the validity of two artificial prescriptions for weak interactions of light nuclei and find that both prescriptions affect the dynamics. Furthermore, there are differences in neutrino luminosities by $\sim10\%$ and in average neutrino energies by $0.25-1\,\mathrm{MeV}$ from those of the fiducial model. We also find that the neutronization burst neutrino signal depends on the progenitor more strongly than on the EOS, preventing a detection of this signal from constraining the EOS. 

\end{abstract}




\section{Introduction} \label{sec:intro}
A star more massive than about $10M_\odot$ will build up an iron core massive enough to collapse under its own gravity. During the few tenths of a second of collapse, the stellar material of ever-increasing density emits a large number of electron neutrinos, making the matter very neutron-rich. Once the core reaches a few times nuclear saturation density, the strong nuclear force quickly stiffens the equation of state (EOS), halting the collapse and driving a shock wave out through the rest of the star. If enough energy is absorbed below the shock, the shock is driven through the rest of the star, resulting in a core-collapse supernova (CCSN, see \cite{2017hsn..book.1095J} for a review). However, the mechanism by which the explosion is driven is not well understood. 

In the absence of direct probes of the internal dynamics, simulations are used in an effort to understand the important physics. CCSN dynamics depend sensitively on the flow of energy, momentum, and lepton number determined by the EOS and weak interaction rates, but both the EOS and weak interactions are very challenging to treat accurately. Properties of matter at super-nuclear densities are poorly understood at the moment due to theoretical difficulties and weak constraints from experiments and observations of neutron stars (see a recent review by \citet{2017RvMP...89a5007O}). One major challenge facing a theoretical description of nuclear abundances relevant to CCSNe is the presence of extremely heavy nuclei at densities and electron fractions where no experimental data are available. The uncertainty of the state of these nuclei obscures the effects of neutrino-nucleus interactions, which can significantly influence CCSNe.

The first theoretical nuclear EOS applicable to CCSNe simulations was developed by \citep{1985ncnd.conf..131H}. Subsequently, Lattimer-Swesty \citep{1991NuPhA.535..331L} and Shen \citep{1998PThPh.100.1013S,1998NuPhA.637..435S} were developed and used extensively in the CCSNe community for decades. The Lattimer-Swesty EOS uses Skyrme-type interactions with multi-body terms for uniform matter and a compressible liquid drop model for non-uniform matter. On the other hand, the Shen EOS  was developed based on the relativistic mean field (RMF) approach with TM1 parameter set for uniform matter and the Thomas-Fermi model for non-uniform matter. During the last few years, extensive studies have been conducted on development of better EOS (see e.g., \citet{2014EPJA...50...46F,2017RvMP...89a5007O}), and new EOS tables are available online (see e.g., stellarcollapse.org, CompOSE EOS database\footnote{\url{https://compose.obspm.fr}}, and NCT\footnote{\url{http://user.numazu-ct.ac.jp/~sumi/eos}}). Some are constructed using the common RMF approach but with different nuclear parameter sets (see e.g., \citet{2012ApJ...748...70H}), while others employ a liquid-drop model with Skyrme type interactions \citep{2017arXiv170701527D}. The Skyrme DFT method proposed by \citet{2016arXiv161208992P} is commonly used to construct EOS, but this method is based on an approximate Hartree-Fock computation that relies on Skyrme-type effective interaction. The variational method (VM) is an alternative to the Skyrme DFT method  that is capable of producing an EOS with a much more realistic treatment of nuclear forces (see Sec.~\ref{sec:VM} for more details).

During the collapse and post-bounce phases of CCSNe, the matter temperature of the inner core exceeds  0.5 MeV. In such a high temperature state, nuclear statistical equilibrium (NSE) is achieved and the nuclear abundances are determined entirely by the density, temperature, and electron fraction. The outer core at large radii is at a lower density and temperature, and a full reaction network would be needed to obtain the most realistic nuclear abundances in this region. In both the Lattimer-Swesty and Shen formulations, the abundances of heavy nuclei are computed under the single nucleus approximation (SNA), in which the ensemble of heavy nuclei is accounted for using a single representative nucleus. However, the average mass and charge number of heavy nuclei in the SNA are significantly different from those computed using a full ensemble of nuclei in some thermodynamical states relevant to CCSNe \citep{2009ApJ...707.1495S,1984ApJ...285..294B}. In addition, the full ensemble of nuclear data is mandatory to evaluate accurate electron capture rates, which is one of the most important input physics for the gravitational collapse of an iron core \citep{2003PhRvL..90x1102L,2003PhRvL..91t1102H}. Indeed, \citet{2016ApJ...816...44S} and \citet{2017JPhG...44i4001F} showed that electron capture rates are dominated by $\sim 100$ nuclei during the collapse phase, which should be included in CCSN simulations to accurately model deleptonization of the inner core.

 Weak interactions other than electron capture also depend on the nuclear species, which means imposing the SNA in simulations of CCSNe carries the risk of making neutrino-matter interaction rates incorrect.  Motivated by the need for a realistic EOS, multi-nuclear EOS were developed to account for an ensemble of atomic nuclei (e.g., \citealt{2011ApJ...738..178F,2010NuPhA.837..210H}). Considering the abundances of a full ensemble of individual nuclei in addition to thermodynamical quantities in the EOS allows the weak interaction rate for each nucleus-neutrino interaction to be computed separately (see also e.g. the Nulib opacity table in \citet{2015ApJS..219...24O}).

In the last decade, observations and nuclear experiments have rapidly progressed and they give some constraints on properties of the EOS at super-nuclear densities (see, e.g., \citet{2017RvMP...89a5007O} and references therein). The precise measurements of neutron star masses of $M_{{\rm NS}}  \sim 2M_\odot$ \citep{2010Natur.467.1081D,2013Sci...340..448A} provide a stringent constraint on lower limit of the maximum mass of a neutron star. More recently, a joint analysis of the gravitational waves and their electromagnetic counterparts from the neutron star merger event GW170817 \citep{2017PhRvL.119p1101A,2017ApJ...848L..13A,2017ApJ...848L..12A} suggests that the maximum neutron star mass is $M_{{\rm NS}} \lesssim 2.2 M_{\sun}$ \citep{2017ApJ...850L..19M,2017PhRvD..96l3012S}. In addition, according to an analysis of the tidal deformabilities of neutron stars obtained from GW170817 by \citet{2017PhRvD..96l3012S}, the neutron star radius for $M_{{\rm NS}} \sim 1.35 M_{\sun}$ should be less than $13$km. This is consistent with other constraints by X-ray observations (see e.g., \citet{2013ApJ...774...17S}). Although ambiguity remains, all of these constraints drastically narrow down possible candidates of the nuclear EOS for uniform matter. The allowable parameter space is so narrow, in fact, that both the Lattimer-Swesty and Shen EOS do not satisfy all constraints. The Shen EOS predicts a cold neutron star radius for $M_{{\rm NS}} = 1.35 M_{\sun}$ of around $14.5$km, which is larger than the observational constraint \citep{2017NuPhA.961...78T}. On the other hand, the symmetry energy (the energy difference between symmetric nuclear matter and pure neutron matter) in the Lattimer-Swesty EOS violates experimental constraints \citep{2017ApJ...848..105T}.

In this paper, we improve on our previous study by using a consistent treatment of nuclear abundances between the EOS and weak interaction rates on heavy and light nuclei. We describe the essence of these improvements and study the impact of the consistent treatment of nuclear weak interactions on CCSNe dynamics and neutrino signals. \citet{2018PhRvC..97c5804N} carried out a comparison between the VM and RMF EOS, but they employ the SNA under the Thomas-Fermi model for inhomogeneous nuclear matter. In our previous paper \citep{2017JPhG...44i4001F}, we investigated the differences between CCSN simulations using the multi-nuclear VM and RMF EOS and a common weak reaction table between them \citep{2010NuPhA.848..454J,2003PhRvL..90x1102L,2000NuPhA.673..481L}\footnote{See also other previous works e.g., \citet{Couch:2012gh,Suwa:2012xd,Pan:2017tpk} for the study of EOS dependence of CCSNe simulations.}. The weak reaction rate table was computed by old nuclear-statistical-equilibrium (NSE) prescriptions \citep{2011ApJ...738..178F,2013ApJ...772...95F}. In this study we improve on this comparison by performing simulations that use both EOS with weak interaction rates consistent with nuclear abundances in each EOS. \citet{2016ApJ...816...44S, 2018JPhG...45a4004T} also recently investigated the sensitivity of CCSNe to electron-captures by heavy nuclei under the consistent treatment of nuclear abundance between EOS and electron capture of heavy nuclei. They focus mainly on the impact of uncertainties of reaction rates but did not discuss the influence of inconsistency of nuclear abundance between EOS and weak reaction rates, which we will address in this paper. In addition, we also study the influence of electron and positron captures by light nuclei on CCSNe, which were also neglected in these previous papers.

This paper is organized as follows. We first introduce the essence of constructing multi-nuclear VM EOS in Sec.~\ref{sec:VM} and new electron and positron capture rates of nuclei in Sec.~\ref{sec:eleposicap}. In Sec.~\ref{numericalsetup}, we summarize numerical set up for CCSN simulations. We present the results of CCSN simulations in Sec.~\ref{sec:results}, which is divided into three subsections. In Sec.~\ref{subsec:VMvsRMF}, we apply our new VM and FYSS EOS to spherically symmetric CCSN simulations of a 11.2 $M_{\sun}$ progenitor in \citet{2002RvMP...74.1015W}, compare the difference of CCSN dynamics between two EOS, and discuss the differences between this and our previous study \citep{2017JPhG...44i4001F}.  In Sec.~\ref{subeq:lightdepe}, we investigate the influence of light nuclei by performing two more simulations with artificial prescriptions of light nuclei and quantify the errors. In Sec.~\ref{subsec:prodepe} we study the progenitor dependence of these results by carrying out simulations for 15, 27 and 40 $M_{\sun}$ progenitors and discuss the "Mazurek's law", mass-radius relation of protoneutron star (PNS), and neutrino signals. Finally we wrap up this paper with summary in Sec.~\ref{sec:summary}.

\section{Multi-nuclear Variational EOS}\label{sec:VM}
In this work, we use a recently-developed EOS based on the variational method (VM) \citep{2013NuPhA.902...53T,2017NuPhA.961...78T}, in which the two-body nuclear potential is expressed with the Argonne v18 potential \citep{1995PhRvC..51...38W} and three-body forces are included to satisfy the experimental constraints of the saturation properties. This EOS was recently extended to include an ensemble of nuclei \citet{2017JPhG...44i4001F}, employing the same framework as in \citet{2017NuPhA.957..188F}. In addition, this new multi-nuclear VM EOS takes into account washout of shell effects in the nuclear-statistical-equilibrium (NSE) computations, which brings a large change to weak interaction rates at high temperatures ($\sim 20(40)\%$ in the early (late) collapse phase, see \citet{2017PhRvC..95b5809F} for more details). With these advances, the VM EOS is a highly realistic nuclear EOS well-suited for simulations of CCSNe. The multi-nuclear VM EOS is publicly available at \url{http://user.numazu-ct.ac.jp/~sumi/eos/.}
 
In this section, we describe the essence of physics in multi-nuclear VM EOS  \citep{2013NuPhA.902...53T,2017NuPhA.961...78T,2017JPhG...44i4001F}. The free energy density is calculated with variational many-body theory. The Hamiltonian consists of the kinetic, the AV18 two-body potential  \citep{1995PhRvC..51...38W} and the UIX three-body potential \citep{1983NuPhA.401...59C,1995PhRvL..74.4396P} terms.  This formulation is extended to non-uniform nuclear matter at sub-nuclear densities to compute the free energy density  $f_{n,p}$ of unbound nucleons that drip from heavy nuclei, as well as the bulk energies $E^{bulk}_{AZ}$ of heavy nuclei. The free energy density of non-uniform nuclear matter is given by
\begin{eqnarray}
f & =& \eta f_{n,p} +  \sum_{AZ}  n_{AZ}  \times \nonumber \\
  & & \left\{ \kappa T \left[ \ln \left(\frac{ n_{AZ}}{g_{AZ} (M_{AZ} T/2\pi \hbar ^2 )^{3/2}  }\right)- 1 \right]  + M_{AZ} \right\},
\label{eq:total}
\end{eqnarray}
where $n_{AZ}$, $M_{AZ}$, and $g_{AZ}$ are the number density, mass, and internal degrees of freedom  for nucleus with the mass number $A$ and atomic number $Z$. The quantities $f_{n,p}$ are defined as the local free energy density for dripped nucleons in the vapor volume $V'$. The latter is calculated as $V'=V-V_N$ where $V$ is the total volume and $V_N$ is the volume occupied by nuclei. The excluded volume effects for nucleons and nuclei are accounted for in 
%$\eta = V_N/V=1/\{1 - \sum_{AZ} ( n_{AZ} A / n_{sAZ}  )  \}$
$\eta = V'/V=1 - \sum_{AZ} ( n_{AZ} A / n_{sAZ} )$
and $\kappa = 1-n_{B}/n_{0}$, respectively, where $n_{0}$ is the nuclear saturation density of symmetric matter.
% The excluded volume effects for nucleons and nuclei are accounted for in $\eta= 1/[1 - \sum_{AZ}( n_{AZ} A/n_{sAZ})]$ and {\bf $\kappa=n_B/n_0$}, respectively, where $n_0$  is the nuclear saturation density of symmetric matter.
We also take into account the dependence of the saturation densities of heavy nuclei $n_{sAZ}$ on the iso-spin of each nucleus, the proton fraction $Z/A$, and temperature.

The mass of a heavy nucleus with  atomic number $6\leq Z \leq 1000$ and mass number $A \leq 2000$ 
is expressed as
\begin{eqnarray}
M_{AZ}&=&E^{bulk}_{AZ} + E^{surf}_{AZ}+E^{Coul}_{AZ}+E^{shell}_{AZ}.
\label{eq:mass}
\end{eqnarray}
The  surface energy $E^{surf}_{AZ}$ and Coulomb energy $E^{Coul}_{AZ}$  depend on number densities of uniformly-distributed dense electrons and dripped nucleons, and on shape changes of heavy nuclei from normal droplets to bubbles just below nuclear normal density.
The shell energies of heavy nuclei $E^{shell}_{AZ}$, which represent quantum effects such as neutron- and proton-magic numbers and pairing, are taken from experimental and theoretical mass data 
\citep{2014NDS...120....1A,2005PThPh.113..305K}.
 The temperature dependencies of  $E^{shell}_{AZ}$ and $g_{AZ}$ are also phenomenologically taken into account. The mass model of the light nuclei with $Z<6$ is given by  $M_{AZ}=M_{AZ}^{data}+\Delta E^{Coul}_{AZ} +\Delta E^{Pauli}_{AZ}+\Delta E^{self}_{AZ}$ where $M_{AZ}^{data}$ is the experimental mass, $\Delta E^{Coul}_{AZ}$ is Coulomb energy shift, and 
 a quantum approach is incorporated
to evaluate Pauli-energy shift $\Delta E^{Pauli}_{AZ}$ and self-energy shift $\Delta E^{self}_{AZ}$  \citep{2009PhRvC..79a4002R,2010PhRvC..81a5803T}.

We also use an EOS computed with relativistic mean field theory (RMF) using the TM1 parameter set \citep{2017NuPhA.957..188F}, which we call the FYSS EOS, as a point of comparison for the VM EOS described above.

\section{Electron and positron captures}\label{sec:eleposicap}
We calculate weak interaction rates of nuclei consistently with the nuclear abundances provided by the EOS. 
The electron capture rates for heavy nuclei are evaluated in the same way as in \citet{2017PhRvC..95b5809F}.
For some nuclei, we use theoretical reaction data in \citet{2000NuPhA.673..481L}, \citet{2003PhRvL..90x1102L},
\citet{1994ADNDT..56..231O} and \citet{1982ApJ...252..715F}, which are based on the shell model or its extension.
 It should be noted, however, that these theoretical computations do not cover the all of nuclei that appear in CCSNe. We adopt an analytical function of the $Q$-value \citep{2003PhRvL..90x1102L} for electron capture rates onto neutron-rich and/or heavy nuclei where data is unavailable\footnote{We refer the reader to Fig.~5 in \citet{2017PhRvC..95b5809F} which displays the corresponding data or formula for each nucleus in (N,Z) plane.}:%
\begin{eqnarray}
 \lambda_{AZ} &=&\frac{({\rm ln}2)B}{K} \left( \frac{T}{m_e c^2} \right)^{5} \times \nonumber \\
&& \left[ F_4(\eta_{AZ}) - 2 \chi_{AZ} F_3(\eta_{AZ}) + \chi_{AZ}^2 F_2(\eta_{AZ})  \right] , \label{eq:appro} 
\end{eqnarray}
where  $K=6146$ sec., $\chi_{AZ}=(Q_{AZ} - \Delta E)/T$, $\eta_{AZ}=(\mu_e +  Q_{AZ} -\Delta E)/T$, $\mu_e$ is the electron chemical potential , and $F_k$ is the relativistic Fermi integral of order $k$. The $Q$-value is defined as $Q_{AZ}=M_{AZ}-M_{A,Z-1}$. It should be noted that our treatment of nuclear masses is different from that used by \citet{2003PhRvL..90x1102L} who use data available for isolated nuclei to set the nuclear masses in dense matter. By contrast, we utilize the original mass formula to take into consideration various in-medium effects such as the surface tension reduction and the shell washout (see also Eq.~(\ref{eq:mass})). The most important in-medium effect is the Coulomb energy shift. The other in-medium effects are of little influence for core-deleptonization because their impact becomes large only at high densities at which point the chemical potentials of electrons are much higher than the energy differences arising from in-medium effects.

We employ the fitting formula developed by \citep{2003PhRvL..90x1102L} only for very heavy nuclei ($A \gtrsim 100$) for which experimental data are unavailable. Although this formula tends to overestimate electron capture rates as pointed out by \citet{2016ApJ...816...44S,2017PhRvC..95b5809F,2018JPhG...45a4004T}, it allows us to obtain electron capture rates for individual nuclei. The individual reaction rates are mandatory for ensuring consistency with nuclear abundances in the EOS. On the other hand, our previous paper employed the FD+RPA prescription in \citet{2010NuPhA.848..454J} which has been already integrated nuclear abundances computed by a specific NSE and EOS model. We use the detailed balance relation to evaluate the rate of the inverse process of neutrino absorption on heavy nuclei.

% As pointed out by \citet{2016ApJ...816...44S,2017PhRvC..95b5809F,2018JPhG...45a4004T}, the present electron capture rates of heavy nuclei covered by the fitting formula \citep{2003PhRvL..90x1102L} are overestimated. With this caveat in mind, we adopt an analytical function of the $Q$-value \citep{2003PhRvL..90x1102L} for electron capture rates onto neutron-rich and/or heavy nuclei where data is unavailable\footnote{We refer the reader to Fig.~5 in \citet{2017PhRvC..95b5809F} which displays the corresponding data or formula for each nucleus in (N,Z) plane.}:%

% which could change the deleptonization of the iron core and impact the neutrino signals at the neutronization burst.

%The discrepancy of reactions rates for very heavy nuclei ($A \gtrsim 100$) also affects to the dynamics of CCSNe (see Sec.~\ref{subsec:VMvsRMF}).

As for light nuclei, we include the following weak interactions in our simulation \citep{2016EPJWC.10906002F},
\begin{eqnarray}
&{\rm (elpp):}  & \hspace{0.5mm} \nu_e + ^{2}{\rm H}  \longleftrightarrow  e^- + p + p , \label{eq:lweak1}\\
&{\rm (ponn): } & \hspace{0.5mm} \bar{\nu_e} + ^{2}{\rm H}  \longleftrightarrow  e^+ + n + n , \label{eq:lweak2} \\
&{\rm (el2h): } & \hspace{0.5mm} \nu_e + n+n  \longleftrightarrow  e^- + ^{2}{\rm H} , \label{eq:lweak3}\\
&{\rm (po2h): } & \hspace{0.5mm} \bar{\nu_e} + p+p  \longleftrightarrow  e^+ + ^{2}{\rm H} , \label{eq:lweak4} \\
&{\rm (el3he): }& \hspace{0.5mm} \nu_e + ^{3}{\rm H}  \longleftrightarrow  e^- + ^{3}{\rm He} , \label{eq:lweak5} \\
&{\rm (po3h): } & \hspace{0.5mm} \bar{\nu_e} + ^{3}{\rm He}  \longleftrightarrow  e^+ + ^{3}{\rm H}. \label{eq:lweak6} 
\end{eqnarray}
For neutrino absorptions on deuterons  (Eqs.~\ref{eq:lweak1}~and~\ref{eq:lweak2}), we use vacuum cross-section data \citep{2001PhRvC..63c4617N}. To account for the medium modifications of deuteron mass (i.e., $\Delta E^{Coul}_{AZ}$,  $\Delta E^{Pauli}_{AZ}$, and  $\Delta E^{self}_{AZ}$), we introduce the shifted neutrino injection energy $E^*_{\nu}=E_{\nu} + m^*_{^{2}{\rm H}} - m_{^{2}{\rm H}}$, where the deuteron mass is given by  $m_{^{2}{\rm H}}$ in vacuum and $m^*_{^{2}{\rm H}}$ in medium. The in-medium deuteron mass is evaluated by the same mass model in the EOS. The neutrino absorption rate of Eq.~(\ref{eq:lweak1}) is then
\begin{equation}
1/\lambda (E_\nu)= n_{^{2}{\rm H}} \int {\rm d} p_e \left[ \frac{{\rm d}\sigma_{\nu ^{2}{\rm H}}}{{\rm d} p_e}(E^*_{\nu}) \right] (1-f_e(E_e)] ,
\end{equation} 
where $ n_{^{2}{\rm H}}$ is deuteron number density and $f_e$ denotes the Fermi-Dirac distribution of electrons, and similarly for Eq.~(\ref{eq:lweak2}). We evaluate the rates of the electron and positron capture on two nucleons forming a deuteron (leftward reactions of Eqs.~(\ref{eq:lweak1})~and~(\ref{eq:lweak2})) through the detailed balance with the absorption rate. We also ignore other minor reactions involving deuterons, such as pair processes and neutral-current breakup reactions, since they are less dominant than the charged current interactions of deuterons as described in Eqs.~(\ref{eq:lweak1}) to (\ref{eq:lweak4}) \citep{2013ApJ...774...78F,2015ApJ...801...78N}.

We evaluate the rates of electron and positron capture on deuterons (Eqs.~(\ref{eq:lweak3})~and~(\ref{eq:lweak4})) under the assumption that the matrix elements of electron and positron captures are equivalent to those of neutrino absorptions (Eqs.~(\ref{eq:lweak1})~and~(\ref{eq:lweak2})). This assumption is reasonable for CCSNe conditions, in which the energy deposited to the relative motion between two nucleons is negligible. The result is that  
\begin{equation}
 \frac{{\rm d} \sigma_{\mathrm{el2h}}} {{\rm d} p_{\nu}}  \approx \frac{1}{2} \frac{{\rm d} \sigma_{\nu ^{2}{\rm H}}} {{\rm d} p_e},
\end{equation}
where the factor of 2 comes from the difference in spin degrees of freedom between neutrinos and electrons. 

The three-nucleon nuclei $^3 \rm H$ and $^3 \rm He$ interact with neutrinos via breakup or charge exchange, the latter of which is the dominant neutrino opacity source.  Therefore, we treat only the charge exchange reaction as described in Eqs.~(\ref{eq:lweak5})~and~(\ref{eq:lweak6}). We calculate those rates as 
\begin{equation}
1/\lambda (E_\nu)= n_{^{3}{\rm H}} \left[\frac{G^2_F V^2_{ud}}{\pi (\hbar c)^4}\right] p_e E_e [ 1-f_e(E_e)] B(GT),
\end{equation} 
where  $ n_{^{3}{\rm H}}$ is  triton number density, $B(GT)=5.87$, and $V_{ud}=0.967$ as in \citet{2016EPJWC.10906002F}. We do not take in-medium effects on nuclear masses into account in these reactions. We do not expect this assumption to have a large effect, but further work would be required to determine the quantitative effects.

We neglect inelastic scatterings between alpha-particles and neutrinos in this study for simplicity and because, similar to the recoil treatment in nucleon-neutrino interactions, the energy exchange by scatterings between neutrinos and alpha-particles is quite small. However, inelastic scattering with alpha particles may still play a non-negligible role to the shock revival if the shock wave has reached close to the revival \citep{2007ApJ...667..375O}, though the precise treatment of such a small energy exchange is technically challenging for mesh-based methods.

\section{Numerical set up}\label{numericalsetup}
\startlongtable
\begin{deluxetable}{cccccc}
\tablecaption{Summary of models \label{tab:model}}
\tablehead{
Model & EOS & ECPH & EPCPL & Progenitor Mass\\
 &  &  &  &   ($M_\odot$)
}
\startdata
V112 & VM & new & new & 11.2  \\
F112 & FYSS & new & new & 11.2 \\
OV112 & VM & old & no & 11.2 \\
OF112 & FYSS & old & no & 11.2 \\
NV112 & VM & new & no & 11.2 \\
DV112 & VM & new & nucleons & 11.2 \\
V15 & VM & new & new & 15 \\
V27 & VM & new & new & 27 \\
V40 & VM & new & new & 40 \\
\enddata
\tablecomments{The ECPH column denotes whether rates of electron capture on heavy nuclei are inconsistent (old) or consistent (new) with nuclear abundances in the EOS. The EPCPL column denotes whether electron and positron captures on light nuclei are included (new), neglected (no), or treated as weak interactions on free nucleons (nucleons). OV112 and OF112 are the same models as in \citet{2017JPhG...44i4001F}.}
\end{deluxetable}

All simulations presented in this paper are performed by our multi-dimensional (multi-D) neutrino radiation hydrodynamics code \citep{2014ApJS..214...16N,2017ApJS..229...42N}. We solve the special relativistic Boltzmann equations for neutrino transport. 
%The hydrodynamics and gravity are treated as Newtonian.
We employ the central scheme \citep{2000JCoPh.160..241K} for numerically solving the Newtonian hydrodynamics equations (see also \citealt{2011ApJ...731...80N}). We solve the Poisson equation for Newtonian gravity using the mass integration method. Details of the formulation and numerical methods in our code were presented in a series of our previous papers \citep{2012ApJS..199...17S,2014ApJS..214...16N,2014PhRvD..89h4073S,2017ApJS..229...42N} and the reliability of our code was established by a detailed comparison with Monte-Carlo transport method in \citet{2017ApJ...847..133R}. For all simulations, we employ the same weak interaction rates as those used in \citep{2018ApJ...854..136N}, except for electron and positron captures on nuclei. In our simulations, we cover the spatial domain of $0 \le r \le 5000 {\rm km}$ on a spherical-polar grid with $384$ radial grid points. The neutrino momentum space is gridded into 20 grid points in energy ($0 \le \epsilon \le 300$MeV) and 10 grid points in polar angle ($0^{\circ} \le \bar{\theta} \le 180^{\circ}$). The momentum-space resolution is almost the same as that in other spherically symmetric Boltzmann simulations (see e.g., \citet{2008ApJ...688.1176S,2012ApJ...747...73L,2012ApJ...760...94L}). As shown in \citet{2017ApJ...847..133R}, however, the limited resolution yields results in errors at some level. For instance, neutrino luminosities at our current resolution are likely underestimated by several percent due to numerical diffusion. Though fully resolved calculations are desirable, we do not expect that qualitative trends in the weak dependence of CCSN dynamics on weak interactions vary strongly with resolution.

%{\bf The momentum-space resolution is almost the same as that in other spherically symmetric Boltzmann simulations (see e.g., \citet{2008ApJ...688.1176S,2012ApJ...747...73L,2012ApJ...760...94L}). As shown in \citet{2017ApJ...847..133R}, however, the resolution yields outcome with errors at some level. For instance, neutrino luminosities in our resolutions would be underestimated serveral percents due to numerical diffusions. It should be noted, however, that the qualitative trend in weak interaction dependence of CCSN dynamics and neutrino signals is not sensitive to the resolution, which was also confirmed by the extensive resolution study in \citet{2017ApJ...847..133R}.}

%that the extensive 

%which will be done in the future work
%the momentum-space resolution is not enough to obtain the defenitive values for some quantities relevant to neutrinos (e.g., the energy flux), we have employed the resolution due to limitations of computational resources. For more quantitative arguments to directly compare to real observations,

Tab.~\ref{tab:model} lists a summary of the models simulated in this work. V112 is our fiducial model which employs 11.2 $M_{\sun}$ progenitor in \citep{2002RvMP...74.1015W}, VM EOS and the improved electron and positron captures on nuclei as described in Sec.~\ref{sec:eleposicap}. F112 is identical to V112, except in that it employs the FYSS EOS. Note that the weak reaction rates in F112 are consistently computed with the nuclear abundances provided by the FYSS EOS. Models OV112 (VM EOS) and OF112 (FYSS EOS) are from our previous work \citep{2017JPhG...44i4001F}. These models employ an old FD+RPA prescription \citep{2010NuPhA.848..454J} for electron captures on heavy nuclei and neglect electron and positron captures on light nuclei. In model NV112, all input physics are the same as V112 except that weak interactions with light-nuclei are neglected. In model DV112, weak interactions with light nuclei except for alpha particles are artificially replaced by weak reactions with their constituent free nucleons, which is similar to the treatment of light nuclei in previous studies that employ the SNA. Models V15, V27 and V40 use the same physics as V112, but use 15$M_\odot$, $27M_\odot$, and $40M_\odot$ progenitors, respectively. We run simulations up to $400$ms after the bounce for V112, and $500$ms for other progenitors. We stop the simulation of V112 earlier than those of other progenitors since the outermost mass shell of the progenitor in our computational domain would reach the shock wave between $400$ms and $500$ms.

%  For 11.2 $M_{\sun}$ progenitor in \citet{2002RvMP...74.1015W}.
%We start our simulations from the time when the central density $\rho_c$ reaches $\sim 6 \times 10^{9} {\rm g/cm^3}$.

\section{Results}\label{sec:results}

\subsection{VM EOS vs. RMF EOS} \label{subsec:VMvsRMF}

\begin{figure}
\vspace{15mm}
\epsscale{1.2}
\plotone{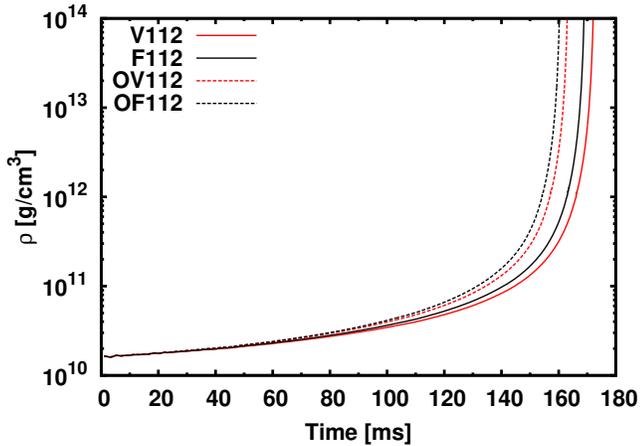}
\caption{Time evolution of central density during the collapsing phase for V112, F112, OV112 and OF112.
\label{graph_tevocentrho_preb}}
\end{figure}

\begin{figure*}
\vspace{15mm}
\epsscale{1.2}
\plotone{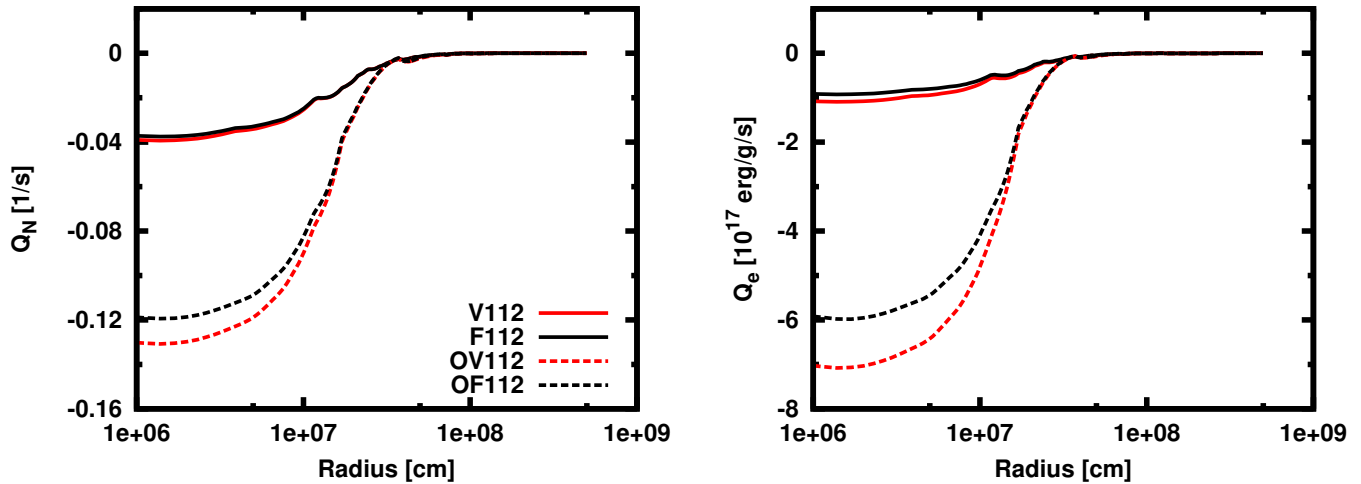}
\caption{Radial profiles of deleptonization rate ($Q_{\rm N}$, left) and neutrino cooling rate ($Q_{\rm e}$, right) during the collapse of a $11.2\,M_\odot$ progenitor when the central density is $\rho_c=6\times10^9\,\,\mathrm{g\,cm^{-3}}$ which corresponds to the time of beginning of our simulation. V112 and F112 employ the VM and FYSS EOS, respectively. OV112 and OF112 are similar, but use an old prescription for weak interaction rates on heavy nuclei that is inconsistent with the nuclear abundances in the EOS. Consistency between weak interactions and nuclear abundances in the EOS results in significantly slower cooling and deleptonization.
\label{IniVMvsRMFcompare}}
\end{figure*}

Figure~\ref{graph_tevocentrho_preb} shows the time evolution of central density during the collapsing phase for V112, F112, OV112 and OF112.
The contraction of inner core in V112 and F112 evolves more slowly than in previous models (OV112 and OF112)\footnote{Hereafter, we sometimes collectively refer V112 and F112 as "new models", and OV112 and OF112 as "previous models."}, which is due to the difference in deleptonization rates. This is apparent in Fig.~\ref{IniVMvsRMFcompare}, which shows that both the rates of cooling and deleptonization are considerably slower in the new models. The stable nuclei with $N = 28$ (e.g., $^{54}$Fe) are very susceptible to electron capture and generate high deleptonization rates (see also Fig.~9 in \citet{2017PhRvC..95b5809F}). Our new NSE model takes the washout of the shell effect into account, which reduces the nuclear abundance of $A \sim 50$ ($N \sim 28$).  As a result, both electron degeneracy and thermal pressure are removed less efficiently in the new models and the collapse is decelerated.

\begin{figure*}
\vspace{15mm}
\epsscale{1.2}
\plotone{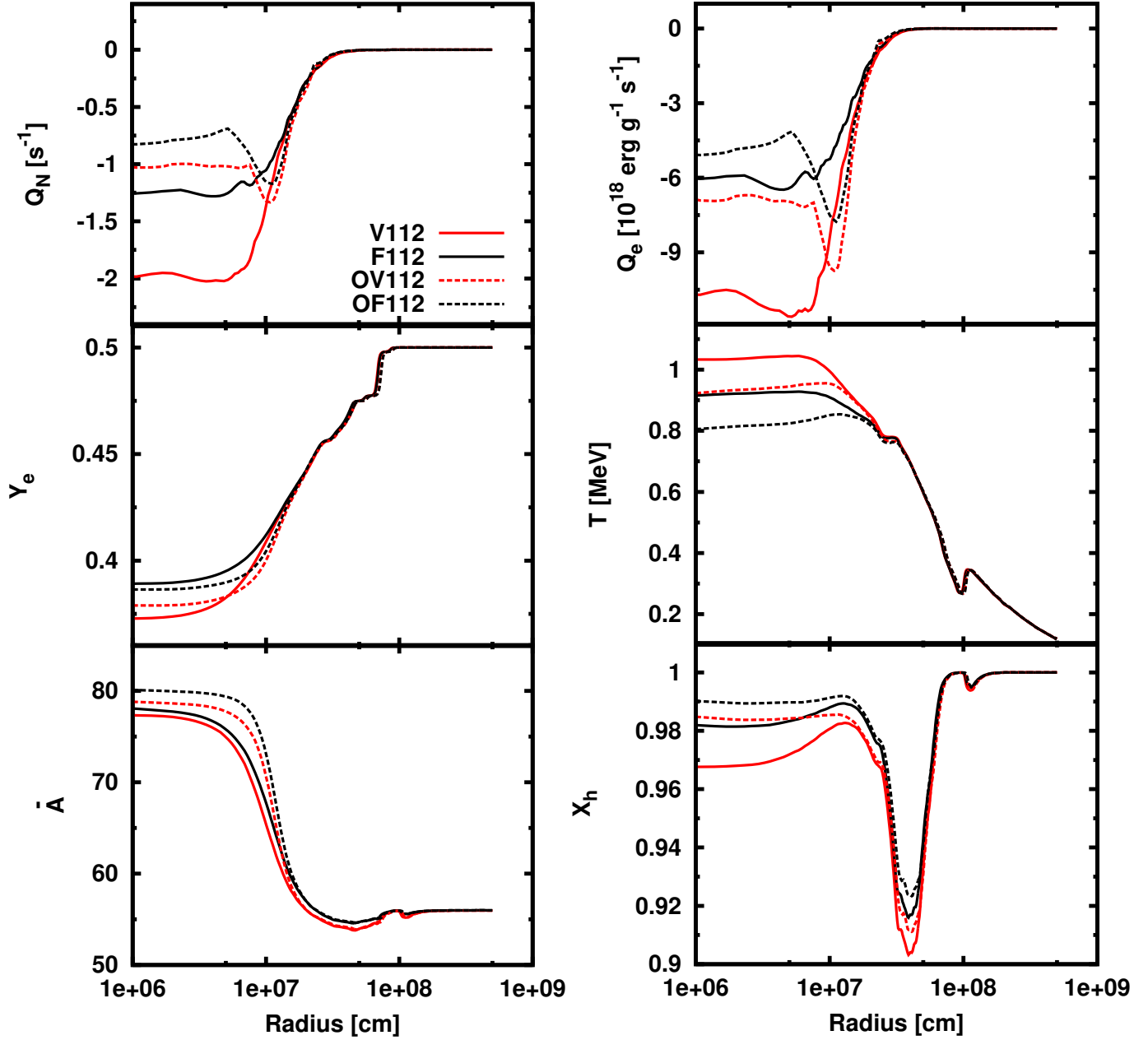}
\caption{Comparison between different EOS and weak interaction assumptions during the collapse phase when the central density is $\rho_c=10^{11} {\rm g/cm^3}$. $Q_N$ is the net leptonization rate, $Q_e$ is heating rate, $Y_e$ is electron fraction, $T$ is the temperature, $\bar{A}$ is average number of nucleons in heavy nuclei, and $X_h$ is the mass fraction of heavy nuclei.
V112 and F112 employ the VM and FYSS EOS, respectively. OV112 and OF112 are similar, but use an old prescription for weak interaction rates on heavy nuclei that is inconsistent with the nuclear abundances in the EOS.
\label{rhoc1e11VMvsRMFcompare}}
\end{figure*}

\begin{figure*}
\vspace{15mm}
\epsscale{1.2}
\plotone{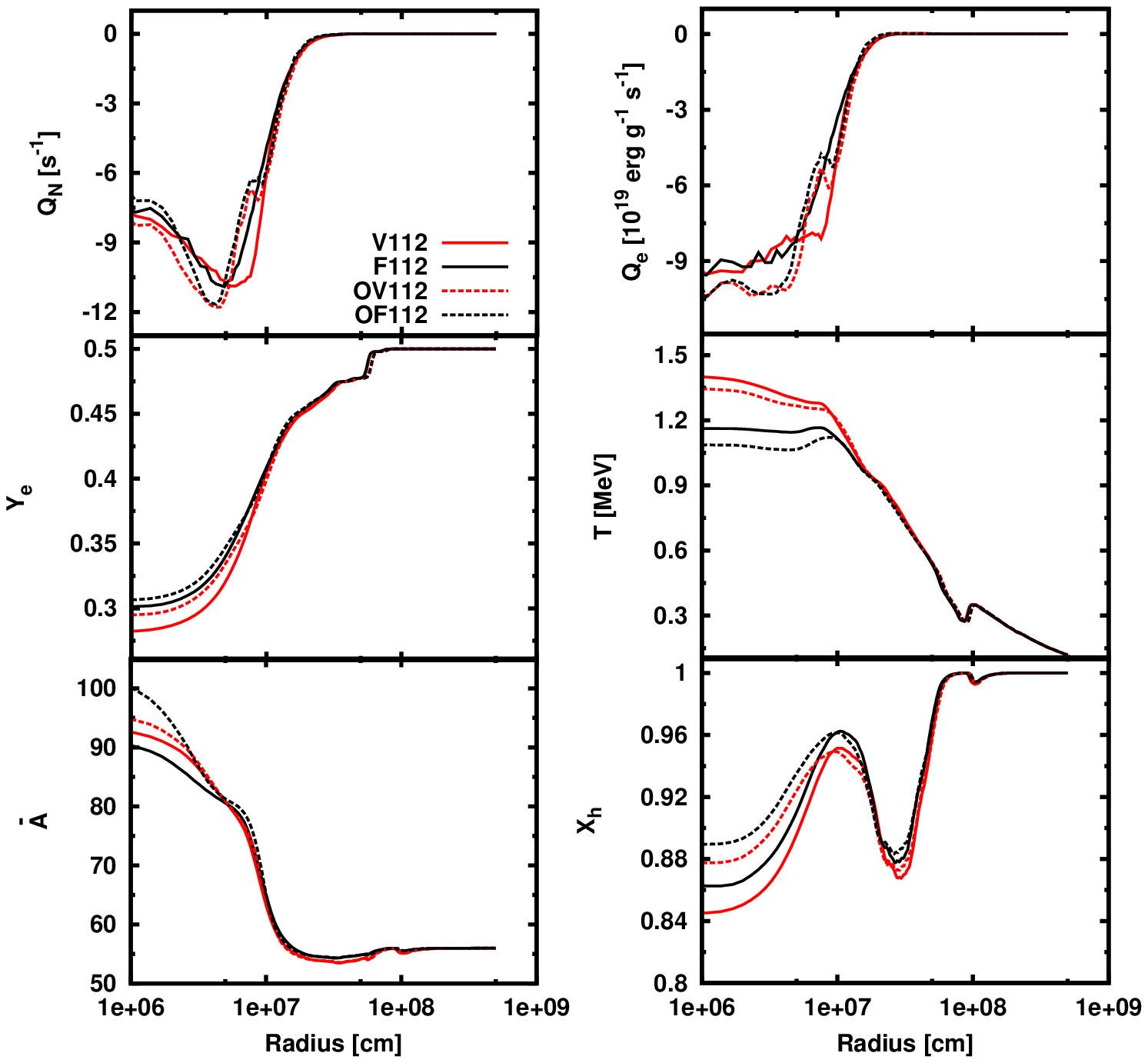}
\caption{Comparison between different EOS and weak interaction assumptions during the collapse phase when the central density is $\rho_c=10^{12} {\rm g/cm^3}$. Quantities and models are the same as in Fig.~\ref{rhoc1e11VMvsRMFcompare}.
\label{rhoc1e12VMvsRMFcompare}}
\end{figure*}

\begin{figure*}
\vspace{15mm}
\epsscale{1.2}
\plotone{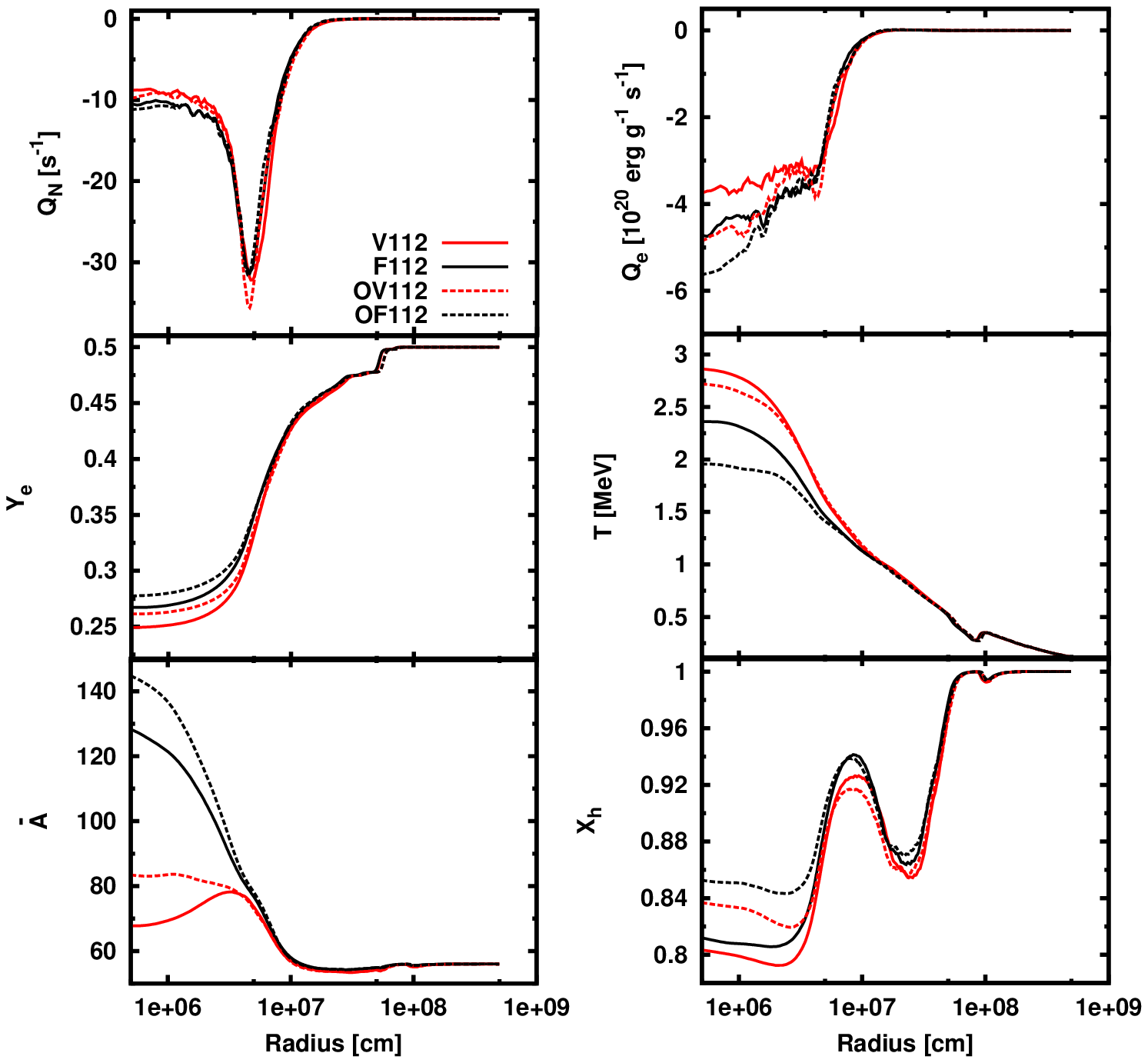}
\caption{Comparison between different EOS and weak interaction assumptions during the collapse phase when the central density is $\rho_c=10^{13} {\rm g/cm^3}$. Quantities and models are the same as in Fig.~\ref{rhoc1e11VMvsRMFcompare}.
\label{rhoc1e13VMvsRMFcompare}}
\end{figure*}

We display radial profiles of several important quantities when the central density reaches $\rho_c=10^{11}\,\mathrm{g\,cm}^{-3}$ in Fig.~\ref{rhoc1e11VMvsRMFcompare}. The top two panels show that both the deleptonization and cooling rates in the inner core are higher in models using the VM EOS than in models using the FYSS EOS. As discussed in \citet{2017JPhG...44i4001F}, this is mainly due to the temperature difference resulting from the different EOSs where the temperature tends to be higher in VM EOS than in FYSS EOS (see e.g., the right middle panel in Fig.~\ref{rhoc1e11VMvsRMFcompare}). The VM EOS has a smaller symmetry energy of nuclei than the FYSS EOS, which yields smaller mass fraction of heavy nuclei. As a result, the entropy per baryon and the adiabatic index tend to be higher in the VM EOS than in the FYSS EOS, which facilitate the increase of temperature during adiabatic contraction.
% higher entropy per baryon due to 
% gives smaller mass fraction of heavy nuclei and then higher entropy per baryon. The adiabatic index in VM EOS is larger than in FYSS EOS, 
% The top two panels show that both the deleptonization and cooling rates in the inner core in models using the VM EOS (V112 and OV112) are higher than those in models that use the FYSS EOS (F112 and OF112). The larger deleptonization rate allowed by the VM EOS over the FYSS EOS reduces electron fraction more efficiently, as seen in the middle left panel. This weakens the electron degeneracy pressure, allowing more compaction, which results in a larger temperature (middle right panel). The higher temperatures reduce the average mass and abundance of heavy nuclei (bottom panels) by dissociating heavy nuclei into lighter ones. These trends persist independent of the treatment of weak interactions on heavy nuclei. This in turn reduces the opacity of coherent neutrino-nucleus scattering.  As a result, neutrinos are more transparent in VM EOS models, which facilitates deleptonization, allowing the difference between two models to spread further.

As displayed in the middle left panel of Fig.~\ref{rhoc1e11VMvsRMFcompare}, the difference of electron fraction at the center between models V112 and F112 is larger than the difference between models OV112 and OF112. This is attributed to the fact that the previous models artificially suppress the difference between EOS by sharing between them a single common table of electron capture rates on heavy nuclei. Both deleptonization and cooling rates  (upper two panels of Fig.~\ref{rhoc1e11VMvsRMFcompare}) for $r\lesssim 10^7\,\mathrm{cm}$ in the new models are higher than those of the previous models, though the opposite is true outside of this region just as in the initial conditions (see Fig.~\ref{IniVMvsRMFcompare}). The difference of deleptonization between new and old models also come from another reason which is the different treatment of electron capture on $A \gtrsim 100$ heavy nuclei. Because of the more rapid deleptonization and neutrino cooling in the new models, the contraction of inner core proceeds faster than those in the previous models at this phase. The models using the VM EOS (V112 and OV112) cool and deleptonize faster than models using the FYSS EOS (F112 and OF112). Note that at this early time, though the deleptonization rate in model OF112 is lower than that in model F112, the central electron fraction in model F112 is larger than in model OF112. This is just a vestige of the opposite ordering of deleptonization rates at the onset of collapse (Fig.~\ref{IniVMvsRMFcompare}).

We also find that models using the VM EOS show a bigger difference of neutrino reactions between new and previous models than models using the FYSS EOS (see Fig.~\ref{IniVMvsRMFcompare} and upper panels in Fig.~\ref{rhoc1e11VMvsRMFcompare}). As we have already mentioned, the models with the VM EOS have higher inner core temperature than do models with the FYSS EOS. Since higher temperature facilitates deleptonization by more electron capture, the difference between new and previous models is more prominent when using the VM EOS.

The dynamics deviate remarkably between V112 and F112  with increasing the central density. In Figs.~\ref{rhoc1e12VMvsRMFcompare} and \ref{rhoc1e13VMvsRMFcompare}, we display the same quantities as in Fig.~\ref{rhoc1e11VMvsRMFcompare}, but at later times during the collapse phase when the central density $\rho_c$ reaches $10^{12}$ and $10^{13} {\rm g\,cm}^{-3}$, respectively. The electron fraction remains lower and the temperature higher in model V112  than in model F112 at these later times. Since both deleptonization and neutrino cooling are suppressed by neutrino absorption and scattering in the high density region, the matter profile becomes less sensitive to the difference of electron capture rates and the scattering opacities for neutrinos. As a result, the difference of electron fraction and temperature between two models remains locked in even in the late phases of collapse. Indeed, the difference in deleptonization rate between V112 and F112 is subtle at the center (see e.g., top left panel in Fig.~\ref{rhoc1e13VMvsRMFcompare}) despite the fact that the scattering opacity in V112 is remarkably lower than F112. This difference in opacity is apparent in the bottom left panel of Fig.~\ref{rhoc1e13VMvsRMFcompare}, noting that  opacity of coherent scattering of neutrinos by heavy nuclei is proportional to $\bar{A}^2$. The difference of $\bar{A}$ is mainly attributed to the difference of temperature. In fact, the ordering of $\bar{A}$ follows the opposite ordering of temperature (see middle right panel of Fig.~\ref{rhoc1e13VMvsRMFcompare}).

\begin{figure*}
\vspace{15mm}
\epsscale{1.2}
\plotone{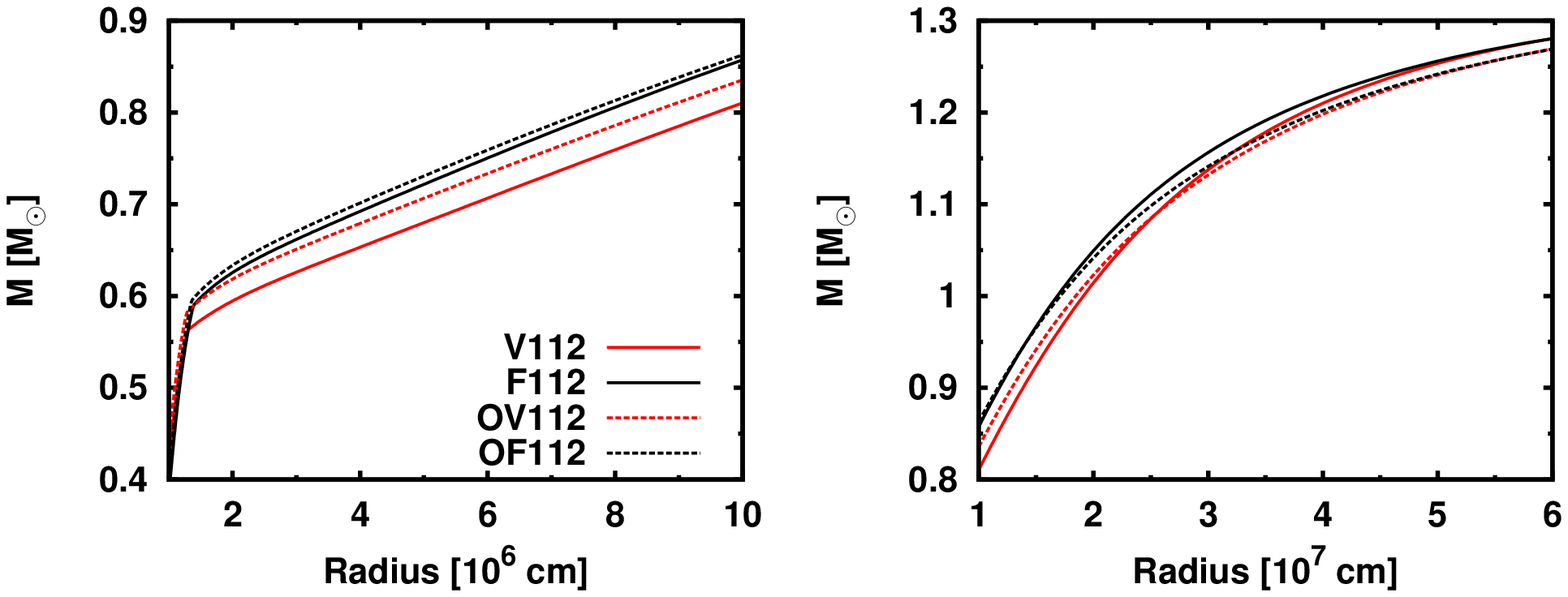}
\caption{Enclosed mass versus radius at the time of core bounce. The left panel shows the region where $10^{6}\,\mathrm{cm} \leq r \leq 10^{7}\,\mathrm{cm}$ and the right panel shows the region where $10^{7}\,\mathrm{cm} \leq r \leq 6 \times 10^{7}\,\mathrm{cm}$.
\label{BounceVMFYSS_MR}}
\end{figure*}

Fig.~\ref{BounceVMFYSS_MR} shows the enclosed mass as a function of radius at the time of core bounce. The outer core in models using the VM EOS (V112 and OV112) are less compact than that those in models using the FYSS EOS (F112 and OF112) for $r \lesssim 6 \times 10^{7}\,\mathrm{cm}$. Above this radius, the compactness ($M/r$) in models using the same neutrino physics converges. However, the new models (V112 and F112) have slightly larger compactness than the previous (OV112 and OF112) models. This is also due to the previously mentioned detail that the new models collapse more slowly. Though this makes the inner core (left panel) less dense and increases the length of time between the onset of collapse and core bounce. The different neutrino interactions between the old and new models have little effect on the matter at large radii (right panel), so an increased collapse time in the new models allows the matter at large radii more time to fall in before core bounce, resulting in higher compactness.

\begin{figure*}
\vspace{15mm}
\epsscale{1.0}
\plotone{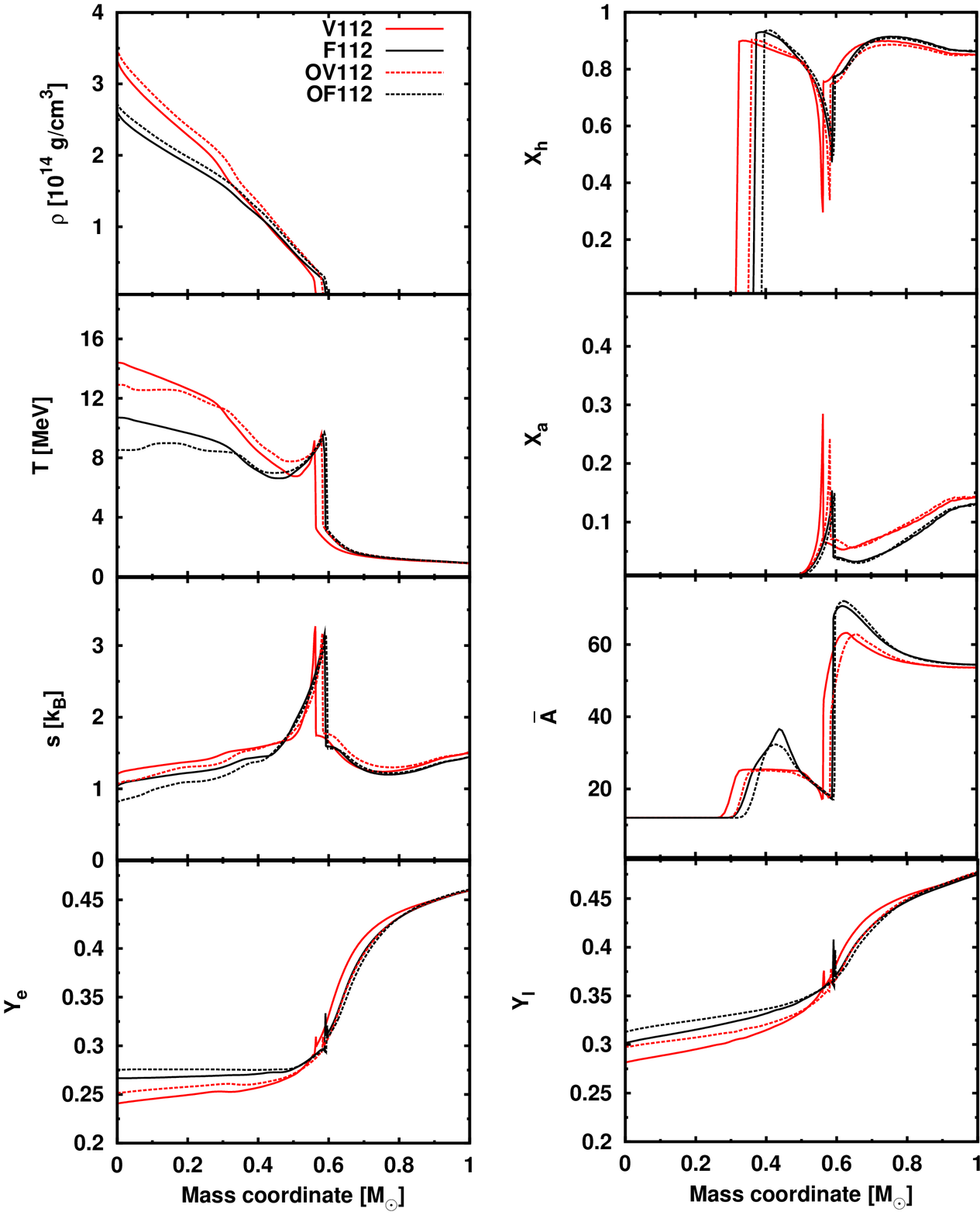}
\caption{Comparison between different EOS and weak interaction assumptions at core bounce. From top to bottom, we plot baryon mass density ($\rho$), temperature ($T$), entropy per baryon ($s$) and electron fraction ($Y_e$) in the left column, while we plot the mass fraction of heavy nuclei ($X_h$), the mass fraction of light nuclei ($X_a$), the average mass number ($\bar{A}$) and the lepton fraction ($Y_l$) in the right column. All quantities are displayed as a function of mass coordinate in this figure.
\label{BounceVMFYSS}}
\end{figure*}

%% \begin{figure*}
%% \vspace{15mm}
%% \epsscale{1.0}
%% \plotone{BounceVMvsRMFcompare_s3_radius.eps}
%% \caption{Same as Fig.~\ref{BounceVMFYSS} but as a function of radius.
%% \label{BounceVMFYSS_radius}}
%% \end{figure*}

We display radial profiles of several important quantities at the time of core bounce (defined as when the post-shock entropy per baryon first reaches $3 k_B$) in Fig.~\ref{BounceVMFYSS} as a function of mass coordinate.
% and in Fig.~\ref{BounceVMFYSS_radius} as a function of radius. 
As was apparent in Fig.~\ref{BounceVMFYSS_MR}, we find that the new models (V112 and F112) have systematically lower central density and electron fraction but higher temperature and entropy than those in the old models (OV112 and OF112). The systematic differences can be interpreted as follows. Except at the onset of collapse, the deleptonization rate is larger in the new models, which yields smaller electron and proton fractions in iron core. The smaller proton fraction makes nuclear matter stiffer since the repulsive force is enhanced by the larger asymmetry between proton and neutron mass fractions. In addition, the smaller electron fraction leads to the higher temperature because of the reduction of electron degeneracy pressure during collapse. Thus, the thermal contribution to the total pressure in the new models is larger than in the previous models. Because of these two effects, the core bounce in the new models takes place at a lower central density than the previous models. The lepton fraction is directly related to the inner core mass at bounce, so the ordering of the mass coordinate of the shock at bounce (entropy plot in Fig.~\ref{BounceVMFYSS}) follows the ordering of the lepton fraction in the inner core.

As shown in the third panel on the right in Fig.~\ref{BounceVMFYSS}, there is a substantial amount of light nuclei just under the shock wave. This tells us that the prompt shock wave is not powerful enough to completely photodisintegrate heavy nuclei into nucleons (see e.g., Figs.~\ref{graphXaXhYeS_lightdepe}). As we will discuss later, various light nuclei are generated in the almost entire region of post-shock flows in the post-bounce phase. The role of light nuclei to CCSN dynamics and neutrino signals will be analyzed in the next subsection.

\begin{figure*}
\vspace{15mm}
\epsscale{1.2}
\plotone{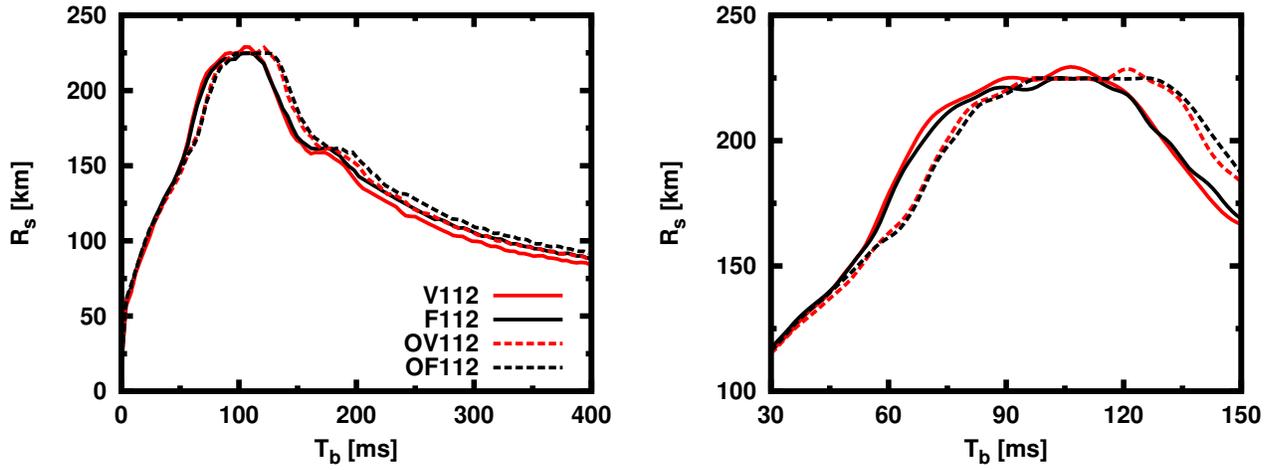}
\caption{Evolution of the shock radius as a function of time after core bounce. The right panel magnifies the peak region. Models V112 and OV112 use the VM EOS, while models F112 and OF112 use the FYSS EOS. Models V112 and F112 use a new prescription for weak interactions on nuclei, while models OV112 and OF112 use an old prescription (see Secs.~\ref{sec:VM} to \ref{sec:eleposicap}).
\label{shockevoVMvsRMF}}
\end{figure*}

\begin{figure*}
\vspace{15mm}
\epsscale{1.2}
\plotone{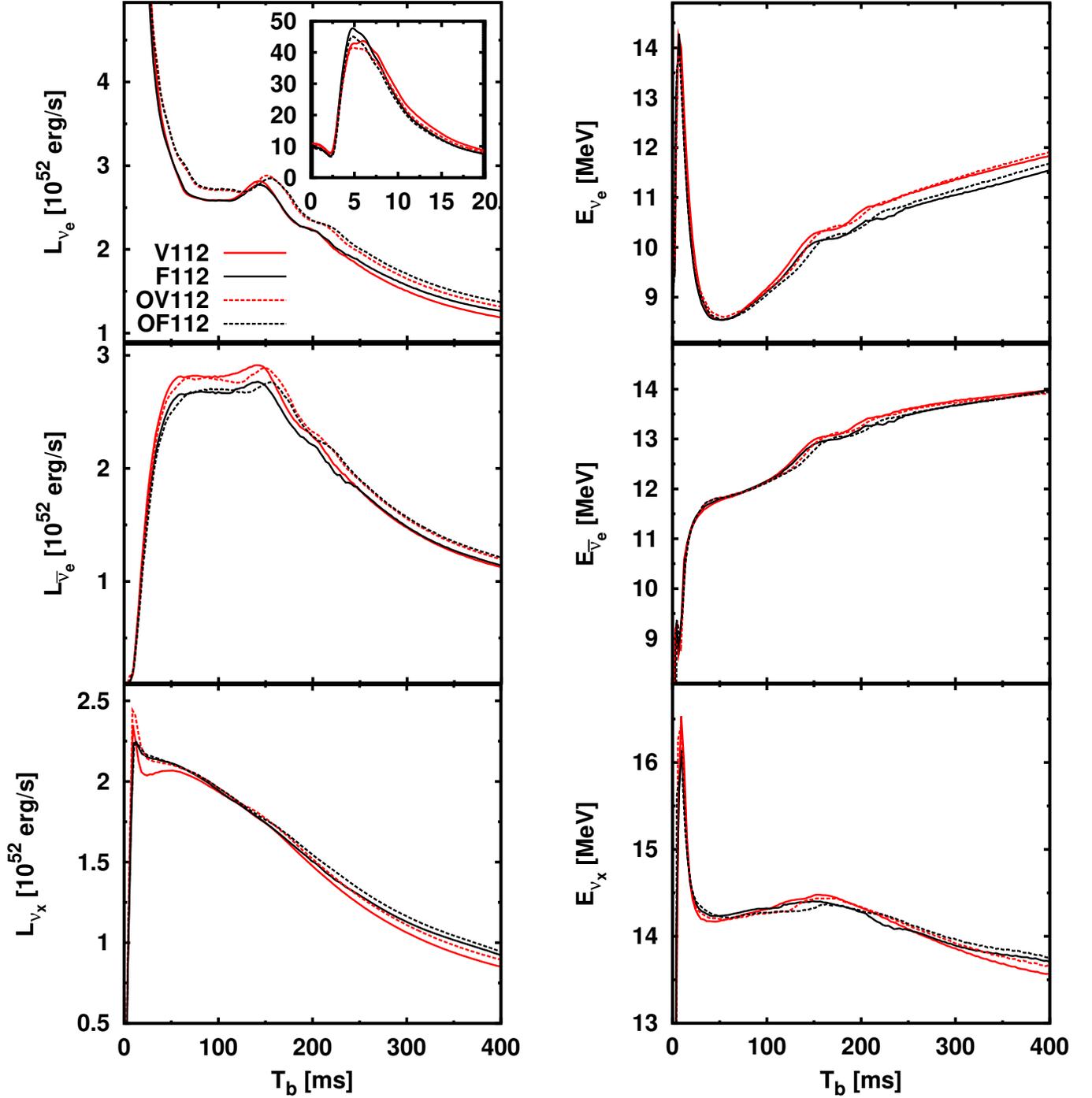}
\caption{Time evolution of the neutrino luminosity (left) and the mean energy (right) for $\nu_e$ (top), $\bar{\nu}_e$ (middle) and $\nu_x$ (bottom). Models V112 and OV112 use the VM EOS, while models F112 and OF112 use the FYSS EOS. Models V112 and F112 use a new, consistent prescription for weak interactions on nuclei, while models OV112 and OF112 use and old prescription. We measure these quantities at $r = 500$km.
\label{Neutrino_VMvsRMFcompare}}
\end{figure*}

Despite the fact that there are many differences between V112 and F112 during the collapse and bounce phases, the shock trajectory is very similar up to $\sim 200$ms after bounce (see Fig.~\ref{shockevoVMvsRMF}). This attributable to the fact that the smaller mass of the iron core (due to the smaller electron fraction) weakens the shock and the higher central density (due to the softness of VM EOS) in V112 works in the opposite direction to strengthen the prompt shock wave, leaving a small effect on the net result. A more significant difference between the two models appears at ${\rm T}_{\rm b} > 200$ms, where the shock wave in V112 recedes faster than F112. This can be understood as follows. In the early post-bounce phase, the thermal contribution (which is insensitive to the difference of EOS) plays an important role to determine the structure of post shock flows and it smears out the EOS dependence. As time advances, neutrinos take away the thermal energy of PNS, and then the EOS dependence appears and spreads gradually in the post-shock region. The softness of the VM EOS makes the structure of post-shock flow more compact, which results in pulling the shock wave back to PNS. On the other hand, one may speculate that the compact structure of the PNS progressively works to push the shock wave out by increasing of neutrino heating since neutrinos possess higher average energy (see the top and middle right panels in Fig.~\ref{Neutrino_VMvsRMFcompare}). However, the neutrino heating works less efficiently to compensate the regression of shock wave since the shock wave has already receded to $R_s < 150$km, which means that the gain region has a small volume. It should be noted that the difference of shock trajectory between two models are at most $10\%$ even in the late phase, which suggests that the dynamics of CCSNe are insensitive to differences between the two EOS.

Here, we put a caution about the above argument. The almost identical shock trajectory between two EOS at the early post bounce phase may be an artifact due to the spherical symmetry, since the symmetry artificially suppresses the prompt convection. We find that the radial profiles of entropy and electron fractions are different between two models, which means that the strength of convection would also be different (see discussions in \citet{2003PhRvL..91t1102H}). As discussed in \citet{2018ApJ...854..136N}, the prompt convection generates outward-traveling acoustic waves, which pushes the bounce shock wave out. In addition, the convection-driven inhomogeneities can be seed perturbations of other fluid instabilities such as the standing accretion shock instability (SASI), which also progressively works to strength the shock wave. 

As shown in the right panel of Fig.~\ref{shockevoVMvsRMF}, we also find that the shock trajectories between new and previous models start to deviate at $\sim 50$ms after bounce.
%The shock waves in the new models evolve faster than the previous models.
%{\bf From $\sim 50$ms, the shock radii in new models expand more rapidly than those in previous models, and then they start receding at $\sim 120$ms which is $\sim 20$ms earlier than in previous models.}
After this time, the shock radii in the new models expand more rapidly than those in previous models. Subsequently, they start receding at $\sim 120$ms, which is $\sim 20$ms earlier than in the previous models.
We attribute this to differences in the arrival time of Si/Si-O composition interface to the shock front. Once the Si/Si-O interface reaches the shock front, the rapid decrease of density reduces the ram pressure of mass accretion, which facilitates the expansion of shock wave. As we have already mentioned, the length of time from the onset of collapse to core bounce in the new models is longer than in the previous models, which means that the Si/Si-O interface reaches a smaller radius at the time of bounce (see also Fig.~\ref{BounceVMFYSS_MR}) and so hits the shock wave sooner after bounce. Importantly, such an arrival time shift of the Si/Si-O interface would play a key role for the shock-revival as was recently reported in up-to-date multi-D simulations (see, e.g., axisymmetric CCSN simulations in \citet{2018MNRAS.477.3091V}).

The neutrino luminosity and mean energy are summarized in Fig.~\ref{Neutrino_VMvsRMFcompare}. The neutronization burst is marked by the sharp increase and drop of electron-type neutrino ($\nu_e$) luminosity, which can be clearly seen in the inset on the top left panel of the figure. Note that we measure the neutrino luminosity at $r=500$km, and it takes $\sim 1.5$ms to reach this radius from the neutrino-breakout radius at $r \sim 60$km. The peak luminosities ($L_{\rm p}$) for all models are in the range $4 < L_{\rm p}/(10^{53}\, {\rm erg\,s^{-1}}) < 5$. As we can see in this figure, model V112 has a smaller $L_{\rm p}$ than those in F112. This may be attributed to the fact that model F112 has a slightly stronger bounce shock wave than model V112. As discussed already, the deleptonization of the iron core during the collapsing phase in F112 is less active than V112, which forms a larger inner core and then generates a stronger shock wave in F112.

After the neutronization burst, the time evolution of $\nu_e$ luminosity is almost identical between V112 and F112 until ${\rm T}_{\rm b} \sim 200$ms. As shown in Fig.~\ref{Neutrino_VMvsRMFcompare}, the luminosity in model V112 is slightly smaller than in model F112. However, the mean energy for $\nu_e$ in model V112 is higher than in model F112. The difference can be interpreted in a similar way as the EOS dependence of the shock trajectory. In the late post-bounce phase, the weakening contribution to the post-shock flow from thermal pressure makes the EOS dependence of the PNS structure more prominent. This makes the post-shock structure in model V112 is more compact than in model F112. As a result, the neutrinosphere in model V112 is at a smaller radius, which results in less luminous neutrino emission but with higher mean energy. We also find a systematic EOS dependence of the peak of $\bar{\nu}_e$ luminosity ($50 \lesssim {\rm T}_{\rm b} \lesssim 150$ms).  Since VM EOS models have slightly higher temperature around the inside of the PNS, the thermal emission of $\bar{\nu}_e$ is stronger than in the FYSS EOS models.

We find a systematic difference in the time evolution of $\nu_e$ luminosity between new and previous models. As shown in the top left panel of Fig.~\ref{Neutrino_VMvsRMFcompare}, the time evolution is a bit faster for the new models. This is again due to the difference of the structure of the outer core. However, the systematic difference between the new and previous models is not so remarkable for electron-type anti-neutrinos ($\bar{\nu}_e$) in the early post bounce phase (see the middle left panel). Since the $\bar{\nu}_e$ neutrinosphere is located at a smaller radius than the $\nu_e$ neutrinosphere, it is more shielded from the accretion of the outer core and the light curve of $\bar{\nu}_e$ in the early post-bounce phase is less affected by the difference of the outer core structure. This systematic difference for $\bar{\nu}_e$ between new and previous models arises from $\sim 230$ms after the bounce, which corresponds to the time when the $1.2 M_\odot$ mass coordinate reaches the shock. The systematic differences in the accretion rate are due to slightly faster collapse in the new models (see the discussion in reference to Fig.~\ref{BounceVMFYSS_MR}). Interestingly, there are no such systematic differences of light curve in heavy leptonic neutrinos ($\nu_x$) between new and previous models (see the bottom left panel in Fig.~\ref{Neutrino_VMvsRMFcompare}). This may be attributed to the fact that the characteristics of $\nu_x$ signals are determined deeper inside of PNS than other species, which is less sensitive to the structure of outer core, at least up to $\sim 400$ms after the bounce.

\subsection{Influence of light nuclei} \label{subeq:lightdepe}

We turn our attention to the impact of light nuclei on CCSNe. In this study, we analyze three models with different treatments of weak reactions with light nuclei. Our fiducial model V112 takes into account electron and positron captures on light nuclei self-consistently with VM EOS. Model NV112 artificially neglects all interactions with light nuclei. Weak reactions with light nuclei in model DV112 are artificially treated as if the light nuclei were dissociated into free nucleons. See Sec.~\ref{numericalsetup} for model details.

In the collapse phase, no remarkable differences are found in the time evolution of fluid dynamics and the neutrino signals among three models, even though the abundance of light nuclei is not small. Indeed, there are some regions with $X_a > 10 \%$ (see e.g., Fig.~11 in \citet{2017JPhG...44i4001F}). Such an insensitiveness to light nuclei in the collapse phase is attributed to the fact that the large weak reaction rates relevant to heavy nuclei such as the electron capture and coherent scattering make the effect of light nuclei imperceptible. The effect of light nuclei becomes non-negligible in the post-bounce phase, when heavy nuclei become less dominant as a result of photodissociation of heavy nuclei in the post-shock region.

Therefore, we only focus on the post-bounce phase below. Before comparing among three models in detail, we first analyze some important characteristics of light nuclei based on the results of V112, and then quantify how these prescriptions change both CCSN dynamics and neutrino signals.

\subsubsection{Characteristics of light nuclei in CCSNe} \label{subsubsec:Characteristicsoflight}

\begin{figure*}
\vspace{15mm}
\epsscale{1.2}
\plotone{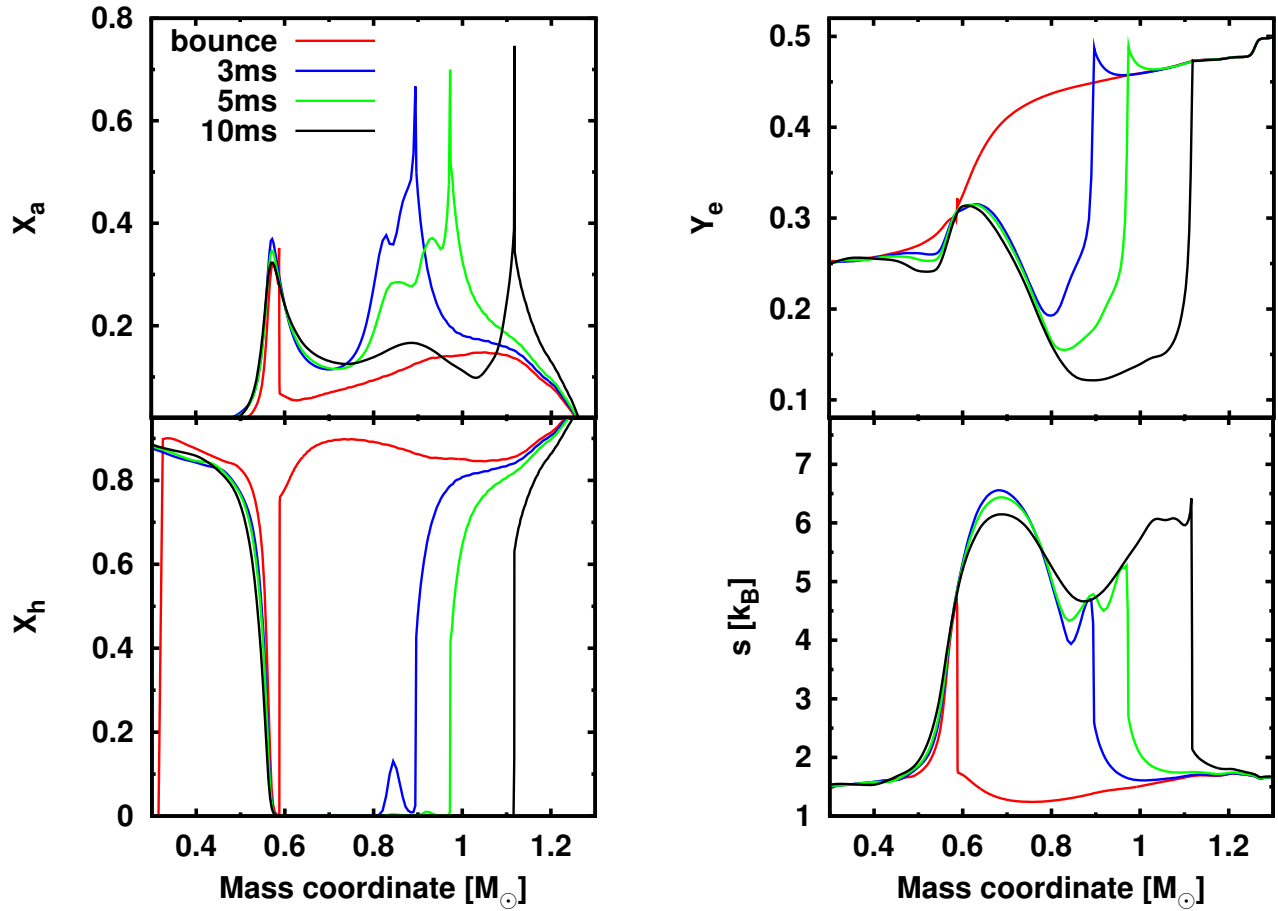}
\caption{Radial profiles of the mass fraction of light nuclei (top left), heavy nuclei (bottom left), electron fraction (top right), and entropy per baryon (bottom right) in model V112. Red, blue, green and black colors denote $0$ms, $3$ms, $5$ms and $10$ms after the bounce, respectively.
\label{graphXaXhYeS_lightdepe}}
\end{figure*}

The radial distributions of mass fractions of light and heavy nuclei in the early post bounce phase (${\rm T}_{\rm b} \le 10$ms) are displayed in left two panels of Fig.~\ref{graphXaXhYeS_lightdepe}. In front of and behind the shock wave, the sharp peak of $X_a$ rises with the rapid drop of $X_h$. We first look at the properties of $X_a$ in the post-shock region. As mentioned already in Sec.~\ref{subsec:VMvsRMF}, the energy of the shock wave is not enough to completely decompose heavy nuclei into nucleons; instead some light-nuclei are formed. In the post-shock flows, the mass fraction of light nuclei decreases inward from the shock, and then again increases around the mass shell of $\sim 0.6 M_{\sun}$. Such an excess of light nuclei is sustained for a long time during the post-bounce phase. The generation and durability of the excess of light nuclei can be understood as follows. At the core bounce, the shock wave is generated around the mass shell of $\sim 0.56 M_{\sun}$ in model V112. In the very early phase, the abundance of light nuclei ($X_a$) behind the shock wave increases with the shock propagation. Once the shock wave is strong enough to destroy heavy nuclei into many nucleons, $X_a$ decreases. As a result, the nascent shock draws the spike profile of radial distributions of light nuclei. Because of the large neutrino opacity, the matter evolves almost adiabatically with time. Although the density increases with time by the contraction of PNS, the constant entropy and proton fraction (right panels in Fig.~\ref{graphXaXhYeS_lightdepe}) work to sustain the same $X_a$. Note that the entropy and proton fraction change in the neutrino diffusion time scale, and so the characteristic time scale of change of the abundance of light nuclei is also dictated by the neutrino diffusion.

Interestingly, the mass fraction of light nuclei in the pre-shock region (in particularly close to the shock wave) drastically change with time in the early post-bounce phase as shown in the top left panel in Fig.~\ref{graphXaXhYeS_lightdepe}. This is due to the fact that many neutrinos are absorbed in the pre-shock matter. At the shock front, neutrinos are absorbed mostly by free protons, but $\sim 10 \%$ and $\sim 1 \%$ contributions by light and heavy nuclei, respectively.

\begin{figure*}
\vspace{15mm}
\epsscale{1.2}
\plotone{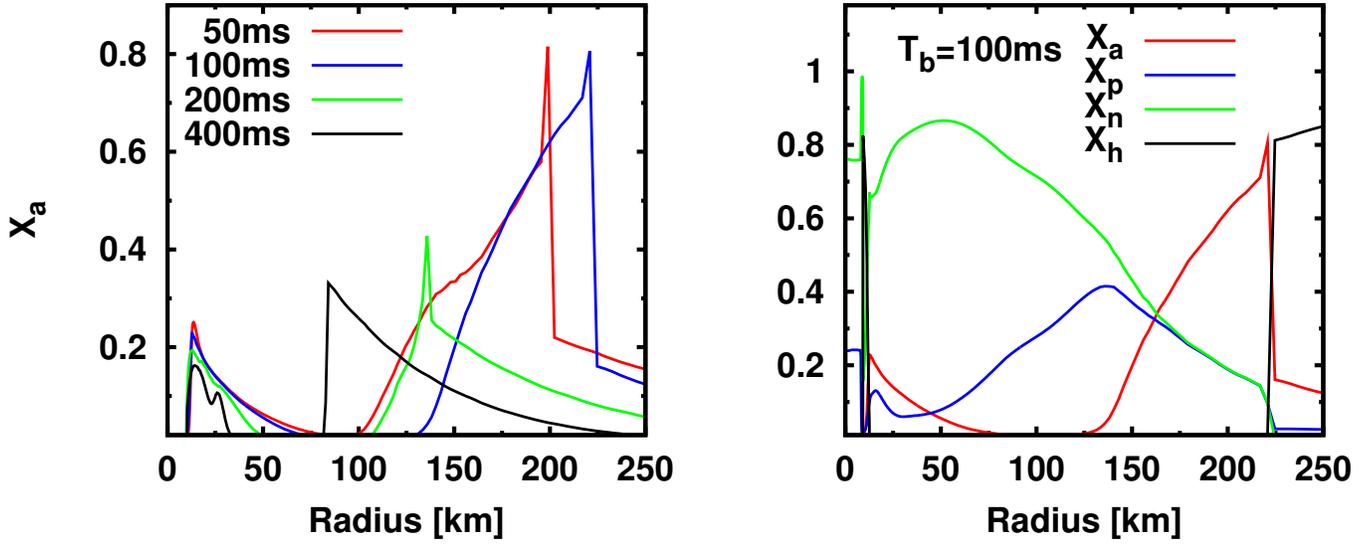}
\caption{Left: Radial distribution of mass fraction of light nuclei in the post-bounce phase for model V112. Color distinguishes the different post-bounce times of $50$ms (red), $100$ms (blue), $200$ms (green) and $400$ms (black) after bounce, respectively. Right: radial distributions of mass fractions of light nuclei (red), free protons (blue), free neutrons (green) and heavy nuclei (black) at the $100$ms after bounce in model V112.
\label{graph_postbouncephase_massfrac}}
\end{figure*}

Fig.~\ref{graph_postbouncephase_massfrac} shows that significant amounts of light nuclei continue to exist in the post-shock region in the mid and late post-bounce phases. Almost all nuclei are photodissociated into light nuclei (mainly alpha-particles) right behind the shock wave, which is clearly visible in the snapshot of $100$ms after bounce in the left panel of Fig.~\ref{graph_postbouncephase_massfrac}. This phase roughly corresponds to the time when the shock wave reaches the maximum radius in V112 (see e.g., Fig.~\ref{shockevoVMvsRMF}). As shown in the right panel of Fig.~\ref{graph_postbouncephase_massfrac}, the mass fraction of light nuclei is more than twice that of nucleons. On the other hand, the abundance of light nuclei decreases as the stalled shock wave recedes, which is displayed in two lines for $200$ms and $400$ms after bounce in the left panel. This is because the kinetic energy per baryons in the pre-shock accretion flows is larger for the smaller shock radius. As a result, the post-shock temperature tends to be high, and then the photodissociation of heavy nuclei become more complete. Therefore, the effect of light nuclei on the dynamics of CCSNe declines with time. It should be noted, however, that the influence of light nuclei would be enhanced in multi-D cases since the shock radius is generally larger than in spherically symmetric simulations.

\begin{figure*}
\vspace{15mm}
\epsscale{1.2}
\plotone{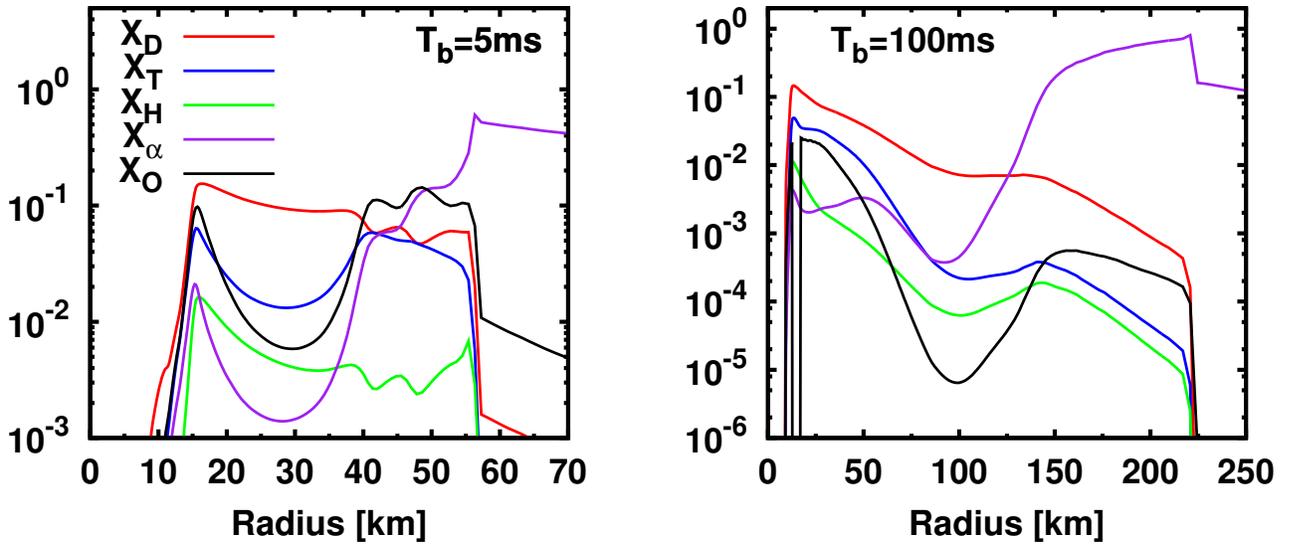}
\caption{Radial profiles of mass fraction of each light nucleus. Different colors denote mass fractions of deuteron ($X_{D}$, red), triton ($X_{T}$, blue), heliton ($X_{H}$, green), alpha ($X_{\alpha}$, purple) and the sum of other light nuclei ($X_{O}$, black), respectively. The left and right panels display ${\rm T}_{\rm b}=5$ms and ${\rm T}_{\rm b}=100$ms, respectively.
\label{graph_postbouncephase_detailedmassfrac}}
\end{figure*}

Next, let us take a look at the abundance of individual components of light nuclei. Their radial distributions are displayed in Fig.~\ref{graph_postbouncephase_detailedmassfrac} at 5ms (left) and 100ms (right) after bounce. As can be seen in both panels, deuterons are the dominant light nucleus around the surface of PNS, but the mass fraction declines with increasing radius. Instead, alpha particles dominate close to the shock wave. It should be noted, however, that the energy transfer to neutrinos by reactions with deuterons is comparable to that with nucleons and is roughly ten times more efficient than that with alpha particles \citep{2008PhRvC..77e5804S,2008PhRvC..78a5806A,2012ApJ...748...70H,2013ApJ...774...78F}. This makes deuterons an effective source of opacity even where the mass fraction is small, giving them a primary role for both cooling and heating.

\begin{figure*}
\vspace{15mm}
\epsscale{1.2}
\plotone{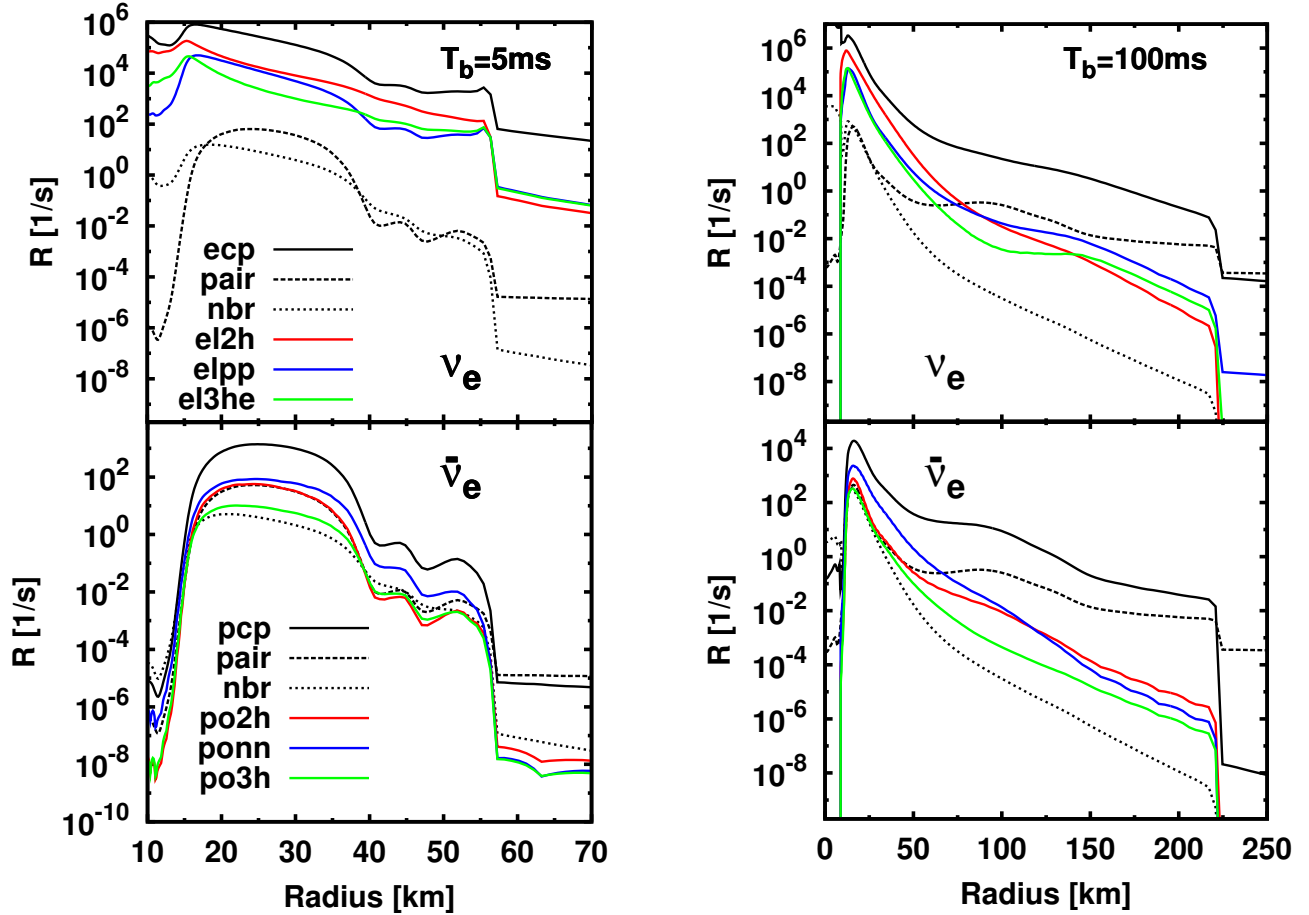}
\caption{Radial profiles of frequency-integrated neutrino emissivities by various interactions. Upper and lower panels are for $\nu_e$ and $\bar{\nu}_e$, respectively. The relevant interactions are electron capture by free protons (ecp), positron capture by free neutrons (pcp), electron-positron pair annihilation (pair), nucleon bremsstralung radiation (nbr), electron neutrino and anti-neutrino absorption on deuterons (elpp and ponn, respectively), electron and positron capture on deuterons (el2h and po2h, respectively), electron capture on $^3$He (el3he), and positron capture on $^3$H (po3h). See Eqs.~(\ref{eq:lweak1})~to~(\ref{eq:lweak6}) for reaction details. The left and right columns correspond to times ${\rm T}_{\rm b}=5$ms and ${\rm T}_{\rm b}=100$ms.
% after bounce.
\label{graphRemistotcompare}}
\end{figure*}

In Fig.~\ref{graphRemistotcompare}, we display radial profiles of frequency-integrated emissivities of neutrinos by weak interactions with light nuclei (Eqs.~(\ref{eq:lweak1})~to~(\ref{eq:lweak6})), along with emission from electron capture, positron capture, electron-positron pair annihilation, and nucleon-nucleon bremsstrahlung radiation. As displayed in these panels, contributions from light nuclei are subdominant to electron and positron capture in the entire post-bounce phase, even though the mass fraction of light nuclei overwhelms that of nucleons in some regions (see also \citet{2016EPJWC.10906002F}). This is attributed to the fact that both electron and positron captures by light nuclei require more energy than those in nucleons since they need to break up or excite from the nuclear bound state, which results in reducing the reaction rates as well as decreasing the average energy of emitted neutrinos. We find that the next-dominant emissivity at ${\rm T}_{\rm b}=5\,\mathrm{ms}$ comes from electron capture on deuterons (el2h), which contributes $\sim 10 \%$ of total $\nu_e$ emissivity at $r \sim 10$km. At ${\rm T}_{\rm b}=100$ms electron neutrino absorption on deuterons (elpp) overwhelms the emissivity of "el2h" in the region $r > 100$km. However, the contribution to the total emissivity is negligibly small ($< 1 \%$). We also find that the "ponn" and "po2h" processes dominate the $\bar{\nu}_e$ emissivities of light nuclei in the cooling and heating regions, respectively. However, they are $< 1 \%$ contribution to the total $\bar{\nu}_e$, which means their impact on CCSNe is weak.

\begin{figure*}
\vspace{15mm}
\epsscale{1.2}
\plotone{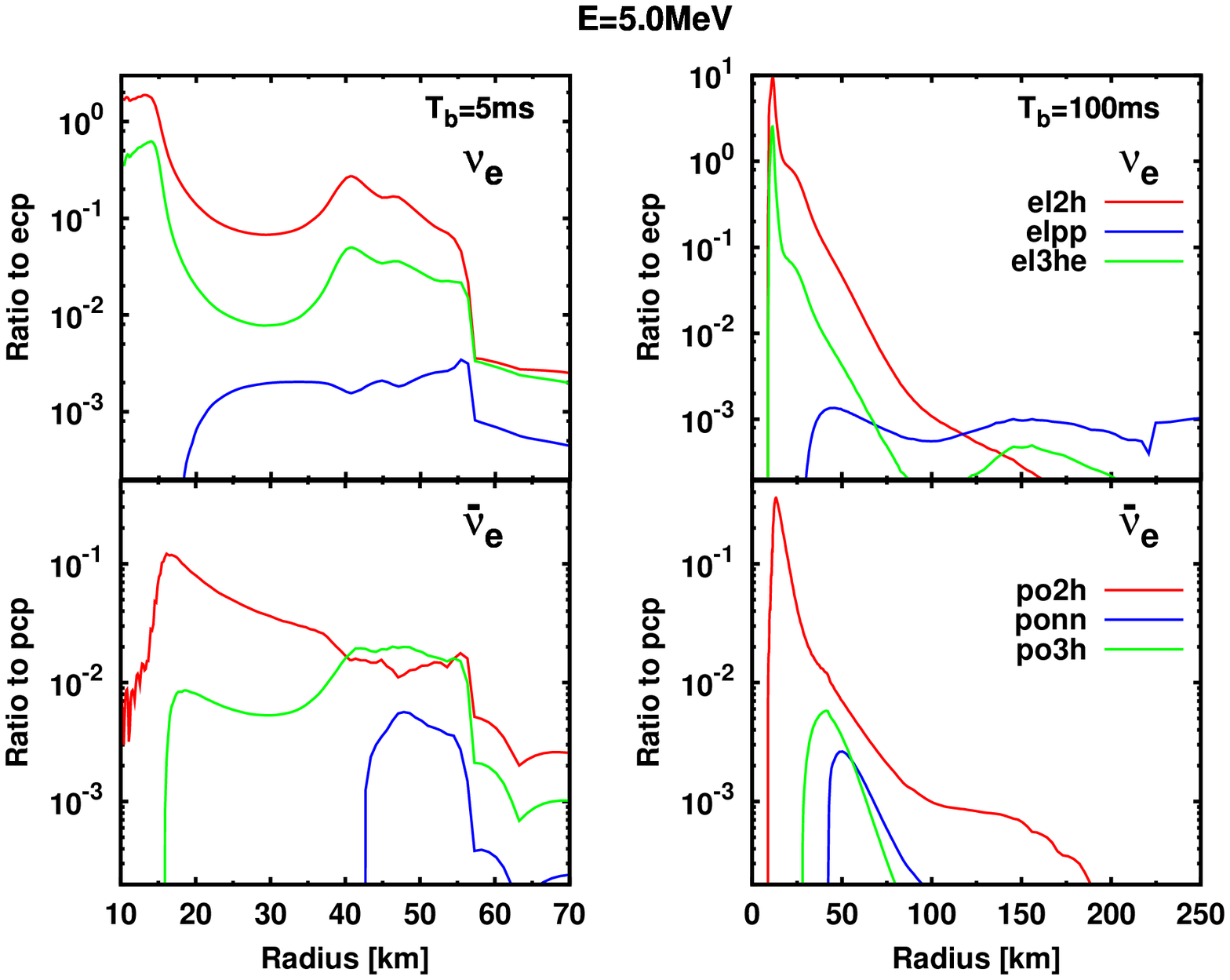}
\caption{Radial profiles of the ratio of emissivities from interactions with light nuclei relative to electron capture (top) or positron capture (bottom) for $5$Mev electron neutrinos (top) and electron anti-neutrinos (bottom). The relevant interactions are electron neutrino and anti-neutrino absorption on deuterons (elpp and ponn, respectively), electron and positron capture on deuterons (el2h and po2h, respectively), electron capture on $^3$He (el3he), and positron capture on $^3$H (po3h) as in Eqs.~(\ref{eq:lweak1})~to~(\ref{eq:lweak6}).
% Note that the "el2pp", "ponn", and "po3h" reaction rates are not shown since they are less than $10^{-4}$ times the electron or positron capture rates as relevant.
\label{graphRemistotcompare_ie5}}
\end{figure*}

%\begin{figure*}
%\vspace{15mm}
%\epsscale{1.2}
%\plotone{graphRemistotcompare_ie1.eps}
%\caption{Radial profiles of the ratio of emissivities from interactions with light nuclei relative to electron capture (top) or positron capture (bottom) for $1$Mev electron neutrinos (top) and electron anti-neutrinos (bottom). The relevant interactions are electron neutrino and anti-neutrino absorption on deuterons (elpp and ponn, respectively), electron and positron capture on deuterons (el2h and po2h, respectively), electron capture on $^3$He (el3he), and positron capture on $^3$H (po3h) as in Eqs.~(\ref{eq:lweak1})~to~(\ref{eq:lweak6}). Note that the "el2pp", "ponn", and "po3h" reaction rates are not shown since they are less than $10^{-4}$ times the electron or positron capture rates as relevant.
%\label{graphRemistotcompare_ie1}}
%\end{figure*}

%\begin{figure*}
%\vspace{15mm}
%\epsscale{1.2}
%\plotone{graphRemistotcompare_ie5.eps}
%\caption{Same as Fig.~\ref{graphRemistotcompare_ie1} but for 5 MeV neutrinos.
%\label{graphRemistotcompare_ie5}}
%\end{figure*}

%\begin{figure*}
%\vspace{15mm}
%\epsscale{1.2}
%\plotone{graphRemistotcompare_ie10.eps}
%\caption{Same as Fig.~\ref{graphRemistotcompare_ie1} but for 18.8 MeV neutrinos.
%\label{graphRemistotcompare_ie10}}
%\end{figure*}

%\begin{figure*}
%\vspace{15mm}
%\epsscale{1.2}
%\plotone{graphRemistotcompare_ie15.eps}
%\caption{
%\label{graphRemistotcompare_ie15}}
%\end{figure*}

Weak reactions with light nuclei should predominantly affect the low-energy end of the neutrino energy spectrum. Indeed, as shown in upper panels of Fig.~\ref{graphRemistotcompare_ie5}, "el2h" overwhelms the emissivity of "ecp" at $r \lesssim 20$km for low energy neutrinos ($\lesssim 20$MeV) . There are two main reasons for this trend. First, the average energy of emissivity from light nuclei tends to be lower than that of free protons, as discussed above. Second, the mass fraction of free protons at $r \sim 10$km is very small ($\sim 0.2$), which suppresses the electron capture by free protons. On the other hand, for $\bar{\nu}_e$ the emissivity from positron captures by light nuclei never overwhelm "pcp", which is simply because the mass fraction of neutrons is remarkably larger than others.

\subsubsection{Impact of artificial prescriptions of light nuclei}
In Sec.~\ref{subsubsec:Characteristicsoflight}, we discuss characteristics of light nuclei and find some evidence that the impacts of electron and positron captures of light nuclei on CCSNe are subdominant. As we will see below, however, artificial prescriptions for weak reactions with light nuclei in models NV112 and DV112 (see Sec.~\ref{numericalsetup}) change the dynamics of CCSNe and neutrino signals.

\begin{figure*}
\vspace{15mm}
\epsscale{1.2}
\plotone{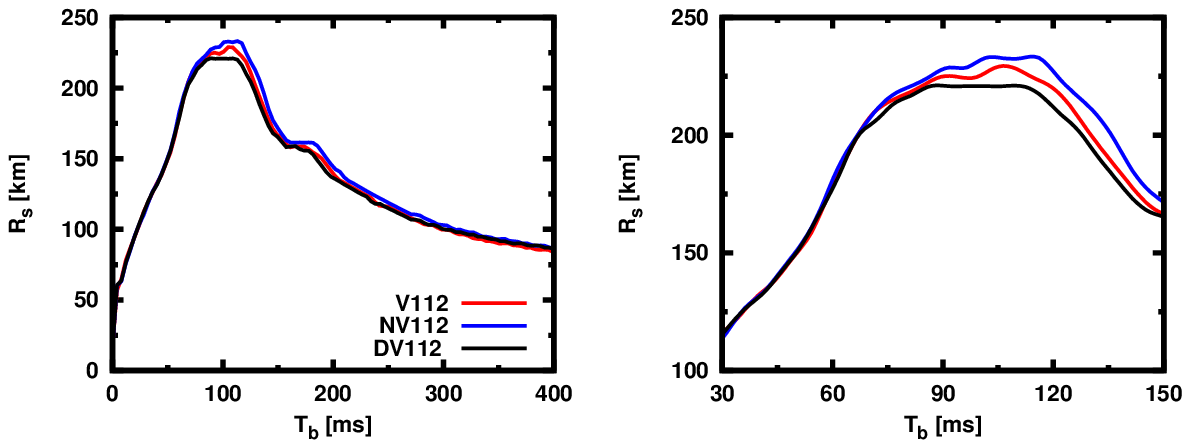}
\caption{Evolution of the shock radius as a function of time after core bounce to see the impact of weak interactions of light nuclei. The right panel magnifies the peak region. Model V112 (red) is the fiducial model and includes weak interactions with light nuclei. Model NV112 (blue) neglects weak interactions with light nuclei. Model DV112 (black) treats nucleons in light nuclei as if they were unbound and free for the purposes of weak interaction rates (see Sec.~\ref{numericalsetup}).
\label{shockevoLightnuclei112}}
\end{figure*}

\begin{figure}
\vspace{15mm}
\epsscale{1.2}
\includegraphics[width=\linewidth]{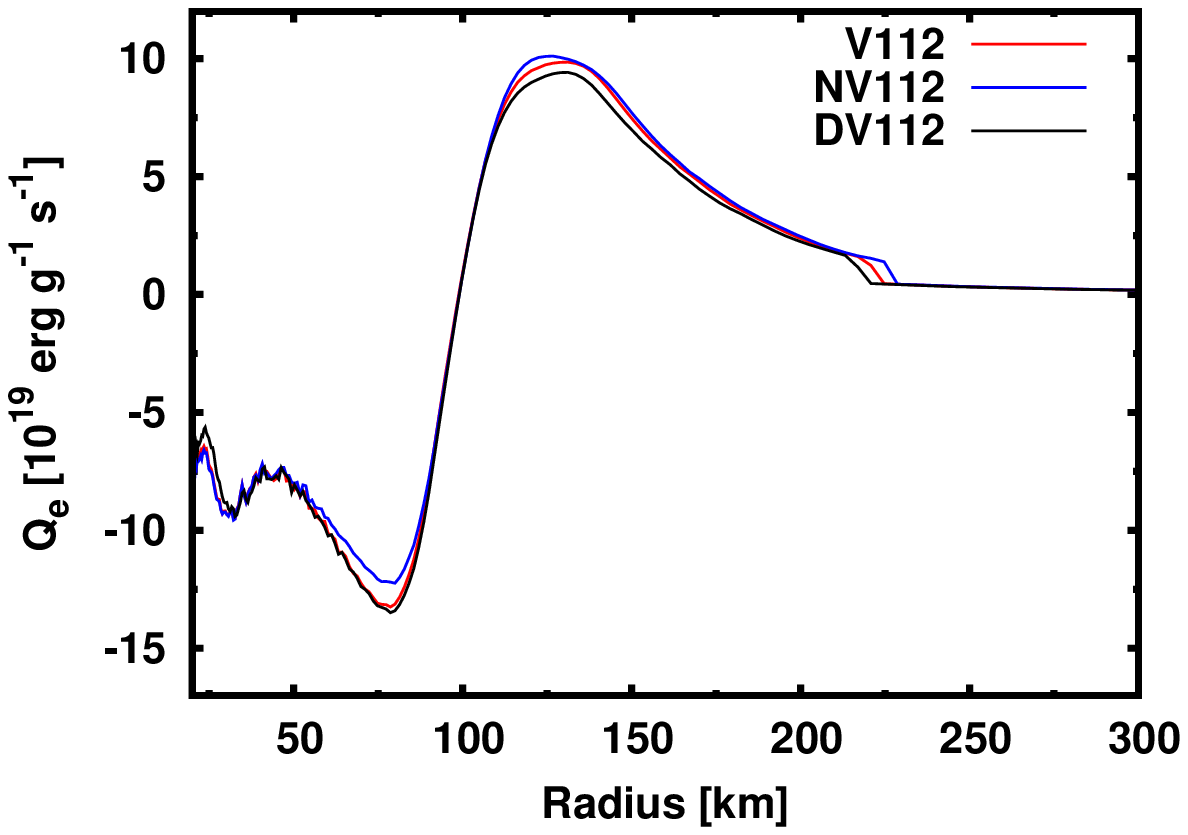}
\caption{Heating rates at ${\rm T}_{\rm b}=100$ms after the bounce for several models. Model V112 (red) is the fiducial model and includes weak interactions with light nuclei. Model NV112 (blue) neglects weak interactions with light nuclei. Model DV112 (black) treats nucleons in light nuclei as if they were unbound and free for the purposes of weak interaction rates (see Sec.~\ref{numericalsetup}).
\label{graphQe_lightnucleicomparev1}}
\end{figure}

We first discuss the shock trajectories in models using different prescriptions for light nuclei displayed in Fig.~\ref{shockevoLightnuclei112}. All three models are almost identical from bounce to ${\rm T}_{\rm b} \sim 100$ms. Perceptible differences arise when the shock wave reaches the maximum radius at that time. Model NV112 (DV112) has a shock radius that is a few percent larger (smaller) radius than that of model V112 (see the right panel in Fig.~\ref{shockevoLightnuclei112}). Fig.~\ref{graphQe_lightnucleicomparev1} shows that the rate of neutrino heating in the gain region (where $Q_e>0$, which lies at $100{\rm km} \lesssim r < R_s$ in this snapshot) for model NV112 (DV112) is always more (less) efficient than in model V112, which naturally explains the trend in the shock radius.

\begin{figure*}
\vspace{15mm}
\epsscale{1.0}
\plotone{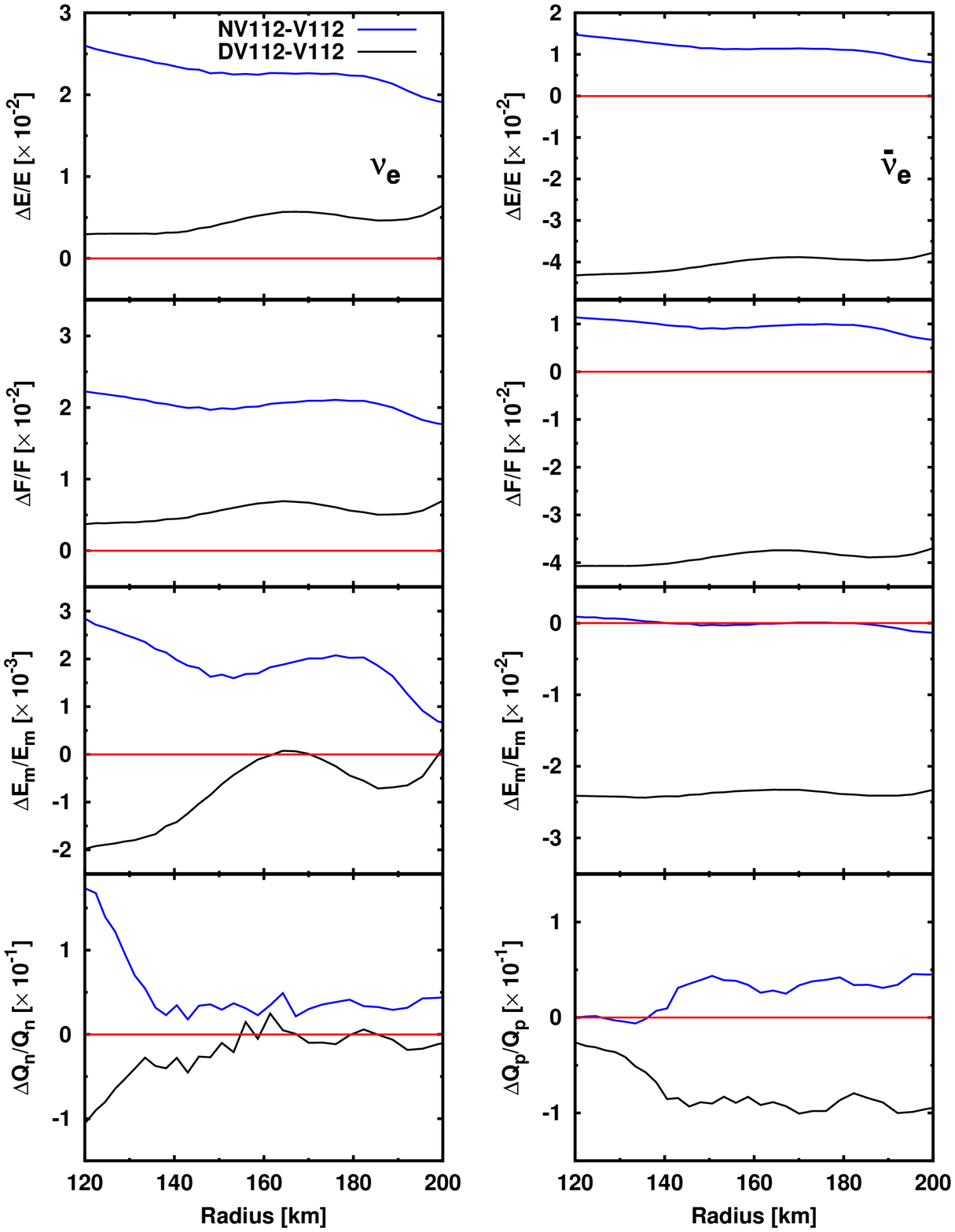}
\caption{Radial profiles of various quantities relevant to neutrino heating in the gain region at ${\rm T}_{\rm b}=100$ms relative to model V112. All quantities are normalized by the values in the fiducial model V112. Model V112 (red lines mark zero on y-axis) is the fiducial model. The left and right columns are for $\nu_e$ and $\bar{\nu}_e$, respectively. From the top, we show the neutrino energy density ($E$), energy flux ($F$) and mean energy ($E_m$), all measured in the fluid rest frame. The bottom panels show the net gain from processes involving $\nu_e$ (denoted $Q_{\rm n}$) and $\bar{\nu}_e$ (denoted $Q_{\rm p}$) from neutrino emission and absorption by free nucleons. Model V112 (red) is the fiducial model and includes weak interactions with light nuclei. Model NV112 (blue) neglects weak interactions with light nuclei. Model DV112 (black) treats nucleons in light nuclei as if they were unbound and free for the purposes of weak interaction rates (see Sec.~\ref{numericalsetup}).
%\caption{Radial profiles of the difference from V112 for various quantities relevant to neutrino heating in the gain region at ${\rm T}_{\rm b}=100$ms. Model V112 (red lines mark zero on y-axis) is the fiducial model. The left and right columns are for $\nu_e$ and $\bar{\nu}_e$, respectively. From the top, we show the neutrino energy density ($E$), energy flux ($F$) and mean energy ($E_m$), all measured in the fluid rest frame. Note that the scale of the y-axis is different for ($E_{\rm m}$) between left and right panels. The bottom panels show the net gain from processes involving $\nu_e$ (denoted $Q_{\rm n}$) and $\bar{\nu}_e$ (denoted $Q_{\rm p}$) from neutrino emission and absorption by free nucleons. Model V112 (red) is the fiducial model and includes weak interactions with light nuclei. Model NV112 (blue) neglects weak interactions with light nuclei. Model DV112 (black) treats nucleons in light nuclei as if they were unbound and free for the purposes of weak interaction rates (see Sec.~\ref{numericalsetup}).
\label{graph_diflightprescript}}
\end{figure*}

In Fig.~\ref{graph_diflightprescript}, we display radial profiles of properties of the neutrino radiation field and neutrino heating rates in the gain region. The left and right panels are for $\nu_e$ and $\bar{\nu}_e$, respectively. In the bottom panels, we also show the energy deposition rate of the reaction ($ e^{-} + p \leftrightarrow \nu_{e} + n$) for  $\nu_e$ and ($e^{+} + n \leftrightarrow \bar{\nu}_{e} + p$) for $\bar{\nu}_e$, which are denoted as $Q_{\rm n}$ and $Q_{\rm p}$, respectively. These reactions are the main processes of neutrino heating in the gain region. Note that we show the difference from model V112 in this figure, and red lines mark zero on y-axis. We first focus on the difference between NV112 and V112. The energy density ($E$), energy flux ($F$) and mean energy ($E_{\rm m}$) for both $\nu_e$ and $\bar{\nu}_e$ are consistently larger than those in V112 because ignoring all interactions with light nuclei reduces the opacity of neutrinos. This facilitates neutrino diffusion from the cooling region, which results in a large neutrino energy density and flux in the gain region. The reduction of neutrino opacity also causes the neutrinosphere to be located farther inside the PNS, which results in an increased $\nu_e$ mean energy in the gain region. As a result, $\nu_e$ energy deposition in the gain region is larger, which can be seen in the bottom panels in Fig.~\ref{graph_diflightprescript}. The mean energy of $\bar{\nu}_e$ in model NV112 is not different from model V112 since the opacity from weak interactions of light nuclei is much smaller than that of nucleon scattering at the PNS surface, which is reflected in a smaller change in heating from $\bar{\nu}_e$ than from $\nu_e$. 

Model DV112 (black lines in Fig.~\ref{graph_diflightprescript}) has the opposite trend as NV112. Recall that light nuclei are treated as if decomposed into free nucleons for the purposes of weak interactions in model DV112. We first look into the trend for $\bar{\nu}_e$. As clearly seen in the right panels, all $E$, $F$, $E_{\rm m}$ and $Q_{\rm p}$ are smaller than those in V112. This is attributed to the fact that the scattering with nucleons dominates the $\bar{\nu}_e$ opacity. Since the abundances of nucleons is artificially increased in model DV112, the scattering opacity is larger and the neutrino sphere shifts outward. This decreases the mean neutrino energy, slows the escape of neutrinos, and decreases neutrino energy flux and density. However, the energy density and flux for low energy $\bar{\nu}_e$ ($\lesssim 10$MeV) are larger than in model V112 (see also the bottom left panel in Fig.~\ref{graphSpectrumcompare_light_100ms}). This can be interpreted as follows. Since nucleon-nucleon bremmstrahlung dominates the $\bar{\nu}_e$ reaction rates deep inside the PNS, the increased mass fraction of nucleons significantly enhances the $\bar{\nu}_e$ emissivity. Since the neutrinospheres for low energy neutrinos is located at a smaller radius than those for higher energy neutrinos, low energy neutrinos in the gain region are more susceptible to the change in bremsstrahlung emissivity in the PNS. The increase of low energy neutrinos also reduces the mean energy of neutrinos as displayed in the right and third row of Fig.~\ref{graph_diflightprescript}.

On the other hand, the response due to the prescription for light nuclei in model DV112 for $\nu_e$ is less sensitive than $\bar{\nu}_e$. Indeed, as displayed in the the first and second panels on the left side of Fig.~\ref{graph_diflightprescript}, the differences of energy density and flux between models DV112 and V112 are much smaller than those for $\bar{\nu}_e$, though the difference is in the same direction as NV112. Since electron capture on free protons dominates the emissivity, the increased mass fraction of free protons enhances the "ecp" process, which causes the larger neutrino energy flux and density in the gain region. By the same token, the increased absorption opacity from the inverse process increases the radius of neutrino sphere, which reduces $E_{\rm m}$.

Note that the difference of $E_{\rm m}$ between two models depends on radius as shown in Fig.~\ref{graph_diflightprescript}. $E_{\rm m}$ in NV112 is smaller than that in V112 for $r \lesssim 160$km, though this difference disappears at $ 160 \lesssim r \lesssim 170$km and reappears again for $r \gtrsim 170$km. Although we do not fully understand the cause of non-monotonic trend of the difference, we speculate that the different matter profiles would account for the complicated structure of the difference of $E_{\rm m}$ between the two models. Indeed, the difference in energy deposition is more sensitive to the differences in matter profiles than to differences in the treatment of light nuclei (see below).
These two effects compete with each other in terms of neutrino heating in the gain region, though reduction in $E_{\rm m}$ ends up winning out and the $\nu_e$ energy deposition is smaller than in model V112 (left panel of Fig.~\ref{graph_diflightprescript}).

%\footnote{ {\bf As shown in Fig.~\ref{graph_diflightprescript}, the difference of $E_{\rm m}$ between two models depends on radius; $E_{\rm m}$ in NV112 is smaller than that in V112 $r \lesssim 160$km; the difference is disappeared around $160 \lesssi\m r \lesssim 170$km; but again $E_{\rm m}$ in NV112 is smaller than that in V112 $r \gtrsim 170$km. Although we do not fully understand the cause of non-monotonic trend of the difference, we speculate that the difference of matter profile would account for the complex property of the difference of $E_{\rm m}$ between two models. Indeed, the difference of energy deposition is more sensitive to the difference of matter profile than that of treatments of light nuclei. See the next paragraph and Fig.~\ref{graphQe_diffromsteady} in more details.}}. 

%At the moment, it is hard to identify what attributes such a non-monotonic relation between two models.

\begin{figure}
\vspace{15mm}
\epsscale{1.2}
\includegraphics[width=\linewidth]{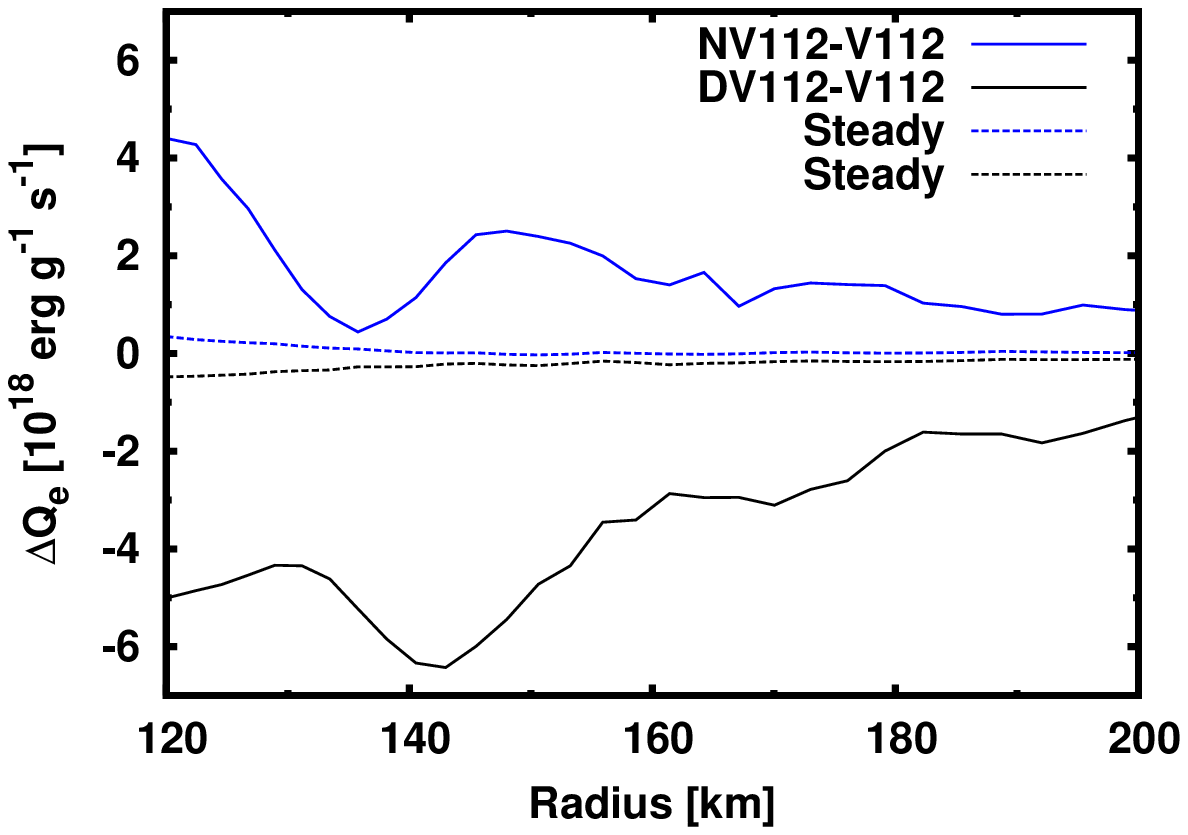}
\caption{Radial profiles of the difference of the net gain by neutrino emission and absorption. Solid lines show the difference in net gain at ${\rm T}_{\rm b}=100$ms between a model with an artificial treatment of light nuclei (NV112 or DV112) and the fiducial model V112. Dashed lines show the difference in net gain between the steady-state radiation fields on the V112 matter background using different prescriptions for light nuclei.  Model NV112 (blue) neglects weak interactions with light nuclei. Model DV112 (black) treats nucleons in light nuclei as if they were unbound and free for the purposes of weak interaction rates (see Sec.~\ref{numericalsetup}). Note that for dynamical models (solid lines), we measure the difference from model V112. On the other hand, for steady state model (dashed lines), we measure the difference from the steady state model with the same prescription of the weak interactions with light nuclei as model V112.
\label{graphQe_diffromsteady}}
\end{figure}

To see the effects of light nuclei in more detail, we calculate the steady-state neutrino radiation field by letting the neutrino field relax on a fixed fluid background. This approach has been frequently used to analyze the qualitative trends of neutrino-matter interactions in CCSNe (see e.g., \citet{2015ApJS..216....5S,2017ApJ...847..133R}). Given fluid distributions at $100$ms after the bounce in model V112, we compute the steady state neutrino radiation fields by the same prescriptions as V112, NV112 and DV112. The radial distributions of the differences in net gain between the dynamical and steady-state radiation fields using the same weak interaction treatments are displayed as dotted lines in Fig.~\ref{graphQe_diffromsteady}. As can be clearly seen in this figure, the differences between different prescriptions of light nuclei in steady state models are much smaller than those in dynamical models. This tells us that the artificial prescription mostly influences CCSN dynamics through feedback to matter. The effect would be more serious in multi-D cases, since the fluid dynamics are more sensitive to small changes in neutrino energy deposition \citep{2016arXiv161105859B}.

\begin{figure*}
\vspace{15mm}
\epsscale{1.2}
\plotone{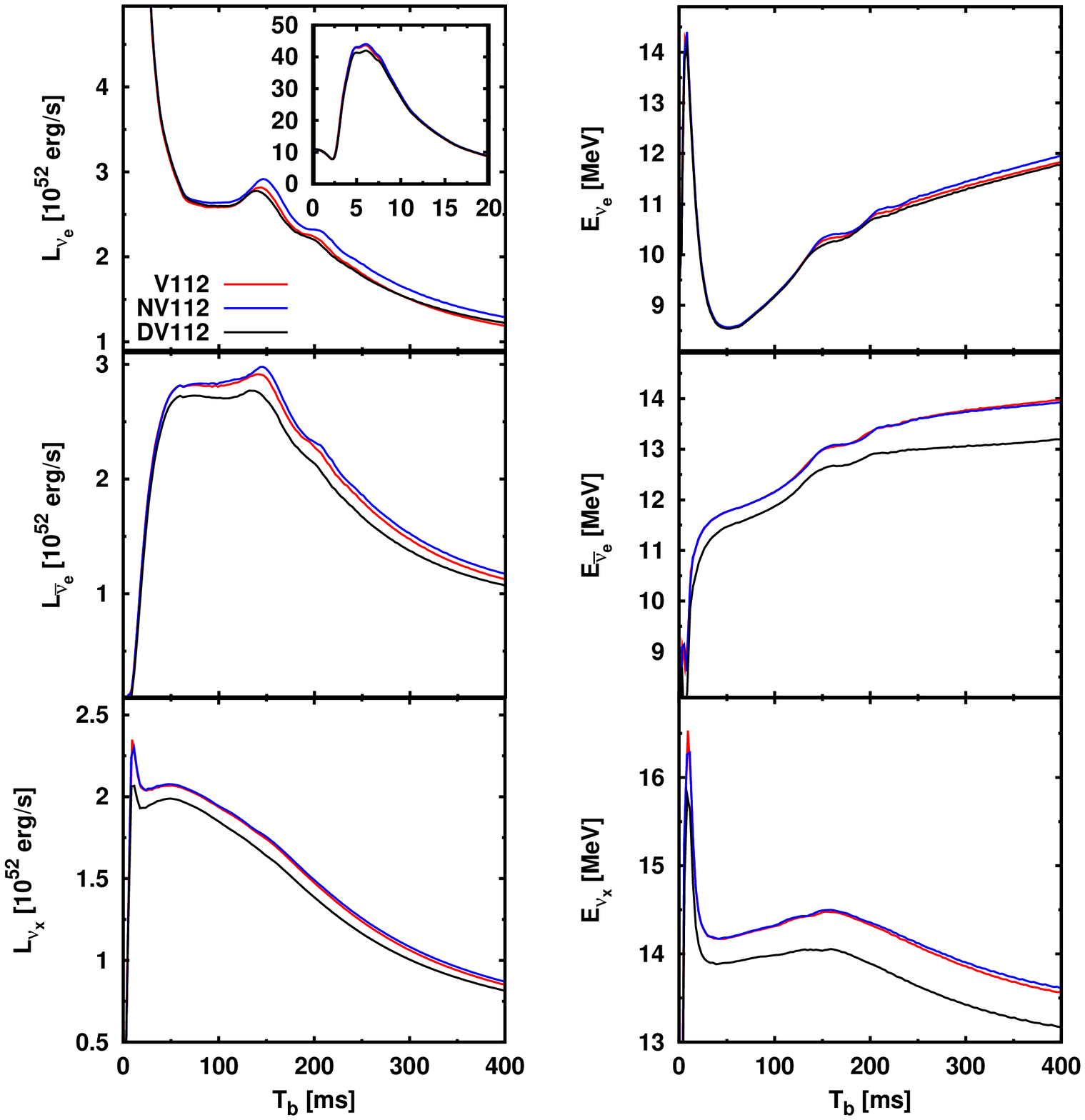}
\caption{Time evolution of the neutrino luminosity (left) and the mean energy (right) for $\nu_e$ (top), $\bar{\nu}_e$ (middle) and $\nu_x$ (bottom) for three models with different treatments of weak reactions with light nuclei. We measure these quantities at $r = 500$km. Model V112 (red) is the fiducial model and includes weak interactions with light nuclei. Model NV112 (blue) neglects weak interactions with light nuclei. Model DV112 (black) treats nucleons in light nuclei as if they were unbound and free for the purposes of weak interaction rates (see Sec.~\ref{numericalsetup}).
\label{Neutrino_Lightucleicompare112}}
\end{figure*}

\begin{figure*}
\vspace{15mm}
\epsscale{1.2}
\plotone{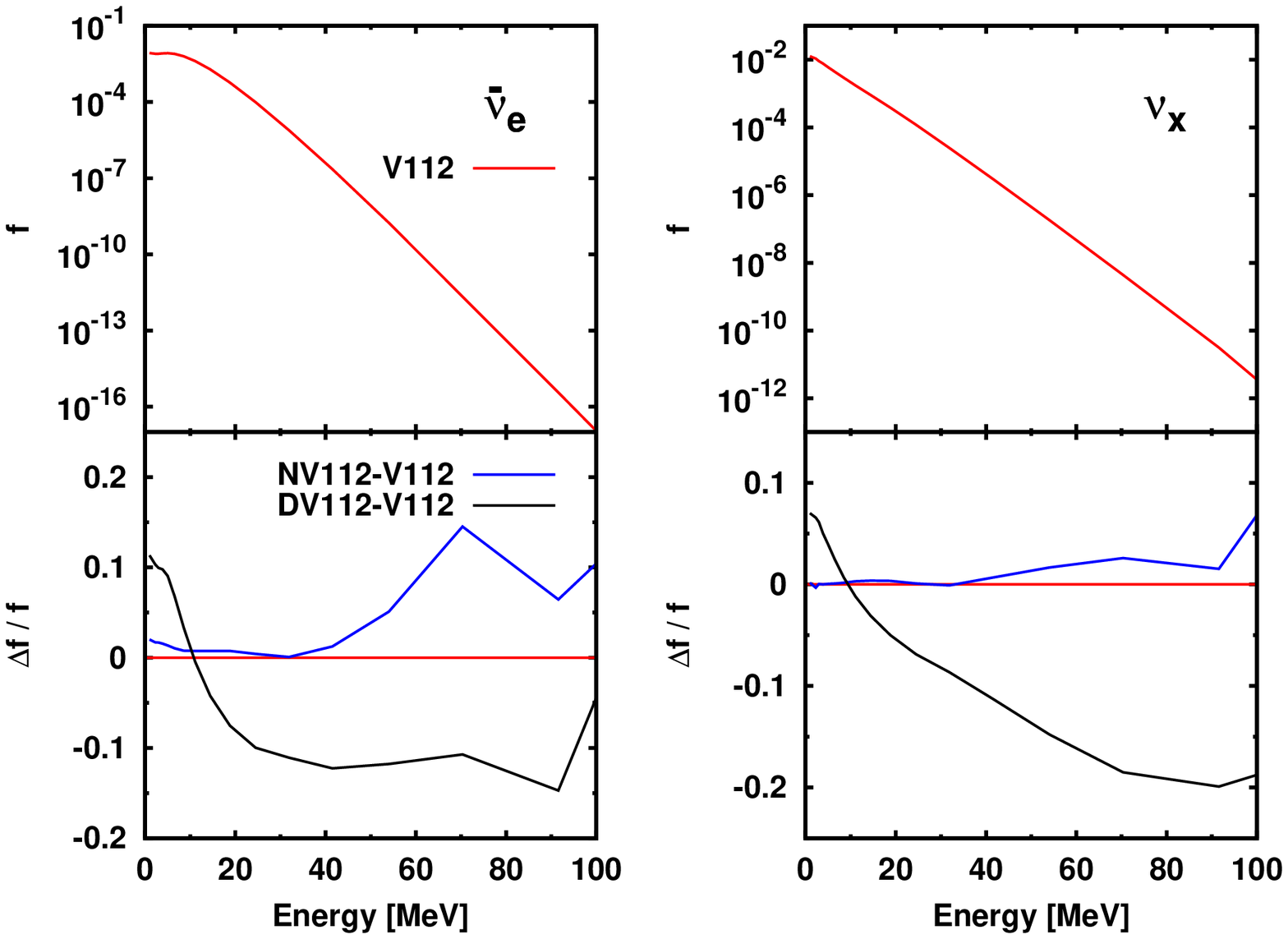}
\caption{Upper: the distribution function of outgoing ($\bar{\theta}=0^{\circ}$) $\bar{\nu}_e$ (left) and $\nu_x$ (right) as a function of energy at fluid rest frame for V112 model. The spectra are measured in $r = 500$km at ${\rm T}_{\rm b}=100$ms. The bottom panels display those in NV112 (blue) and DV112 (thick) but subtracting the V112. We also normalized the difference by V112. The red line marks at zero on y-axis.
\label{graphSpectrumcompare_light_100ms}}
\end{figure*}

Finally, we discuss the influence of the two artificial prescriptions of light nuclei on the neutrino signal. We display the time evolution of neutrino luminosities and mean energies in Fig.~\ref{Neutrino_Lightucleicompare112}. As shown in the upper panels, both neutrino luminosity and mean energy for $\nu_e$ in model NV112 are overestimated compared to model V112. Meanwhile the same quantities in model DV112 are not so different from those in model V112. These trends are in consistent with those discussed above in reference to the left column of Fig.~\ref{graph_diflightprescript} due to a higher opacity and emissivity from electron capture reactions. It should be noted, however, that the peak neutrino luminosity at the neutrino burst in model DV112 is roughly $\sim 5 \%$ smaller than in model V112.
% Since scattering opacity on light nuclei is not included, turning light nuclei into nucleons results in an artificial increase in the scattering opacity and a smaller peak luminosity. Including scattering on light nuclei would decrease this discrepancy.
%Although the small difference between  peak $\nu_e$ luminosity in models V112 and NV112 indicates that $\nu_e$ emissivity by weak interactions in light nuclei are unimportant for the burst signal, the artificial increase of mass fraction of nucleons in model DV112 $\nu_e$ suppresses the peak luminosity via a larger opacity for absorption onto free neutrons.  %%NOTE - replaced this sentence with the current final sentence in above paragraph.

For $\bar{\nu}_e$ and $\nu_x$, the artificial prescription in DV112 changes the time evolution of both the luminosity and mean energy. Again, the difference between DV112 and V112 can be understood in the same way as in previous discussions (see right column in Fig.~\ref{graph_diflightprescript}). That is, the artificial increase in the number of free nucleons increases the opacity from nucleon scattering and the emissivity from nucleon-nucleon bremsstrahlung. Fig~\ref{graphSpectrumcompare_light_100ms} shows the distribution function $f$ for outgoing ($\bar{\theta}=0^{\circ}$) neutrinos at $r=500$km and ${\rm T_b}=100$ms in model V112 (top panel), along with the differences from this spectrum in models NV112 and DV112 (bottom panel). $f$ is a function of the spacetime coordinates $x^{\mu}$ and the neutrino four momentum $p^i$. Since the latter satisfies the condition $p^{\mu} p_{\mu} = - m_{\nu}^2$, where $m_{\nu}$ denotes the mass of the neutrino (which is assumed to be zero in this paper), only three of the four momentum components are independent. The differences between model DV112 and V112 shows similar trends for $\bar{\nu}_e$ and $\nu_x$. The neutrino excess in the low energy side in model DV112 is mostly due to the artificial increase of emissivity of nucleon-nucleon bremsstrahlung. On the other hand, the neutrino number depletion in the high-energy side for model DV112 is due to the artificial increase of opacity from nucleon scattering.

Here is a short summary of the impact of artificial prescriptions of light nuclei to CCSNe. As discussed in Sec.~\ref{subsubsec:Characteristicsoflight}, electron and positron captures by light nuclei are subdominant weak interaction processes in CCSNe. However, the artificial prescriptions quantitatively change the dynamics (as measured by shock trajectory) of CCSNe. Ignoring light nuclei (as in model NV112) tends to produce an artificially larger shock radius due to the overestimation of neutrino energy deposition in the gain region (see Figs.~\ref{graphQe_lightnucleicomparev1}~and~\ref{graphQe_diffromsteady}). On the contrary, another artificial prescription which the abundances of light nuclei are decomposed into nucleons and counted as free nucleons (as in model DV112) underestimates both the shock radius and neutrino heating. The primary cause of the difference of neutrino heating is that ignoring weak interactions in light nuclei enhances escaping neutrinos from the cooling region by reducing the opacity, which results in enhanced neutrino absorption in the gain region (for model NV112). On the contrary, the artificial increase of the mass fraction of free nucleons in model DV112 artificially increases the nucleon scattering and nucleon-nucleon bremsstrahlung opacities. They change the neutrino flux and the spectrum, which reduces the neutrino absorption in the gain region. The impact on neutrino heating rate in the gain region is $\sim 5 \%$. It should be noted, however, that this could lead to non-negligible errors in multi-dimensional simulations. We also find that the impact of artificial prescriptions in neutrino signals in steady-state comparisons are much smaller than in dynamical comparisons (see Fig.~\ref{graphQe_diffromsteady}). This fact tells us that the steady state radiation transport gives us only the qualitative trend and is not capable of measuring the error quantitatively.

\subsection{Progenitor dependence} \label{subsec:prodepe}

The so-called "Mazurek's law" during the collapse phase (see e.g., \citet{1985ApJS...58..771B,2003NuPhA.719..144L}) states that in strongly interacting systems, such as the inner core (i.e., the part of the core that is sonically connected), changes in the model inputs are compensated for and lead to small differences in the output. This, combined with similar properties of the stellar core at the onset of collapse due to the Chandrasekhar criterion, leads to universal inner core properties in the collapse of different progenitors. This is because both the distributions of electron fraction and entropy should be self-regulated by electron captures during collapse. A higher electron fraction enhances the rates of electron capture onto free protons due to the larger number of free protons and electrons, which facilitates the deleptonization and then results in reducing the electron fraction. Similarly, a high entropy state enhances electron capture rates, which facilitates neutrino cooling and reduces the excess of entropy. In previous studies, we have observed such a self-regulation mechanism when electron captures of free protons dictate the time evolution of both deleptonization and neutrino cooling (see e.g., \citet{2005PhRvD..71f3003K,2013ApJ...762..126O}). However, if the rates of electron captures onto heavy nuclei overwhelm those of captures onto free protons, it is not so clear that collapse of different progenitors should show the same universal structure (see e.g., \citet{2003NuPhA.719..144L}). This is because the deleptonization and cooling rates are no longer a simple function of electron fraction and temperature, but rather sensitively depend on the abundances of heavy nuclei. As a matter of fact electron captures onto heavy nuclei are always dominant during the entire collapse phase regardless of the progenitor \citep{2003PhRvL..90x1102L,2010NuPhA.848..454J,2016ApJ...816...44S} (and see also Fig.~\ref{graphprodepeECPratio}).

%\begin{figure}
%\vspace{15mm}
%\epsscale{1.0}
%\includegraphics[width=\linewidth]{graphinidensityprofile.eps}
%\caption{Density profiles of progenitors at the beginning of our simulation for models V112 (red), V15 (blue), V27 (green), and V50 (black).
%\label{graphinidensityprofile}}
%\end{figure}

%\begin{figure*}
%\vspace{15mm}
%\epsscale{1.0}
%\plotone{graphinientropyYeprofile.eps}
%\caption{Radial profiles of electron fraction (left) and entropy per baryon (right) at the beginning of our simulations for models V112 (red), V15 (blue), V27 (green), and V50 (black).
%\label{graphinientropyYeprofile}}
%\end{figure*}

\begin{figure}
\vspace{15mm}
\epsscale{1.0}
\includegraphics[width=\linewidth]{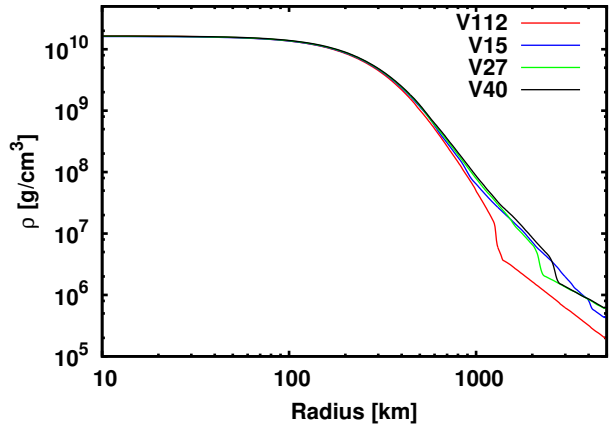}
\caption{Density profiles of progenitors at the initial collapse phase ($\rho_c=1.6 \times 10^{10} {\rm g/cm^3}$) for models V112 (red), V15 (blue), V27 (green), and V50 (black).
\label{graphrhoc16e10densityprofile}}
\end{figure}

\begin{figure*}
\vspace{15mm}
\epsscale{1.0}
\plotone{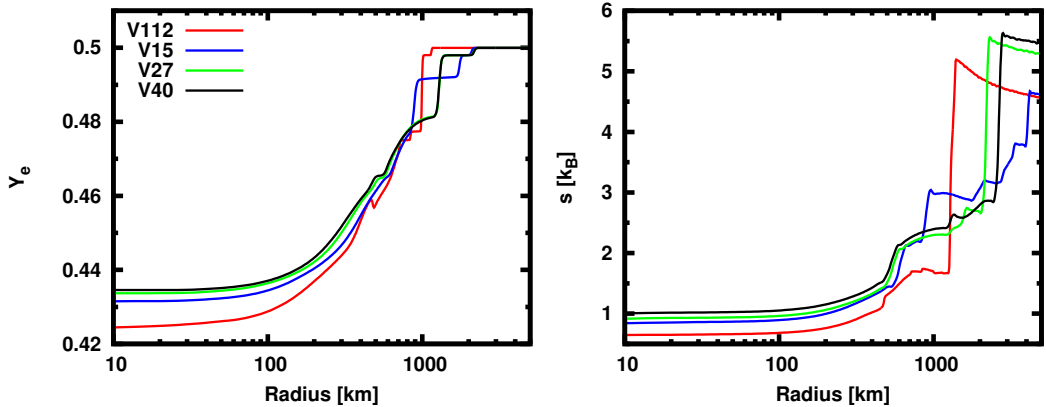}
\caption{Radial profiles of electron fraction (left) and entropy per baryon (right) at the initial collapse phase ($\rho_c=1.6 \times 10^{10} {\rm g/cm^3}$) for models V112 (red), V15 (blue), V27 (green), and V50 (black).
\label{graphrhoc16e10entropyYeprofile}}
\end{figure*}

In this light, there are many previous works that discuss the progenitor dependence of dynamics and neutrino signals in CCSNe (see e.g., \citet{1987ApJ...318..288M,2008ApJ...688.1176S,2013ApJS..205....2N,2003NuPhA.719..144L,2005PhRvD..71f3003K,2013ApJ...762..126O,2016ApJ...825....6S,2016ApJ...818..123B,2017PhRvD..95f3019R,2017ApJ...850...43R,2017arXiv171201304O}). This study, however, is a fist attempt to understand the progenitor dependence of the CCSNe in the context of up-to-date rates of electron capture on heavy and light nuclei consistently with a multi-nuclear EOS with a realistic nuclear force. To do this, we compare four progenitor models with ZAMS masses of $11.2\,M_\odot$ (V112), $15\,M_\odot$(V15), $27\,M_\odot$ (V27), and $40\,M_\odot$ (V40), whose matter profiles at the initial collapse phase are shown in Figs.~\ref{graphrhoc16e10densityprofile} and \ref{graphrhoc16e10entropyYeprofile}\footnote{The time snapshot in these figures corresponds to the beginning of our simulations for $11.2\,M_\odot$ but 160ms, 172ms and 197ms for $15\,M_\odot$, $27\,M_\odot$ and $40\,M_\odot$, respectively.}. As shown in these figures, the matter distributions in the different progenitors are quite different from each other. Among our selected progenitors, higher-mass progenitors show a higher core electron fraction and higher core entropy per baryon. This means that more massive progenitors are expected to have more efficient deleptonization and neutrino cooling in the inner core, allowing the different models to be driven closer to the same inner core structure as collapse proceeds.

% also shows the radial profiles of the electron fraction and entropy per baryon. In the latter figure we match the time for all progenitors as the central density reaches $\rho_c=1.6 \times 10^{10} {\rm g/cm^3}$ (which corresponds to the time of beginning of simulation for $11.2\,M_\odot$) in order to compare quantities at the same evolution point for all progenitors.

\begin{figure*}
\vspace{15mm}
\epsscale{1.0}
\plotone{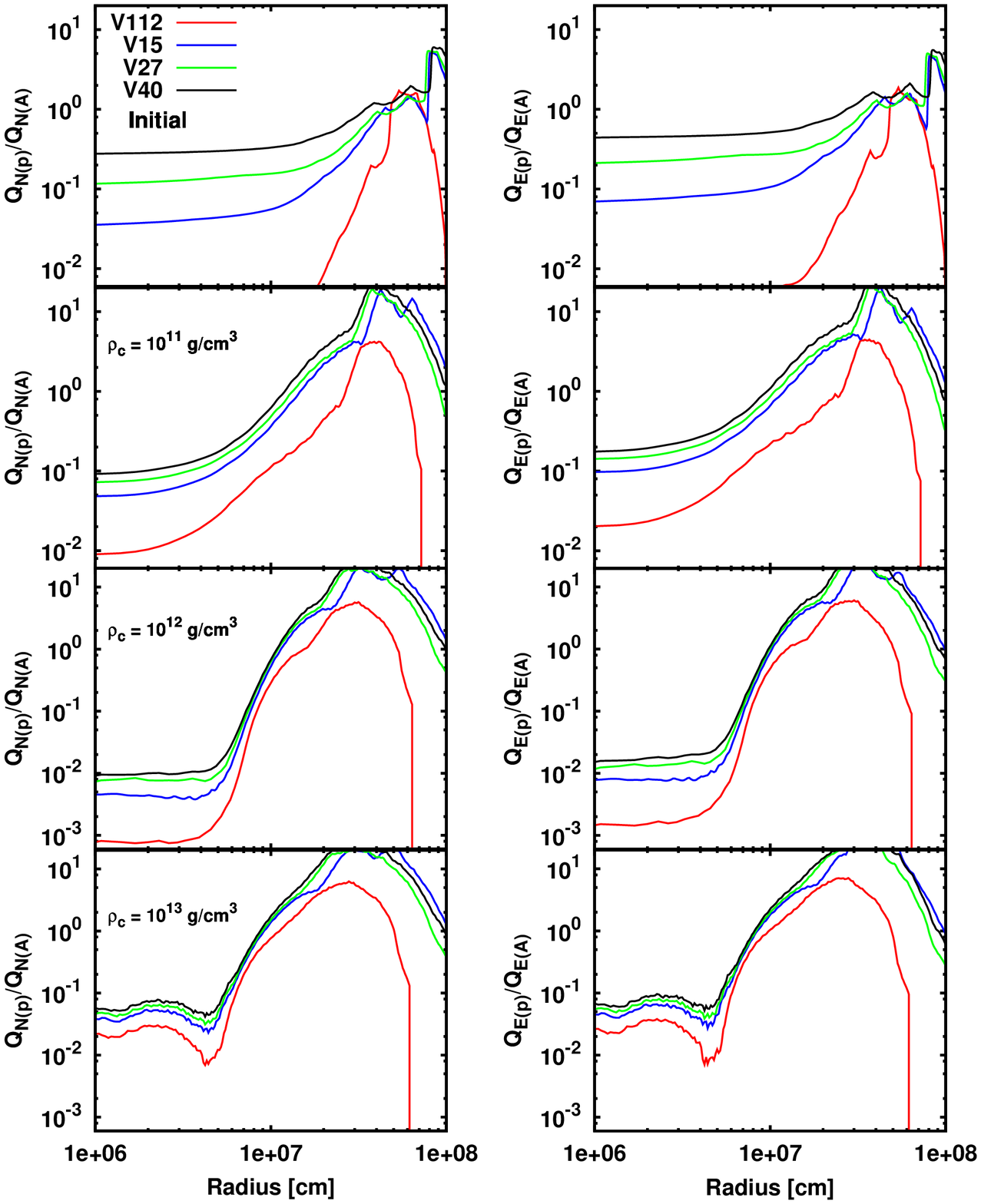}
\caption{Radial profiles of the ratio of the rates of electron capture onto free protons to those onto heavy nuclei. Left panels show deleptonization rates and right panels show cooling rates. From top to bottom, we display the initial profile, followed by profiles at times when the central density is $\rho_c = 10^{11} {\rm g/cm^3}$, $\rho_c = 10^{12} {\rm g/cm^3}$, and $\rho_c = 10^{13} {\rm g/cm^3}$.
\label{graphprodepeECPratio}}
\end{figure*}

We first take a look whether electron captures by free protons or heavy nuclei are dominant for deleptonization and cooling of the core during the collapse phase. Figure~\ref{graphprodepeECPratio} shows that in the inner core (quasi-horizontal segments of the curves on the left sides of the plots), electron capture by heavy nuclei dominates that by free protons at all times during the collapse phase. However, this is not the case outside of the inner core. In general, the electron chemical potential should be larger than the mass difference $\Delta_{np}$ between neutrons and protons for electron capture by free protons, and larger than the nuclear Q value for capture by heavy nuclei. The nuclear Q value includes not only $\Delta_{np}$ but also the biding energy difference between parent and daughter nuclei, which means that Q value is larger than  $\Delta_{np}$. In the outer core, although the electron chemical potential exceeds $\Delta_{np}$, i.e., meeting the requirement of electron captures by free protons, it is not large enough compared to the nuclear Q value due to the low temperature and density. As  a result, electron capture by heavy nuclei is strongly suppressed. It is also important to note that the electron fraction in the outer core is larger than in the inner core, which creates a supports a larger amount of free protons and results in a higher electron capture rate by free protons than by heavy nuclei. At later times during collapse (i.e., lower panels), the ratio of electron captures on heavy nuclei to protons approach the same value between different simulations in the inner core. These trends are true for both the rate of change of lepton number (left plots) and for the rate of change of internal energy (right plots). Even in the late collapse phase ($\rho_c > 10^{13} {\rm g/cm^3}$), the electron capture of heavy nuclei remains dominant in the inner core, which supports the need for a consistent treatment of nuclear abundances in electron capture rates \cite{2003PhRvL..90x1102L,2010NuPhA.848..454J,2016ApJ...816...44S}.

\begin{figure*}
\vspace{15mm}
\epsscale{1.0}
\plotone{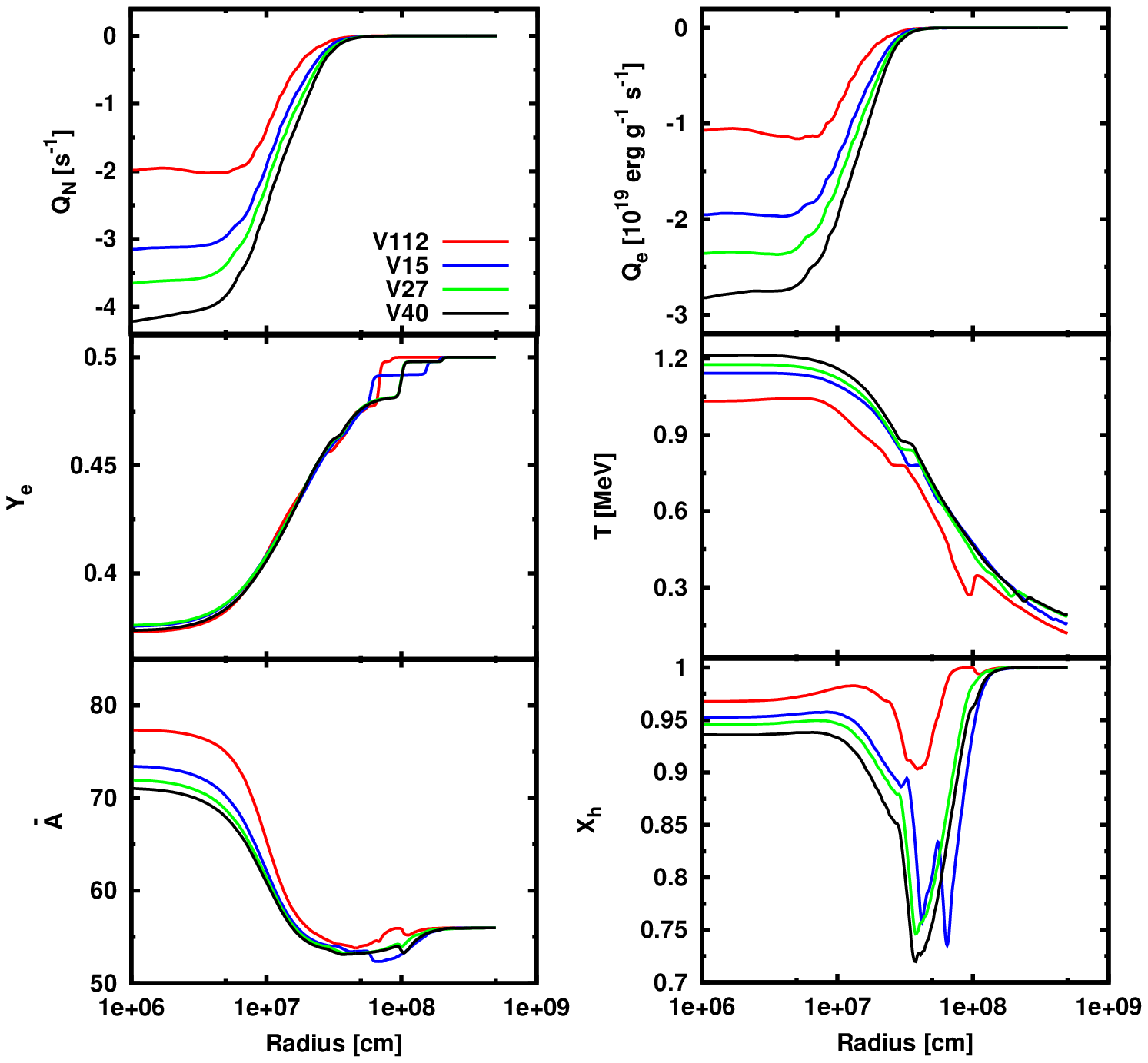}
\caption{Comparison between different progenitors during the collapse phase when the central density is $\rho_c=10^{11} {\rm g/cm^3}$. $Q_N$ is the net leptonization rate, $Q_e$ is heating rate, $Y_e$ is electron fraction, $T$ is the temperature, $\bar{A}$ is average number of nucleons in heavy nuclei, and $X_h$ is the mass fraction of heavy nuclei. Results are shown for models V112 (red), V15 (blue), V27 (green), and V50 (black). 
\label{rhoc1e11progenitorcompare}}
\end{figure*}

\begin{figure*}
\vspace{15mm}
\epsscale{1.0}
\plotone{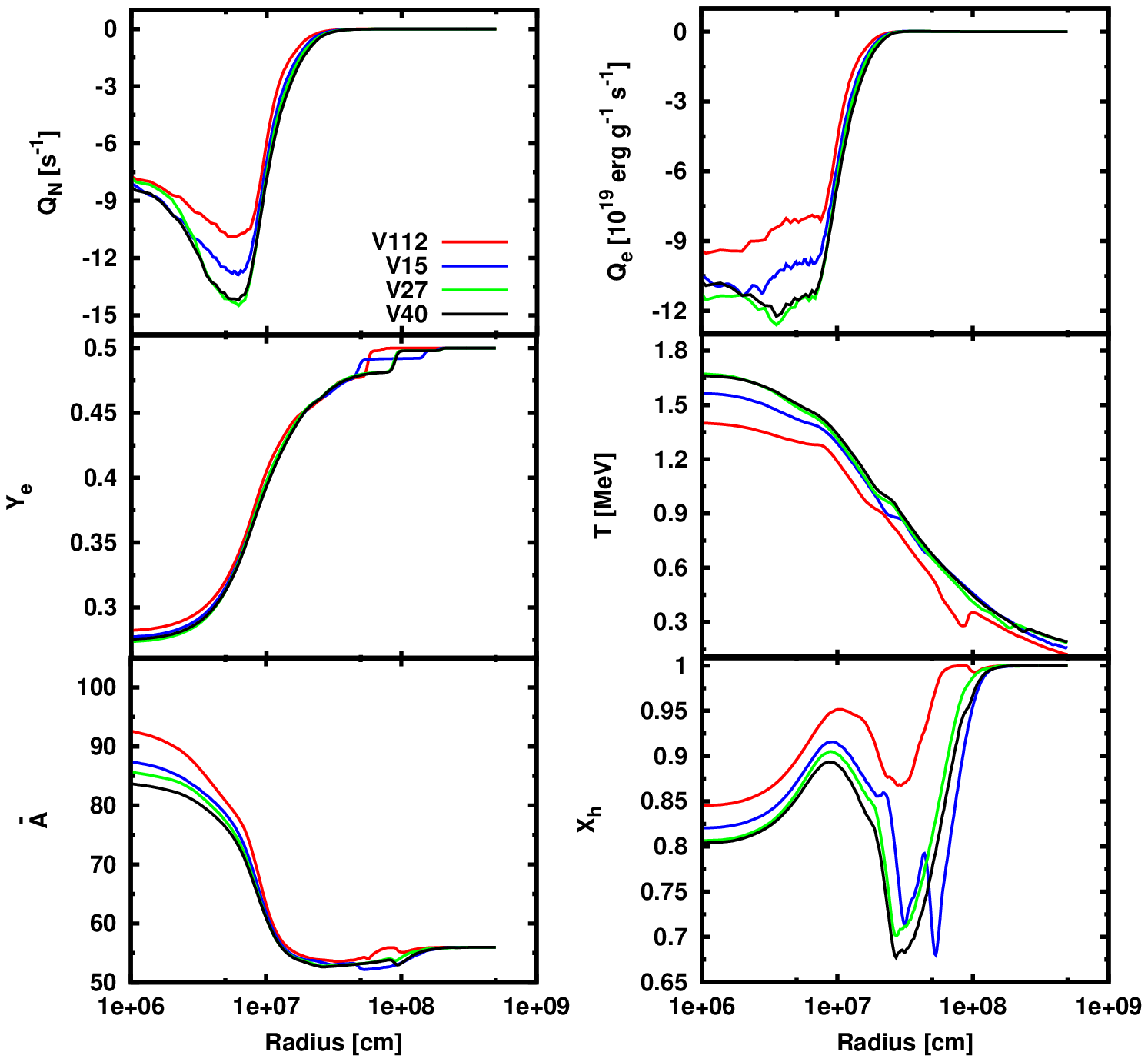}
\caption{Same as Fig.~\ref{rhoc1e11progenitorcompare} but when $\rho_c = 10^{12} {\rm g/cm^3}$.
\label{rhoc1e12progenitorcompare}}
\end{figure*}

\begin{figure*}
\vspace{15mm}
\epsscale{1.0}
\plotone{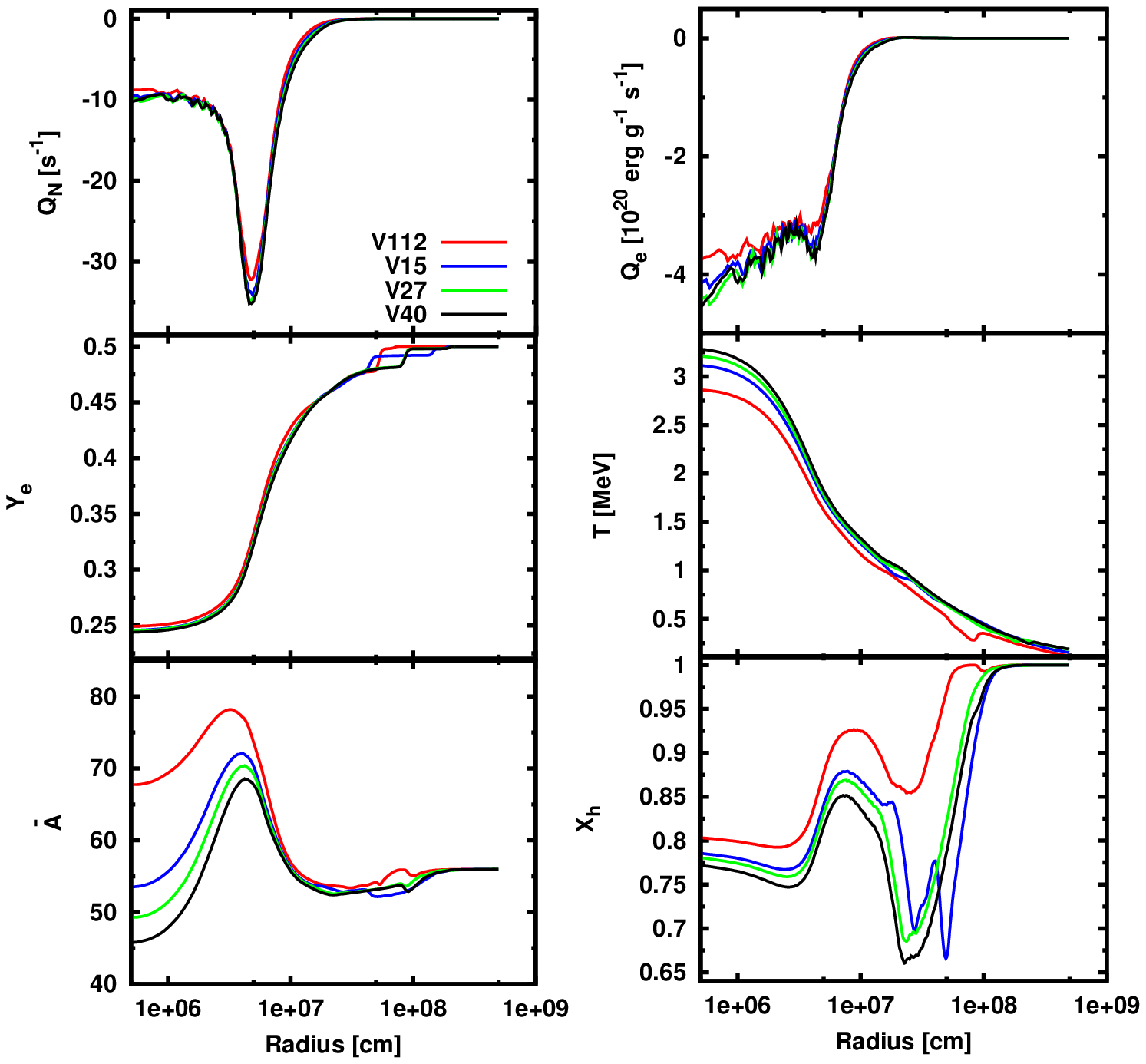}
\caption{Same as Fig.~\ref{rhoc1e11progenitorcompare} but when $\rho_c = 10^{13} {\rm g/cm^3}$.
\label{rhoc1e13progenitorcompare}}
\end{figure*}

Larger mass progenitors are much more dominated by captures onto free protons than lower mass progenitors. This is due to the fact that there is a higher entropy per baryon in the calculations starting from high-mass progenitors (visible as a higher temperature in Figs.~\ref{rhoc1e11progenitorcompare} to \ref{rhoc1e13progenitorcompare}). Because of the temperature difference, the average mass (bottom left panels) and abundance of heavy nuclei (bottom right panels) is higher in models using lower-mass progenitors. Interestingly, despite the differences in cooling rates, deleptonization rates, temperatures, and nuclear abundances, the difference of electron fraction in the inner core among models almost disappears by the time when the central density reaches $10^{11} {\rm g/cm^3}$ (see the middle left panel in Fig.~\ref{rhoc1e11progenitorcompare}). Even in the late collapse phase, the time evolution of the central electron fraction is almost identical between the models starting from different progenitors,  just as in models using more approximate treatments of electron capture (e.g., \cite{Liebendoerfer:2005gm}). Deleptonization and neutrino cooling rates become insensitive to these differences after the time at $\rho_c \sim 10^{12} {\rm g/cm^3}$ (see top panels), since neutrinos become trapped in the inner core.

\begin{figure*}
\vspace{15mm}
\epsscale{1.0}
\plotone{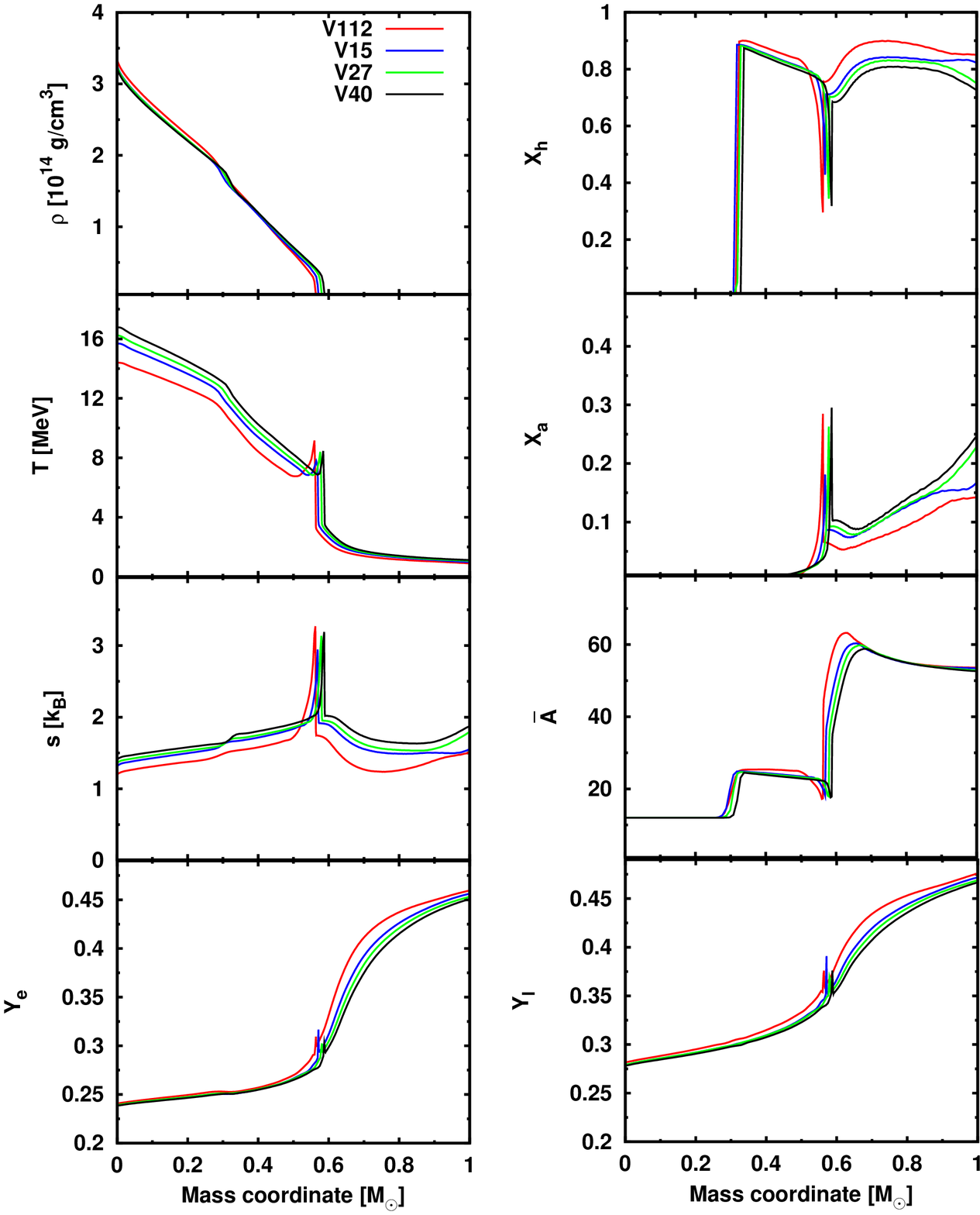}
\caption{Comparison at core bounce between models starting from different progenitors. From top to bottom, we plot baryon mass density ($\rho$), temperature ($T$), entropy per baryon ($s$) and electron fraction ($Y_e$) in the left column, while we plot the mass fraction of heavy nuclei ($X_h$), the mass fraction of light nuclei ($X_a$), the average mass number ($\bar{A}$) and the lepton fraction ($Y_l$) in the right column. All quantities are displayed as a function of mass coordinate in this figure. Models are V112 ($11.2 M_\odot$, red), V15 ($15 M_\odot$, blue), V27 ($27 M_\odot$, green), and V40 ($40 M_\odot$, black).
\label{Bounceprogenitorcompare_s3}}
\end{figure*}

 At the time of core bounce, the shock wave is generated for all progenitors at the mass shell between $0.56 M_{\sun}$ and $0.6 M_{\sun}$ (see Fig.~\ref{Bounceprogenitorcompare_s3}). The minor progenitor dependence arises due to the difference in the thermal component of the pressure. As discussed above, model V112 has the lowest temperature in the inner core among the models using different progenitors, which means that the thermal pressure is the weakest and the total mass of inner core becomes the smallest. On the other hand, the weaker thermal pressure support allows a higher central density (top left panel), so the inner core is the most compact. As during the collapse phase, the electron fraction and lepton fraction (bottom panels) are consistent among the models. At bounce, the density is large enough in the inner core that there are no light or heavy nuclei, though the abundances of nuclei outside the inner core follow the same ordering as during collapse (i.e., larger progenitors have more light nuclei and fewer heavy nuclei).

\begin{figure*}
\vspace{15mm}
\epsscale{1.0}
\plotone{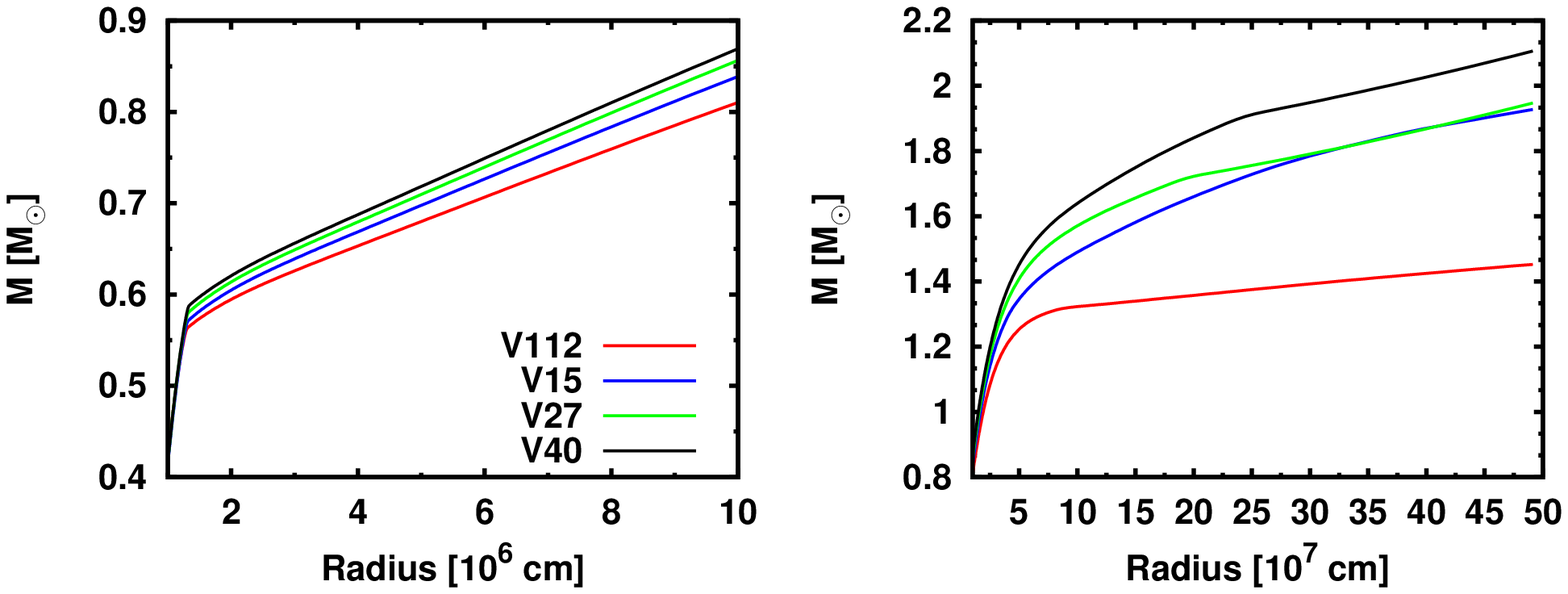}
\caption{Enclosed mass versus radius at the time of core bounce for models V112 (red), V15 (blue), V27 (green) and V40 (black). The left panel shows the region where $10^{6}\,\mathrm{cm} \leq r \leq 10^{7}\,\mathrm{cm}$ and the right panel shows the region where $10^{7}\,\mathrm{cm} \leq r \leq 6 \times 10^{7}\,\mathrm{cm}$. Larger progenitors lead to more compact density profiles at bounce.
\label{Bounceprogenitorcompare_s3_MR}}
\end{figure*}

\begin{figure}
\vspace{15mm}
\epsscale{1.0}
\includegraphics[width=\linewidth]{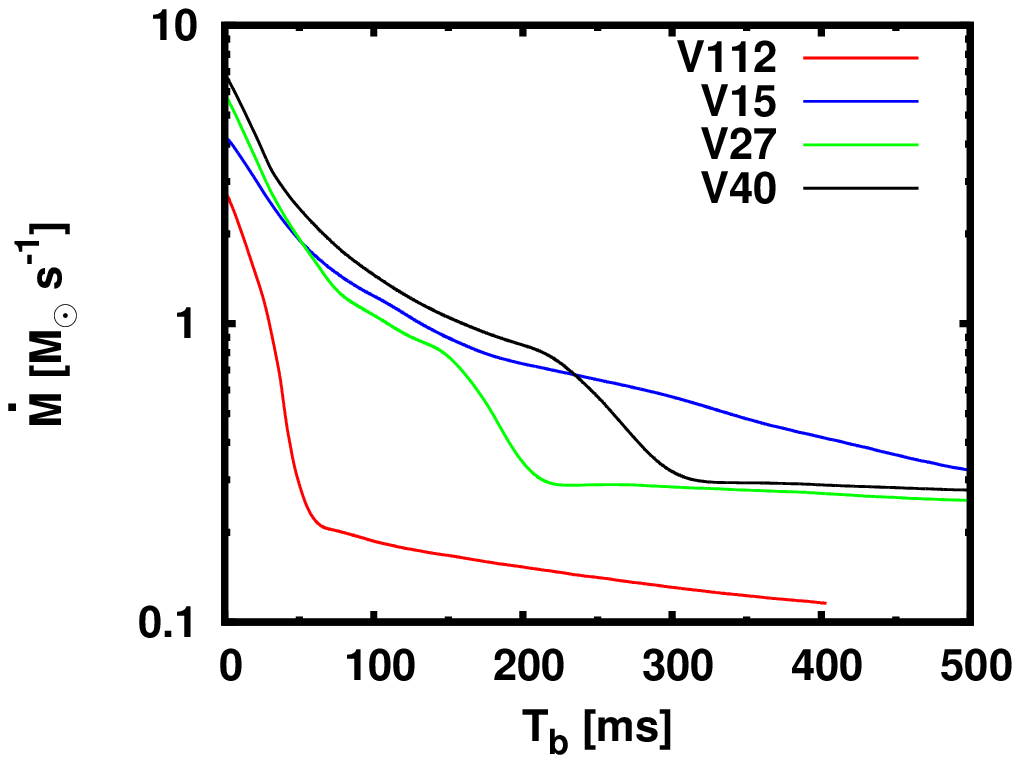}
\caption{Mass accretion rates measured at $r=500$km as a function of time for models using different progenitors. The more compact density profiles of larger progenitors generally cause higher accretion rates at early times.
\label{graphMdotprogenitor}}
\end{figure}
Contrary to that of the inner core, the compactness of outer core increases with progenitor mass (see Fig.~\ref{Bounceprogenitorcompare_s3_MR}). The trend simply reflects the density profiles in the different progenitors at $r \gtrsim 500$km in the initial collapse phase (see Fig.~\ref{graphrhoc16e10densityprofile}). The compactness in the outer core dictates the time evolution of the mass accretion rate in the post-bounce phase (see in Fig.~\ref{graphMdotprogenitor}). Changes in the slopes in Fig.~\ref{Bounceprogenitorcompare_s3_MR} result in sudden change in the accretion rate when that part of the star passes through the shock, reducing the ram pressure, but also reducing the neutrino luminosity.

\begin{figure*}
\vspace{15mm}
\epsscale{1.2}
\plotone{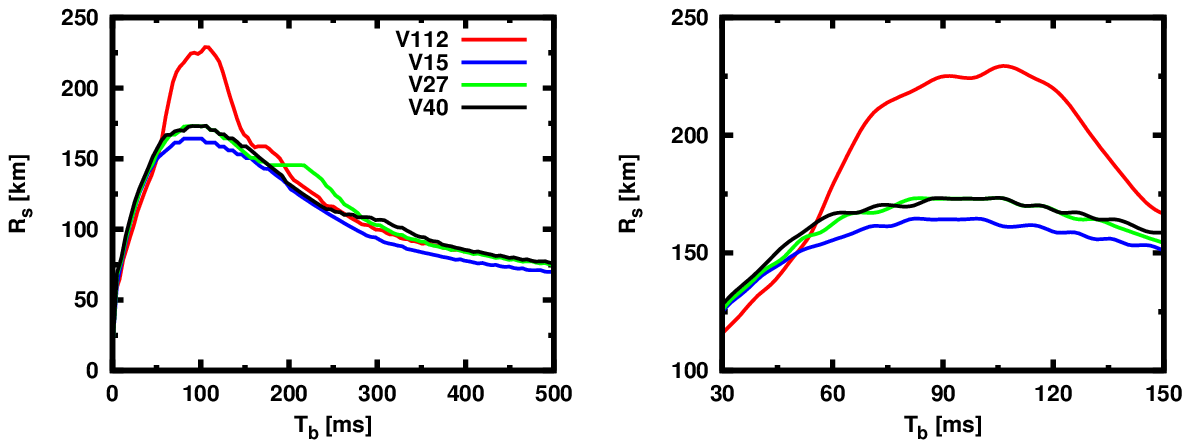}
\caption{Same as Fig.~\ref{shockevoVMvsRMF} but for V112 (red), V15 (blue), V27 (green) and V40 (black).
\label{shockevoProgenitor}}
\end{figure*}

The trajectory of shock wave in each model is shown in Fig.~\ref{shockevoProgenitor}. For all models, the shock wave reaches the maximum radius at similar time after the bounce (${\rm T}_{\rm b} \sim 100$ms). On the other hand, the maximum radius depends on the progenitor. In particular, there is quite a large deviation in V112 from other progenitors, though it quickly returns to match the shock radii of the other models by $\sim 150 {\rm ms}$ after core bounce. The primary cause of the large deviation in V112 is that the Si/Si-O composition interface reaches the shock front at ${\rm T}_{\rm b} \sim 50$ms while the shock is still expanding. The sudden decrease of mass accretion rate allows the shock expansion to accelerate. The characteristics of the earliest arrival of the Si/Si-O interface in V112 is predictable from the density distribution at the beginning of collapse, where the sharp density decline is clearly visible at $r \sim 1200$km in Fig.~\ref{graphrhoc16e10densityprofile}. For other progenitors, it can be seen around $4200$km, $2400$km and $2800$km for V15, V27 and V40, respectively. Except for V15, the Si/Si-O interface passes through the shock wave by the end of our simulations, which also correlates with the sudden decrease of the mass accretion rate in Fig.~\ref{graphMdotprogenitor}. The shock wave in V27 and V40 suspends the recession for a while once the Si/Si-O interface hits the shock front, which are shown in the left panel of Fig.~\ref{shockevoProgenitor}.

\begin{figure*}
\vspace{15mm}
\epsscale{1.2}
\plotone{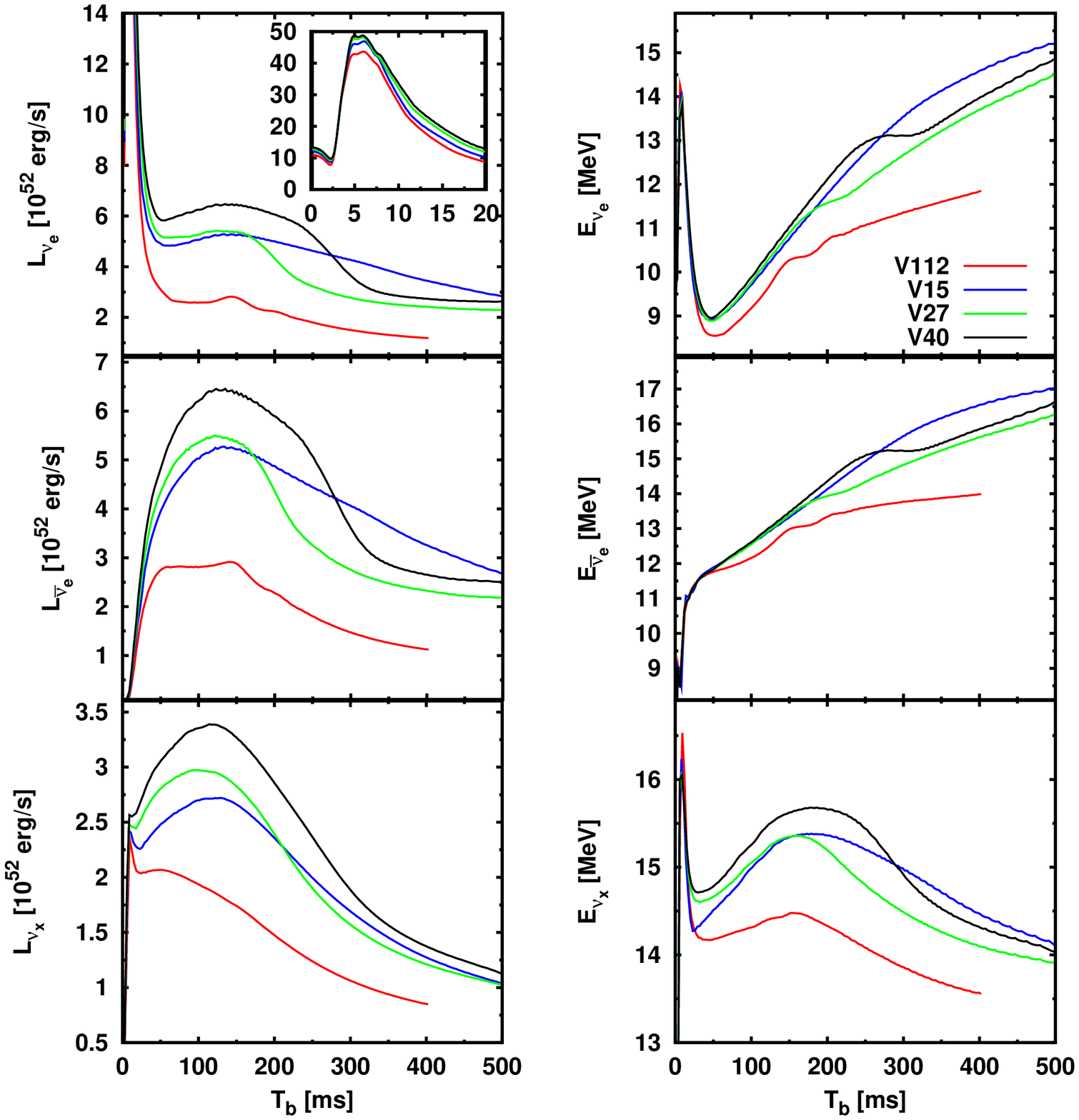}
\caption{Time evolution of the neutrino luminosity (left) and the mean energy (right) for $\nu_e$ (top), $\bar{\nu}_e$ (middle) and $\nu_x$ (bottom). The progenitors used for model V112, V15, V27, and V40 have masses of $11.2 M_\odot$, $15 M_\odot$, $27 M_\odot$, and $40 M_\odot$, respectively.
\label{Neutrino_progenitorcompare}}
\end{figure*}

\begin{figure*}
\vspace{15mm}
\epsscale{1.0}
\plotone{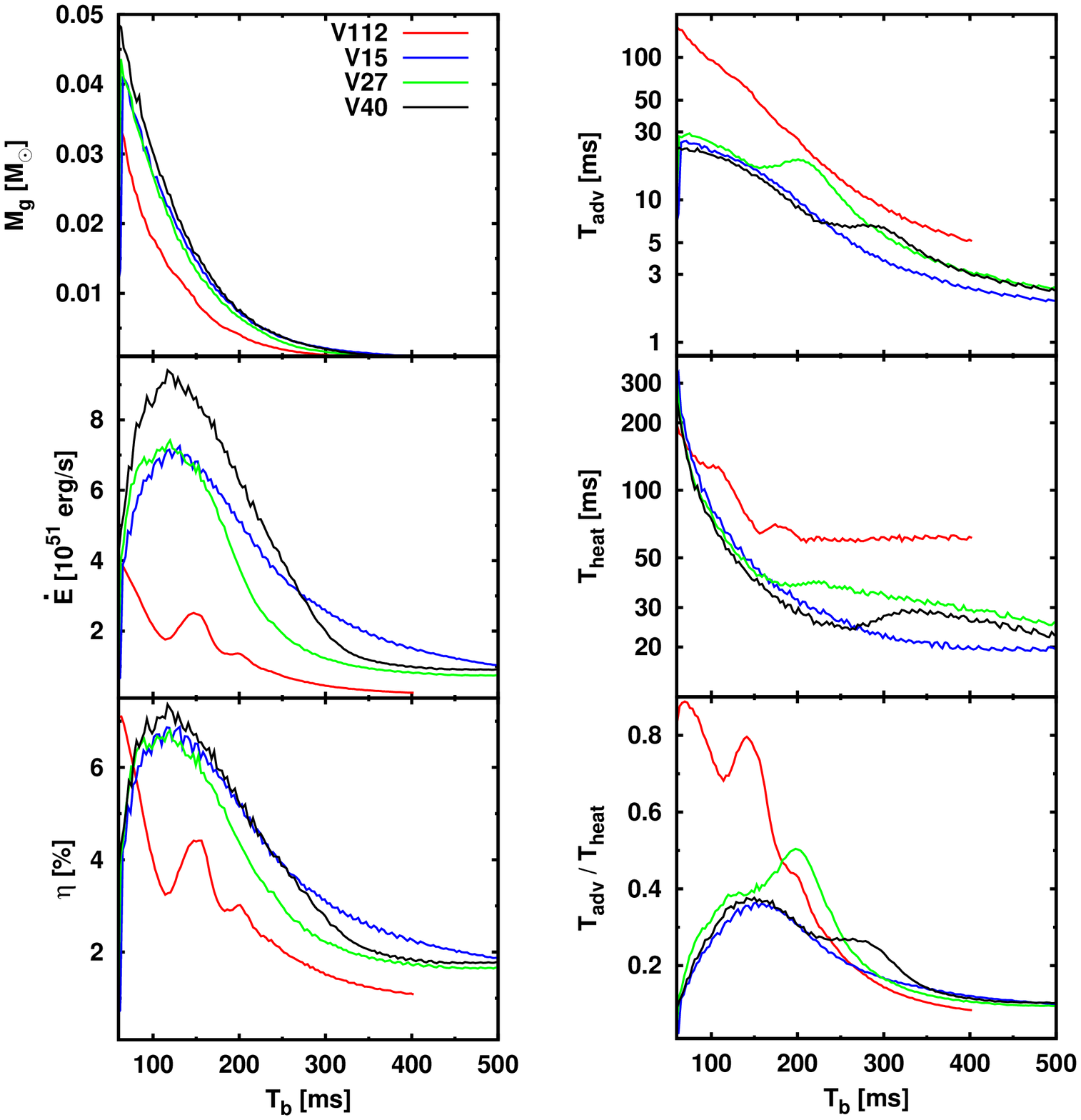}
\caption{Time evolution of several diagnostic quantities for CCSNe simulations. In the left column, the mass contained in the gain region (top), the energy deposition rate (middle) and the heating efficiency (bottom). In the right panel, the advection time scale in the post-shock flow (top), the heating time scale (middle) and ratio of the former to the latter (bottom). Colors are the same as Fig.~\ref{shockevoProgenitor}.
\label{graph_basicquant_progenitorcompare}}
\end{figure*}

The driving force of shock propagation during the early shock expansion phase is the core bounce, which is insensitive to the differences in the outer core structure between different models. Interestingly, except for model V112, the shock trajectory at later times also depends weakly on the progenitor structure. For instance, the difference in maximum shock radius between models V15, V27 and V40 is less than $6 \%$. In the late phase (beyond $\sim 150\,\mathrm{ms}$ after core bounce), the shock trajectory is roughly identical for all four progenitor models. Such a weak dependence is mainly attributed to the competition between mass accretion rates and neutrino heating. The large mass accretion rate in model V40, for example, pushes the shock wave back with a large ram pressure. On the other hand, the shock wave is pushed out by the enhancement of neutrino heating in the gain region, which is due to the increase of both accretion components of neutrino luminosity and the baryon mass in the gain region (see left panels in Fig.~\ref{Neutrino_progenitorcompare} for the time evolution of neutrino luminosity and the left column in Fig.~\ref{graph_basicquant_progenitorcompare} for relevant quantities to neutrino heating).

The sudden decrease of mass accretion rate when the Si/Si-O interface accretes through the shock simply reduces the ram pressure, which results in rapid expansion of the shock wave. This is visible as a bump in the shock trajectory in all models in Fig.~\ref{shockevoProgenitor}, but is exaggerated in model V112 because the drop in ram pressure occurs while the prompt shock is still expanding. On the other hand, after the shock stagnates, the time evolution of shock wave is dictated by the balance between the ram pressure from mass accretion and neutrino heating from hot accreted mass, as described above. This negative feedback softens the impact of the accretion rate drop in models V15, V27, and V40, and returns the shock in model V112 to a location similar to the other three models by $~150\,\mathrm{ms}$ after bounce.

The time evolution of the luminosity and the mean energy are displayed in Fig.~\ref{Neutrino_progenitorcompare}. The peak $\nu_e$ luminosity at the neutronization burst varies by $\sim 20 \%$ among the models, with higher luminosities coming from higher mass progenitors (top left panel). Among our models, large mass progenitors lead to more compact outer cores\footnote{Note that the trend between mass and compactness would not be monotonic outside of this mass range in reality. See below in more details} (see also Fig.~\ref{Bounceprogenitorcompare_s3_MR}), which means that the post-shock flow also becomes also compact. As a result, the post-shock flow is more opaque for neutrinos models using more massive progenitors, which causes the neutrinos to decouple with matter at the larger radius. Such an increase of radius of neutrino sphere causes the enhancement of neutrino luminosity.

As shown in the left column of Fig.~\ref{Neutrino_progenitorcompare}, the larger mass of our models tend to have a higher neutrino luminosity for all species of neutrinos due to the larger accretion rate. Importantly, the information of density structure of outer core is imprinted in the neutrino signal. For instance, the sudden decrease of luminosities which are seen in $\nu_e$ and $\bar{\nu}_e$ for V15 and V27 at ${\rm T}_{\rm b} \sim 200$ms and $300$ms, respectively, correlates with the sharp decline of mass accretion rate by Si/Si-O interface (see Fig.~\ref{graphMdotprogenitor}). We also find that the increase of average energy of neutrinos is suppressed at the same time (right panels in Fig.~\ref{Neutrino_progenitorcompare}). As discussed already, once the Si/Si-O interface is engulfed by the shock wave, the shock wave slightly expands (or recedes more slowly), which makes the post-shock flow less compact and decreases the average energy of neutrinos.

%{\bf It should be noted that the monotonic relation between neutrino luminosity in the early post-bounce phase ($\lesssim 100 {\rm ms}$) and progenitor mass is an artifact due to a selection bias. As shown in \citet{2011ApJ...730...70O,2016ApJ...818..124E,2016ApJ...821...38S}, the compactness of progenitors is not a monotonic function of progenitor mass but rather stochastic in particular for the mass range $20M_{\sun} \lesssim M \lesssim 22 M_{\sun}$ or $M \gtrsim 40 M_{\sun}$. Neutrino signals, hence, can not be a direct tool to determine the progenitor mass in real observations. However, we regard that neutrino signals are still useful to narrow down the progenitor candidate since they would provide us the information of density structure of the core at the onset of collapse. The information will be compared to theoretical results computed by stellar evolutions, which would give some constraints on progenitors.}

It should be noted that the monotonic relation between neutrino luminosity in the early postbounce phase ($\lesssim 100$ms) and progenitor mass in our results is an artifact due to a selection bias in our choice of progenitors. As shown in \citet{2011ApJ...730...70O,2016ApJ...818..124E,2016ApJ...821...38S}, the compactness of progenitors is not a monotonic function of progenitor mass and is rather stochastic in particular for the mass ranges $18M_{\sun} \lesssim M \lesssim 22 M_{\sun}$ or $M \gtrsim 40 M_{\sun}$. Hence, neutrino signals cannot be a direct tool to determine the progenitor mass in real observations. However, the neutrino signals could be still useful for narrowing down the range of progenitor candidates in particular for low mass progenitors ($M \lesssim 18M_{\sun}$). Indeed, according to \citet{2016ApJ...821...38S}, these progenitors roughly monotonically correlate with the compactness with their mass.

%Despite the fact,

One of the standard diagnostics for assessing the effectiveness of the neutrino heating mechanism is the ratio of the advection to the heating time scales (see e.g., \citep{Murphy:2008dw,Janka:2016fox}).  We define the advection time scale $T_{\rm adv}=M_g/\dot{M}$ where $M_g$ and $\dot{M}$ denote the mass of gain region and mass accretion rate at $r=500$km, respectively. The heating time scale is $T_{\rm heat}=|E_{\rm tot}|/\dot{Q}$, where $E_{\rm tot}$ and $\dot{Q}$ denote the total energy of matter (the sum of gravitational, kinetic and thermal energy) and neutrino heating rate in the gain region, respectively. Naturally, we find that this ratio is correlated to the shock trajectory. As shown in the bottom right panel in Fig.~\ref{graph_basicquant_progenitorcompare}, the ratio in model V112 is remarkably larger than in other models, especially at ${\rm T}_{\rm b} < 200$ms. This is mainly due to the large advection time scale in model V112, which is more than three times as large as other models at the same time (see the top right panel in Fig.~\ref{graph_basicquant_progenitorcompare}). Though none of the models explode, this diagnostic suggests that model V112 was closest to explosion, followed by models V27, V40, and V15. 

\begin{figure*}
\vspace{15mm}
\epsscale{1.0}
\plotone{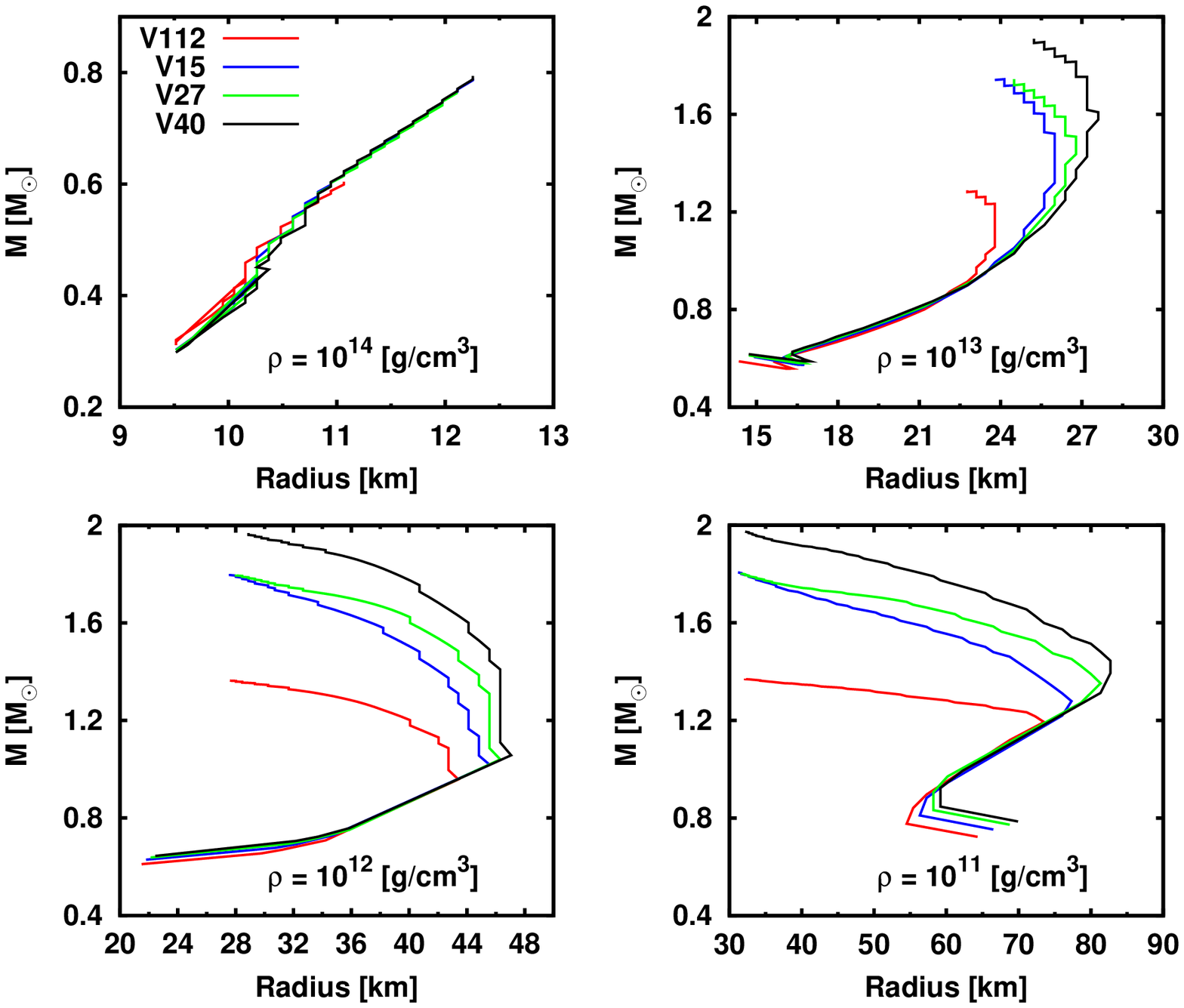}
\caption{Radius of and mass enclosed by density contours during the post-bounce evolution in models V112 (red), V15 (blue), V27 (green) and V40 (black). In general, time increases from the bottom of the curves to the top. Each panel shows a different density contour, as labeled in the plot.
\label{graphMRfigure}}
\end{figure*}

To see the progenitor dependence of PNS structure, we plot the radius of and mass enclosed by several density contours in Fig.~\ref{graphMRfigure}. Plotting mass-radius relations in this way allow us to analyze the EOS dependence of CCSNe in a way that naturally includes thermal effects that are generally not negligible in PNSs. The mass-radius relation for the iso-density surface of $\rho = 10^{14} {\rm g/cm^3}$ (top left panel) is almost identical among all progenitors because this contour is at all times deep within the PNS, the structure of which is rather universal (see Fig.~\ref{Bounceprogenitorcompare_s3}). The matter distribution at this contour evolves quasi-steadily and adiabatically. We also see the universality in the mass-radius relation at the lower density contours at early times, though the universality does not hold at later times (i.e., the curves recede in radius and deviate from each other). Note also that the timing of the appearance of the turnover depends on the model. Model V112 deviates from the rest at the lowest value of enclosed mass (e.g., a remarkable deviation can be seen around the point of $M \sim 0.9 M_{\sun}$ and $r \sim 23$km for the $\rho = 10^{13} {\rm g/cm^3}$ contour).

\begin{figure*}
\vspace{15mm}
\epsscale{1.0}
\plotone{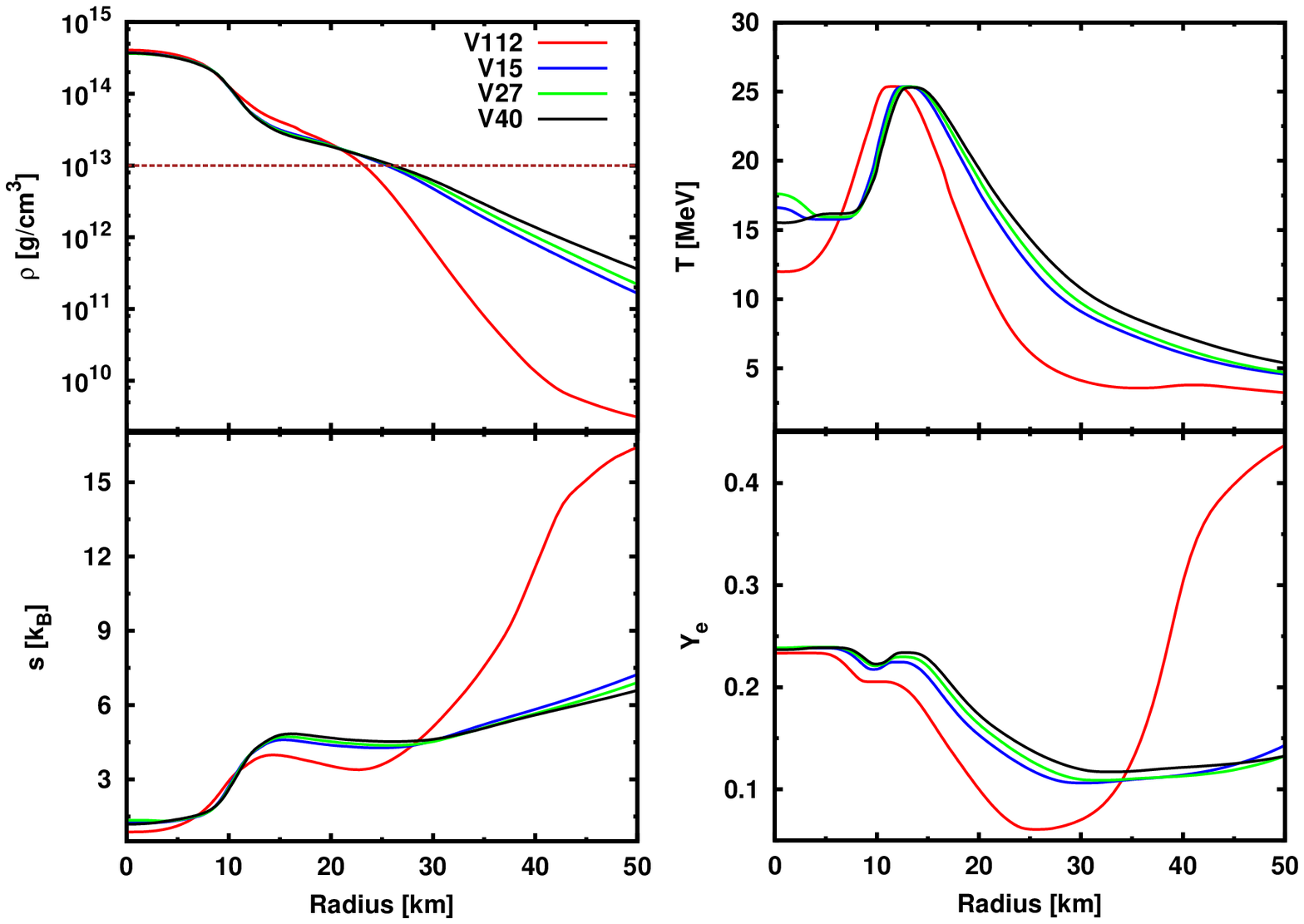}
\caption{Radial profiles of density (top left), temperature (top right), entropy per baryon (bottom left) and electron fraction (bottom right) in the post-bounce phase of V112, V15, V27 and V40 at times ${\rm T}_{\rm b} = 335$ms, $186$ms, $171$ms, and $154$ms, respectively. For all models, the enclosed mass at the radius of $\rho = 10^{13} {\rm g/cm^3}$ is $1.25 M_{\sun}$. The brown doted line in the top left panel guides the eye to $\rho = 10^{13} {\rm g/cm^3}$.
\label{graphM2_5e33_rhoc1313_snap}}
\end{figure*}

To understand the cause of the turn-over and the progenitor dependence thereof, we compare the matter profile among models at the time when the enclosed mass for the iso-density surface of $\rho = 10^{13} {\rm g/cm^3}$ reaches $M = 1.25 M_{\sun}$, which are displayed in Fig.~\ref{graphM2_5e33_rhoc1313_snap}. A large mass accretion rate prevents the PNS from cooling (i.e., the PNS evolves adiabatically). When the accretion rate drops, the PNS is able to cool and condense on a diffusion timescale. Since less massive progenitors experience a drop in accretion rate sooner, they break off the universal adiabatic curve first. Indeed, both the entropy (bottom left panel) and electron fraction (bottom right panel) at $ r \sim 20$km in model V112 are lower than those of other models, suggesting that neutrinos have carried away more leptons and thermal energy than in other models. In other words, the PNS has already entered in the cooling phase. On the other hand, other models are still accompanied by a massive outer envelope at $r > 20$km (see top left panel in Fig.~\ref{graphM2_5e33_rhoc1313_snap}), which works to prolong the adiabatic evolution. As shown in these panels, the smaller mass progenitors deviate from the universal line earlier, i.e., the compactness of outer core (also the mass accretion rate) dictates the timing of the turn-over. For lower iso-density surfaces, on the other hand, the turn-over point appears at earlier times than those in $\rho = 10^{13} {\rm g/cm^3}$. This is simply due to the fact that neutrino diffusion timescale is shorter for the less opaque layer.

% Indeed, as shown in the top two panels, the PNS in model V112 shrinks and is more compact than other models.
%Also, since higher density contours are farther inside the PNS, they have a longer diffusion time and depart from the universal adiabatic line later.

\section{Conclusions and Discussion}\label{sec:summary}

We present spherically symmetric CCSNe simulations with full Boltzmann neutrino transport under self-consistent treatment of nuclear abundances between a multi-nuclear EOS and weak interactions.
In most of our simulations, we employ the newly developed VM EOS, which is one of the most up-to-date nuclear EOS, originally developed by \citet{2013NuPhA.902...53T,2017NuPhA.961...78T} and further extended to multi-nuclear treatments by \citet{2017JPhG...44i4001F}. Given baryon density, temperature and proton (electron) fraction, our new EOS table provides us not only other thermodynamical quantities, but also a full distribution of nuclear abundances in nuclear statistical equilibrium. We use these abundances to construct a new weak reaction rate table including electron and positron captures by light nuclei that is consistent with the nuclear abundances in the EOS.

We then carry out CCSN simulations with these detailed physics inputs and study the EOS dependence (in Sec.~\ref{subsec:VMvsRMF}), influence of light nuclei (in Sec.~\ref{subeq:lightdepe}) and progenitor dependence (in Sec.~\ref{subsec:prodepe}) of CCSN dynamics and neutrino signals. The key findings in this study are summarized as follows.
\begin{enumerate}
\item Inconsistent treatments of electron capture rate with nuclear abundances in EOS weaken the EOS dependence of deleptonization and neutrino cooling in the early part of the collapse phase (see radial profiles of $Q_N, Q_e$ and $Y_e$ in Fig.~\ref{rhoc1e11VMvsRMFcompare}).
\item As pointed out by previous studies \citep{2003PhRvL..90x1102L,2003PhRvL..91t1102H}, the appropriate treatment of electron capture by heavy nuclei is important to determine the structure of the supernova core during the collapse. We find that our incomplete-treatments of electron captures of heavy nuclei change the shock radius by up to a few percent (see Fig.~\ref{shockevoVMvsRMF}). Neutrino luminosity and average energy for both $\nu_e$ and $\bar{\nu_e}$ also changed by $\sim 10 \%$ and $\sim 5 \%$, respectively (see Fig.~\ref{Neutrino_VMvsRMFcompare}). 
\item The influence of light nuclei on fluid dynamics in the collapse phase is minimal due to the overwhelming contribution to weak reactions by interactions between neutrinos and heavy nuclei. In the post-bounce phase, the mass fraction of light nuclei starts to be dominant in some post-shock regions and they start to influence both the stalled shock location (by $\sim5\%$), neutrino luminosities (by $\sim 10\%$) and neutrino average energies (by $0.25-1\,\mathrm{MeV}$), which are quantified by comparing between models using different treatments of weak reactions on light nuclei (models V112, NV112 and DV112). Ignoring weak reactions with light nuclei entirely (model NV112) overestimates the shock radius. On the contrary, setting rates of weak reactions with light nuclei (except for alpha particles) to that of their constituent free nucleons (model DV112) underestimates the shock radius. These artifacts are mainly induced by artificial decrease (for NV112) or increase (for DV112) of neutrino opacity. We also find that the influence of light nuclei on neutrino signals is more significant for low energy neutrinos (see Figs.~\ref{graphRemistotcompare_ie5}~and~\ref{graphSpectrumcompare_light_100ms}).
%\item By comparing between models using different treatments of weak reactions on light nuclei (models V112, NV112 and DV112), we show that these prescriptions influence the shock trajectories and neutrino signals. Ignoring weak reactions with light nuclei entirely (model NV112) overestimates the shock radius by a few percent. On the contrary, setting rates of weak reactions with light nuclei  (except for alpha particles) to that of their constituent free nucleons (model DV112) underestimates the shock radius by a few percent. These artifacts are mainly induced by artificial decrease (for NV112) or increase (for DV112) of neutrino opacity. We also quantify the influence on neutrino signals by artificial prescriptions as shown in Fig.~\ref{Neutrino_Lightucleicompare112}.
\item To look into the influence of light nuclei more in detail, we additionally perform steady state simulations of neutrino radiation field given the fluid background. We show that the difference between heating from the dynamical neutrino field and one that is allowed to relax to a steady state on a single fluid snapshot is much smaller than differences in the neutrino heating imparted by artificial prescriptions of light nuclei (see Fig.~\ref{graphQe_diffromsteady}). This must be interpreted carefully, since the large differences from artificial prescriptions of light nuclei are largely due to the fact that they cause differences in the evolution of the matter, while the steady-state calculations use the same matter profile as the dynamical ones.
\item Different progenitors evolve to a common inner core density and electron fraction structure during the collapse phase, despite the fact that the reaction rate is dominated by the electron capture of heavy nuclei and different progenitors have different abundances of heavy nuclei in the inner core. It should be noted, however, that the temperature difference continues to remain up to core bounce, which causes a slight difference in the location for the formation of the shock at core bounce (see Fig.~\ref{Bounceprogenitorcompare_s3}). This universality persists beyond core bounce until deleptonization and neutrino cooling influence on the structure of PNS. Departures from the universal structure appear earlier at larger radii and in models with less compact outer cores (see Sec.~\ref{subsec:prodepe}).
\item The time evolution of the shock wave in the $11.2 M_\odot$ model (V112) between 50 and 150ms after bounce is very different from other more massive progenitor models. This is attributed to the fact that the Si/Si-O interface in model V112 hits the shock wave while the prompt shock is still expanding. On the other hand, for other massive progenitor models, the interface passes through the shock wave after the shock stagnates. At this phase, the time evolution of shock wave is dictated by the balance between the ram pressure from mass accretions and neutrino heating. The sudden drop of mass accretion rate due to the arrival of the Si/Si-O interface at shock wave causes not only weakening the ramp pressure by mass accretions but giving the negative feedback to neutrino luminosity, which results in less impact to the shock dynamics than that in V112. It should be noted that the shock wave in V112 quickly returns to match the shock radius of the other progenitors at $\sim 150 {\rm ms}$ after the bounce. See a main text in Sec.~\ref{subsec:prodepe} for more details.
\item The EOS dependence of neutronization burst is overwhelmed by the progenitor dependence. On the other hand, the strong progenitor dependence of neutrino signals appears in the late phase (see Fig.~\ref{Neutrino_progenitorcompare}). 
\end{enumerate}

One exciting possibility is that neutrino signals could be used to extract information about the nuclear EOS. However, we show in this paper that the progenitor dependence of the $\nu_e$ luminosity in the neutronization burst is comparable to or even overwhelms the difference in $\nu_e$ luminosity between models using the VM and FYSS EOSs (see Sec.\ref{subsec:VMvsRMF}). It is important to emphasize that such a quantitative argument for the neutrino signals is only possible by using simulations with a consistent treatment of the EOS and nuclear weak interactions. Indeed, the smaller progenitor dependence of the neutonization burst in previous studies (see e.g., \citet{2005PhRvD..71f3003K}) may be due to the artificially common weak reaction treatments. Thus observations of the neutronization burst alone, even if the effects of neutrino oscillations can be disentangled from the supernova dynamics, are unlikely to shed light on properties of the nuclear EOS.

It should be noted, however, that the neutrino signals shown in Fig.~\ref{Neutrino_progenitorcompare} signals could potentially be useful for extracting the matter profile of progenitors, since the luminosities and average energies trace the accretion history. For instance, model V112 shows a remarkably lower neutrino luminosity and average neutrino energy during the middle and late post-bounce phase. According to the recent multi-D simulations (see e.g., \citet{OConnor:2015rwy,2017ApJ...850...43R}), this dependence of neutrino signals on progenitor properties is retained at least up to the time of shock revival. Therefore, observations of the luminosity and average energy could help pin down the mass and compactness of the progenitor, but more work is required to assess whether this trend is robust and well-junderstood enough to interpret an isolated CCSN signal. 
%If more realistic models show similar trends, 

 However, care must be taken to disentangle neutrino signal changes due to the Si/Si-O interface passing through the shock from those due to shock revival, since they have similar characteristics (see also \citet{2016ApJ...825....6S,2018arXiv180400689S}). To understand the degeneracy, we need accurate multi-dimensional models that exhibit successful shock revival, along with an understanding of the effects neutrino oscillations have on the neutrino signal at Earth and the characteristics of neutrino detectors. We are currently running axisymmetric CCSNe simulations with our up-to-date input physics and address these issues. The results of this study will lay the foundation of analyzing these complicated multi-dimensional modeling of CCSNe.

Last but not least, a consistent treatment of the EOS and nuclear weak interactions is important for carrying out accurate nucleosynthesis computations in particular for the $\nu$-process which occurs during PNS cooling phase. This process causes spallations of nucleons from heavy nuclei including r-process elements, which influences the abundances of ejected isotopes (see e.g., \citet{1990ApJ...356..272W}). We will study the dependence of nucleosynthesis on the EOS and weak interactions more quantitatively in the future work.

\acknowledgments 
 We acknowledge to Adam Burrows, Christian D. Ott, Andre da Silva Schneider, Tomoya Takiwaki and Satoshi X. Nakamura for fruitful discussion. The numerical computations were performed on the supercomputers at K, at AICS, FX10 at Information Technology Center of Tokyo University. We also acknowledge to Cray XC40 at YITP of Kyoto University, and SR16000 at KEK under the support of its Large Scale Simulation Program (16/17-11), Research Center for Nuclear Physics (RCNP) at Osaka University and the PC cluster at the Center for Computational Astrophysics, National Astronomical Observatory of Japan for code development and checks. Large-scale storage of numerical data is supported by JLDG constructed over SINET4 of NII. H.N. was supported in part by JSPS Postdoctoral Fellowships for Research Abroad No. 27-348, Caltech through NSF award No. TCAN AST-1333520 and Princeton University through DOE SciDAC4 Grant DE-SC0018297 (subaward 00009650). S.R. was supported by the N3AS Fellowship through the National Science Foundation, Grant PHY-1630782, and the Heising-Simons Foundation, Grant 2017-228. This work was also supported by Grant-in-Aid for the Scientific Research from the Ministry of Education, Culture, Sports, Science and Technology (MEXT), Japan (15K05093, 24103006, 24740165, 24244036, 25870099, 26104006, 16H03986, 17H06357, 17H06365), HPCI Strategic Program of Japanese MEXT and K computer at the RIKEN (Project ID: hpci 160071, 160211, 170230, 170031, 170304, hp180179, hp180111) and the RIKEN iTHEMS Project.
\bibliography{bibfile}

\end{document}